# Hellings and Downs correlation of an arbitrary set of pulsars

Bruce Allen

*Max Planck Institute for Gravitational Physics (Albert Einstein Institute), Leibniz Universität Hannover, Callinstrasse 38, D-30167, Hannover, Germany*

Joseph D. Romano

*Department of Physics and Astronomy, Texas Tech University, Lubbock, Texas 79409-1051, USA*

Pulsar timing arrays (PTAs) detect gravitational waves (GWs) via the correlations they induce in the arrival times of pulses from different pulsars. We assume that the GWs are described by a Gaussian ensemble, which models the confusion noise produced by expected PTA sources. The mean correlation $h^2 \mu_u(\gamma)$ as a function of the angle $\gamma$ between the directions to two pulsars is predicted by Hellings and Downs in 1983. The variance $\sigma_{tot}^2(\gamma)$ in this correlation was recently calculated [B. Allen, Variance of the Hellings-Downs correlation, Phys. Rev. D **107**, 043018 (2023)] for a single noise-free pulsar pair at angle $\gamma$, which shows that after averaging over many pairs, the variance reduces to an intrinsic cosmic variance $\sigma_{cos}^2(\gamma)$. Here, we extend this to an *arbitrary* set of pulsars at specific sky locations, with pulsar pairs binned by $\gamma$. We derive the linear combination of pulsar-pair correlations which is the optimal estimator of the Hellings and Downs correlation for each bin, illustrating our methods with plots of the expected range of variation away from the Hellings and Downs curve, for the sets of pulsars monitored by three active PTA collaborations. We compute the variance of and the covariance between these binned estimates, and show that these reduce to the cosmic variance and covariance $s(\gamma, \gamma')$ respectively, in the many-pulsar limit. The likely fluctuations away from the Hellings and Downs curve $\mu_u(\gamma)$ are strongly correlated/anticorrelated in the three angular regions where $\mu_u(\gamma)$ is successively positive, negative, and positive. We also construct the optimal estimator of the squared strain $h^2$ from pulsar-pair correlation data. Remarkably, when there are very many pulsar pairs, this determines $h^2$ with *arbitrary* precision because (in contrast to LIGO-like GW detectors) PTAs probe an *infinite* set of GW modes. To assess if observed deviations away from the Hellings and Downs curve are consistent with predictions, we propose and characterize several $\chi^2$ goodness-of-fit statistics. While our main focus is ideal noise-free data, we also show how pulsar noise and measurement noise can be included. Our methods can also be applied to future PTAs, where the improved telescopes will provide larger pulsar populations and higher-precision timing.

## I. INTRODUCTION

A pulsar timing array (PTA) is a galactic-scale gravitational-wave (GW) detector. It searches for low-frequency (nanohertz) GWs by precisely monitoring the arrival times of pulses from a set of pulsars [1]. Gravitational waves (e.g., from inspiraling supermassive black-hole binaries in the centers of merging galaxies) influence the pulse arrival times in a way that is correlated between different pulsars. The mean correlation between a pair of pulsars depends upon the angular separation $\gamma$ between the lines of sight to each member of the pair, as seen from Earth [2].

The mean correlation was calculated by Hellings and Downs [3] for a unit amplitude, isotropic and unpolarized GW background. It has the simple analytic form [4]

$$\mu_u(\gamma) = \frac{1}{3} - \frac{1}{6}\left(\frac{1-\cos\gamma}{2}\right) + \frac{1-\cos\gamma}{2}\ln\left(\frac{1-\cos\gamma}{2}\right),$$

(1.1)

where $\gamma$ is the angular separation between a pair of pulsars, and "u" means "unpolarized." Observation of a correlation proportional to this Hellings and Downs curve is the "smoking gun" signature that a PTA has detected GWs [5].

Several groups are searching for such correlations. These groups report strong statistical evidence for fluctuations in the *individual* pulsar arrival times that share the same "red" spectrum [6–9] that a GW background is expected to produce [10]. However, there is currently little evidence for the Hellings and Downs angular dependence, which makes it difficult to claim that a GW background is responsible.

Could the lack of evidence for the Hellings and Downs angular dependence be a statistical fluctuation? To answer this, it is important to understand what variations away from the Hellings and Downs predicted mean might be expected. The size of such fluctuations is quantified by the variance of the Hellings and Downs correlation.

### A. Total variance, pulsar variance, and cosmic variance

Recent work [11] calculates this variance for several GW source models, neglecting all sources of noise. The most important model contains $N$ unpolarized point sources, uniformly distributed (statistically) in space, radiating GWs at the same frequency but with independent random phases. For large numbers of sources, this creates stationary "confusion noise." It is a Gaussian stochastic process [12–14] provided that the amplitude of the individual sources vanishes as $N \to \infty$.

The (total) variance $\sigma_{\text{tot}}^2$ computed in [11] is a sum of "pulsar variance" and "cosmic variance" $\sigma_{\text{cos}}^2$. Both types of variance have been observed in simulations based on synthetic catalogs of sources [15]. Pulsar variance arises because different pairs of pulsars separated by the same angle $\gamma$ have correlations that differ from the average, in a way that depends (unpredictably) upon their sky positions. Reference [11] computes the total variance (there denoted by $\sigma^2$ *without* a subscript) for a *single randomly selected pulsar pair* separated by angle $\gamma$. In contrast, the cosmic variance is the variance of the correlation *after* the correlation has been averaged over all possible locations/ orientations of the pair. This corresponds to employing an *infinite number of pulsar pairs* separated by angle $\gamma$, uniformly distributed about the sky, and is called *pulsar averaging.*

In the confusion-noise model (Sec. III A in [11]), cosmic variance arises because, even after pulsar averaging, the correlation depends upon the relative phases of the GW sources. Each realization of the Universe has different phases, and thus exhibits different pulsar-averaged correlations. The cosmic variance is the amount by which the pulsar-averaged correlation curve is expected to differ from the Hellings and Downs prediction in an ideal world containing an infinite number of noise-free pulsars. Unlike the pulsar variance, it cannot be reduced: the cosmic variance is a fundamental limit to the precision with which the Hellings and Downs predicted mean might be observed at a particular angle $\gamma$.

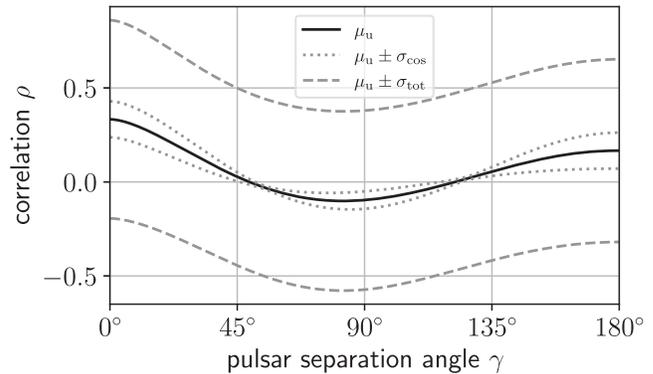

FIG. 1. The mean Hellings and Downs correlation $\mu_{\text{u}}$ as a function of the angular separation $\gamma$ between a pair of pulsars. Also plotted are $\mu_{\text{u}} \pm \sigma_{\text{tot}}$ and $\mu_{\text{u}} \pm \sigma_{\text{cos}}$, where $\sigma_{\text{tot}}^2$ is the total variance for a single pulsar pair and $\sigma_{\text{cos}}^2$ is the cosmic variance. (For this plot, we have set $\hbar^4/h^4 = 1/2$ and $h^2 = 1$; see text for details of the GW source model.) Note that although the correct term is "standard deviation," we sometimes call $\sigma$ the "variance".

Figure 1 illustrates the total and cosmic variance, as offsets away from the mean $\mu_{\text{u}}$ of the Hellings and Downs correlation. The total variance at angle $\gamma$ is the uncertainty associated with the determination of the Hellings and Downs correlation when a single (randomly selected) pulsar pair at angle $\gamma$ is used to estimate that correlation. The cosmic variance at angle $\gamma$ is the uncertainty that remains when an infinite number of uniformly distributed pulsar pairs at angle $\gamma$ are used to do the estimation. *In this paper, we study the transition between these two limits, when a finite set of pulsars at specific sky locations are used to estimate the correlation.* This reflects observational reality because PTA pulsars are nonuniformly distributed on the sky, so the pairs formed from them have no separation angles in common.

The plots in Fig. 1 assume that the GWs arise from the incoherent sum of many weak sources, giving rise, via the central-limit theorem, to a Gaussian ensemble [12–14]. For such sources, the scaling relation between the (squared) mean and the variance is described in Appendix B; these plots take $\hbar^4/h^4 = 1/2$ and $h^2 = 1$. This corresponds to the large source-number limit of the "narrow-band" discrete confusion-noise model (Sec. III A in [11]), where the source frequency is assumed to be commensurate with the inverse observation time.

### B. Variance of the Hellings and Downs correlation for an arbitrary set of pulsars

We assume that the reader is familiar with Ref. [11], and extend that analysis in five ways:

(i) First, we define the Hellings and Downs correlation [3] for *many* pulsar pairs. We have found only one sensible way to generalize the standard "single-pulsar-pair" definition. We partition the angular range $\gamma \in [0, 180°]$ into nonoverlapping intervals called bins. Then, we define the correlation associated with a bin $[\gamma_{\text{low}}, \gamma_{\text{high}}]$ to be an average of the Hellings and Downs correlations $\rho_{ab}$ for the $n_{\text{pairs}}$ pulsar pairs that have $\gamma_{\text{low}} < \gamma_{ab} < \gamma_{\text{high}}$, where the subscripts $a$ and $b$ label pulsars, and $\gamma_{ab}$ are their angular separation on the sky. We say that these pulsar pairs "fall in" or "lie in" or "are in" the bin. If there is only a single pulsar pair in the bin, then our definition reduces to the standard Hellings and Downs definition.

What type of average should we employ? As shown in Fig. 2, a uniform average, where the correlation of each pulsar pair gets the same weight, $1/n_{\text{pairs}}$, is not the best choice. We define the Hellings and Downs correlation for a given bin to be the weighted sum of the pulsar-pair correlations $\rho_{ab}$ in that bin which is (1) unbiased and (2) minimizes the expected variance.

In this paper, that weighted average (for a particular bin) is called the "optimal estimator." In several places, we also consider a special case, which we call the "narrow-bin limit." This is the (mathematical) limit in which the width (in $\gamma$) of an angular-separation bin vanishes, while the number of pulsar pairs in that bin remains constant or grows.

Conditions (1) and (2) lead immediately to a simple formula for the weights. The weights depend upon the sky directions to all of the pulsars which contribute to a particular bin, and are easily computed for any specific set of directions. In the limit of low pulsar noise, the resulting optimal estimator tells an observer (in a universe described by the Gaussian ensemble) the best way to combine measured pulsar-pair correlations to estimate the Hellings and Downs correlation in a particular bin.

As a byproduct, we also obtain simple formulas for (a) the variance of the optimal estimator for a given angular separation bin, and (b) the covariance of the optimal estimators for any pair of bins. *By definition* these are the variance and covariance of the Hellings and Downs correlation. In particular, the variance, which by condition (2) is as small as possible, quantifies the uncertainty in the Hellings and Downs correlation for any particular bin. The variance depends upon the sky directions to all of the pulsars that contribute to that bin.

To illustrate this, Fig. 3 shows the variance obtained by using the 88 pulsars currently employed by three active PTAs, after dividing the $88 \times 87/2 = 3828$ possible distinct pulsar pairs among 30 evenly spaced 6° bins. If the Gaussian ensemble is a good description of the GWs in our Universe and the correlation measurements are noise-free, then this shows the expected deviations away from the Hellings and Downs mean. Thus, it is an upper limit on the ability to which that set of 88 pulsars could be expected to recover the Hellings and Downs curve. Similar plots for the three individual PTAs (employing only "their" pulsars) are given in Sec. V B.

(ii) Second, we extend [11] by showing how the variance and covariance of the expected Hellings and Downs correlation decrease as pulsar pairs are added. For example, the variance of the optimal estimator for narrow bins drops from the total variance $\sigma_{\text{tot}}^2(\gamma)$ for a single pulsar pair to the cosmic variance $\sigma_{\text{cos}}^2(\gamma)$ for an infinite number of pairs (distributed uniformly on the sky). This transition is illustrated later in Fig. 6.

Our approach to deriving the cosmic variance is different than that given in Sec. IV of [11], and provides a useful alternative interpretation. Because different pulsar-pair correlations are correlated with one another, as pulsar pairs are added to a bin, the variance does not approach zero. The reason is straightforward: once there are a sufficient number of pulsar pairs in the bin, adding more pairs to that bin provides negligible additional information about the GW-induced correlations. (Section III E demonstrates this behavior with a simple example.)

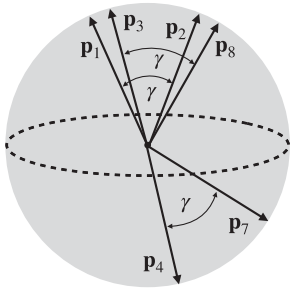

FIG. 2. The bin at angular separation $\gamma$ contains $n_{\text{pairs}} = 3$ pulsar pairs with $\gamma \approx \gamma_{12} \approx \gamma_{38} \approx \gamma_{47}$. The naive correlation estimator $\rho_{\text{naive}} = (\rho_{12} + \rho_{38} + \rho_{47})/3$ weights them uniformly, but is suboptimal. The optimal estimator $\rho_{\text{opt}} = 0.3\rho_{12} + 0.3\rho_{38} + 0.4\rho_{47}$ gives more weight to the correlation $\rho_{47}$. This is because the 12 and 38 pairs are close on the sky, so $\rho_{12}$ and $\rho_{38}$ give nearly redundant estimates. The optimal estimator *defines* the Hellings and Downs correlation of that bin. Note: for large numbers of pairs uniformly distributed about the sky, rotational symmetry implies that all pairs have the same weight.

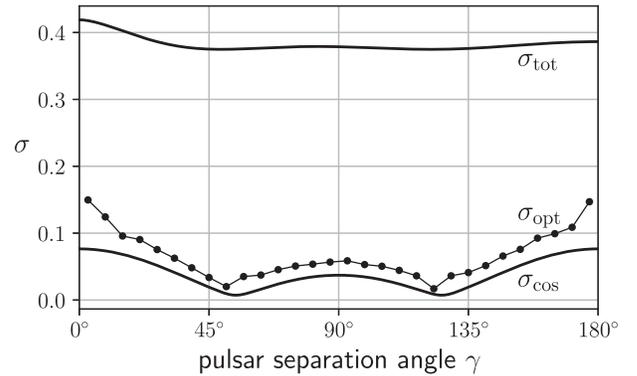

FIG. 3. The dots (lines are to guide the eye) show the predicted GW-induced variance $\sigma_{\text{opt}}^2$, for the current set of 88 IPTA pulsars (assumed free of noise) after optimally combining timing-residual correlations in $30 \times 6°$ angular-separation bins. This reduces the variance far below the single pulsar pair (total) variance $\sigma_{\text{tot}}^2$, bringing it close to the cosmic variance $\sigma_{\text{cos}}^2$. The dots indicate the scale of the expected fluctuations away from the Hellings and Downs curve, for a Gaussian ensemble GW background with the binary-inspiral spectrum described in Appendix B. This plot has the same scale and assumptions as the IPTA plot in Fig. 9, i.e., $\alpha = 1$ for timing residuals and $\hbar^2 \approx 0.5622$. See Sec. V B and Appendix H for more details.

We also derive an elegant analytic expression for the covariance of the optimal estimators in narrow angular separation bins labeled by $j$ and $k$, in the limit where each contains an infinite number of pulsar pairs. This "cosmic covariance matrix" is not given in [11], nor have we found it elsewhere in the literature. We give it both in "position-space" form and in harmonic space, where it is a diagonal sum of products of Legendre polynomials, dominated by the quadrupole plus a few additional multipoles. The cosmic covariance has a surprisingly simple structure, showing that the expected fluctuations away from the Hellings and Downs mean correlation are *strongly correlated/anticorrelated* in the three $\gamma$-angle regions where the Hellings and Downs curve is successively positive, negative, and positive.

(iii) Third, we address the question: "What is the strength of the GW stochastic background in the PTA band?" For simplicity, we assume that the *frequency-dependent shape* of the GW spectrum is known (e.g., $\propto |f|^{-7/3}$ for binary-inspiral sources). The goal is then to estimate the overall scale, which is the squared GW strain $h^2$.

We construct optimal estimators of $h^2$ that are unbiased and have minimum variance. These estimators are linear combinations of the measured pulsar-pair correlations $\rho_{ab}$. We consider three different possible choices of the correlation set: (a) auto+cross: use all possible pairs of pulsars; (b) cross only: use all pulsar pairs where the two pulsars differ; and (c) auto only: use all pulsar pairs where the two pulsars are the same. We then derive the variance (squared uncertainty) in these estimators of $h^2$. As the number of pulsars increases, these uncertainties get smaller. Their scaling behavior, as a function of the number of pulsars, depends upon which correlation set is used.

We show that if cross-correlations are included in the squared-strain estimator, then the variance tends to zero as $N_{\mathrm{pul}} \to 0$. Initially, we were surprised by this: in the same limit, the Hellings and Downs correlation has a nonzero cosmic variance, as do local estimates of the squared strain, such as $h_{\mu\nu} h^{\mu\nu}$. We show that this remarkable behavior arises because a pulsar's response to GWs allows a PTA to probe infinitely many modes of the GW field.

(iv) Fourth, to enable straightforward tests for consistency with the predictions of the Gaussian ensemble, we construct a set of $\chi^2$ goodness-of-fit statistics. These can be used to assess if observational measurements of the Hellings and Downs correlation are consistent with the expected deviations away from $h^2 \mu_{\mathrm{u}}(\gamma)$, based on our predictions of the variance and covariance for the Gaussian ensemble. In defining these $\chi^2$ statistics, there are three possible choices for which pulsar-pair correlations to employ (as explained above in the context of estimating $h^2$) and two possible choices for how the overall scale of the GW strain is determined. By taking all possible combinations, we arrive at six different statistics.

To set the scale of the GW strain, there are two options: (1) assume that $h^2$ has been obtained via other methods/data, or (2) obtain $h^2$ from the goodness-of-fit statistic, via a process we call "projection," which minimizes $\chi^2$. The justification for this choice of name, and the six different options, are discussed in detail in Sec. VIII. Note that the projection process leads to the same $h^2$ estimators as discussed in item (iii) above.

The $\chi^2$ statistics are a "sum of squares"; we denote them with the same symbol as used for the traditional $\chi^2$ test, whose distribution describes the sum of squares of normal Gaussian variables. However, the distributions of our $\chi^2$ statistics and the standard $\chi^2$ statistics differ in an important way: the random variables whose squares are summed are not Gaussian, even though we assume that the GW stochastic background is described by the Gaussian ensemble. The $\chi^2$ statistics that we define have the same mean values as the standard distribution, but they have much larger variances because the GW background creates non-Gaussian fluctuations in the pulsar-pair correlations.

(v) Our final extension to [11], which assumes noise-free data, is to also consider the effects of pulsar and measurement noise. While most of our paper considers ideal data, noise is easily included. We do this starting in Sec. IX, assuming that the noise is uncorrelated between different pulsars. The noise modifies the pulsar-pair correlations and thus adds additional terms to the pulsar-pair expectation value, covariance matrix, and higher-order moments. This means that our entire approach and methodology can still be used in the presence of noise.

The noise contributions to the pulsar-pair correlations are fundamentally different than those which are induced by the GWs. For the GW-induced correlations, as the number of pulsar pairs increases, the variance in the Hellings and Downs correlation approaches the (nonzero) cosmic variance. In contrast, since we assume that the different pulsars have independent noise, the contribution of their noise to the variance of the Hellings and Downs correlation *vanishes* as the number of pulsar pairs increases. Thus, there are reasons to hope that during the coming decades, as PTAs add additional pulsars and improve their measurements, the limit of the cosmic variance can be achieved in practice.

## C. Outline

Here, we give a brief outline of the paper, and introduce the most important notation. Note that our methods and results apply both to pulsar redshifts (set $\alpha = 0$) and to pulsar timing residuals (set $\alpha = 1$). See Appendix A for details.

We begin in Sec. II by defining the time-averaged correlation $\rho_{ab}$ between the redshifts of two pulsars $a$ and $b$, where the subscript denotes the pulsar pair $ab$. For the Gaussian ensemble, in Sec. II A we calculate the first moment $\langle \rho_{ab} \rangle$ of the correlation. For two distinct pulsars (meaning $a \neq b$) this is proportional to the Hellings and Downs curve $\mu_u(\gamma_{ab})$ given in (1.1), where $\gamma_{ab}$ denotes the angle between the directions to the two pulsars. To gracefully accommodate the case $a = b$, for which the "pulsar term" doubles the correlation, we also define the closely related quantity $\mu_{ab}$, cf. (2.4). In Sec. II B, we use this to compute the "second moment" $\langle \rho_{ab}\rho_{cd} \rangle$. From the moments, we construct the covariance matrix $\mathbb{C}_{ab,cd} \equiv \langle \rho_{ab}\rho_{cd} \rangle - \langle \rho_{ab} \rangle \langle \rho_{cd} \rangle$, where $cd$ denotes a second pulsar pair constructed from pulsars $c$ and $d$. Third and fourth moments of $\rho_{ab}$ are computed in Sec. II C and Appendix F.

In this paper, angle brackets $\langle Q \rangle$ denote the average of a function or functional $Q$. This average is always normalized so that $\langle 1 \rangle = 1$. If not explicitly stated otherwise, then this is an average over the Gaussian ensemble of GW realizations [12–14]. However, in some places we (state that we) use it differently. For example, in (9.2) it denotes an average over a Gaussian ensemble of noise sources *and* GW sources. In (4.4) it denotes an average over pulsar pairs at fixed separation angles (indicated by the subscript $cd \in \gamma_k$ on the rightmost angle bracket).

The covariance matrix $\mathbb{C}$ is the most important quantity in our analysis, and is used throughout. It is a square, real, positive-definite symmetric matrix. We denote it by $\mathbb{C}$ when referring to the matrix as a whole, and use indices $\mathbb{C}_{ab,cd}$ to indicate a particular entry of the matrix. In Sec. II D, we derive two identities satisfied by the entries $\mathbb{C}_{ab,cd}$, which are used later to obtain the cosmic variance limit.

The way that $\mathbb{C}$ is laid out, i.e., the indexing of its rows and columns, is discussed in Sec. II E. Ordering pulsar pairs $ab$ by their angular separation $\gamma_{ab}$ puts the covariance matrix $\mathbb{C}$ into "block-matrix form", cf. (2.14). The different blocks correspond to different angular bins, which we label with integers $j$ and $k$. The submatrix $C_{jk}$ (we use the term "block" or "matrix block") of the covariance matrix has rows (columns) corresponding to all pulsar pairs $ab$ ($cd$) that lie in the $j$th ($k$th) angular bin.

Pulsar pairs $ab$ and $cd$ that both lie in the same angular-separation bin $j$ have particular importance. The covariance between such pairs is described by blocks which lie along the diagonal of $\mathbb{C}$. To simplify the formulas and calculations for these, we often drop the indices, denoting that matrix (block) by $C = C_{jj}$.

In Sec. III, we turn our attention to defining the Hellings and Downs correlation for an arbitrary collection of pulsars. This is defined as the optimal estimator of the correlation in a particular angular-separation bin, which is a weighted average of pulsar-pair correlations in that bin: $\rho_{\text{opt}} = \sum_{ab} w_{ab}\rho_{ab}$. In Sec. III A, we derive the weights $w_{ab}$ by requiring that the estimator be unbiased and have the minimum possible variance. In Sec. III B, we calculate the covariance $B_{jk}$ between the optimal estimates in two angular separation bins $j$ and $k$. (Later, this plays an important role in Sec. VIII, where we define $\chi^2$ goodness-of-fit tests. These tests assess observed correlations, to see if they are consistent with the likely fluctuations away from the Hellings and Down mean.) In Sec. III C and III D, we examine the narrow-bin limit of the optimal estimator, and show how the variance of the optimal estimator simplifies if the vector $\mathbb{1}$ containing all ones is an eigenvector of the corresponding block $C$ of the full covariance matrix $\mathbb{C}$. We conclude this section with a simple pedagogical example: Sec. III E shows how an infinite collection of measurements can have a nonzero variance.

Using this framework, Sec. IV investigates the behavior of the binned variance and covariance matrix $B_{jk}$, as the number of pulsar pairs grows. In [11], the variances at the two extremes are called the "total variance" (one pair) and the "cosmic variance" (many pairs). We begin in Sec. IV A by averaging the entries $\mathbb{C}_{ab,cd}$ of $\mathbb{C}$ over a large number of pulsar pairs $ab$ and $cd$ lying in narrow angular-separation bins $j$ and $k$, respectively. In the limit of an infinite number of pairs distributed uniformly on the sky, we obtain, in Sec. IV B, an analytic expression, (4.11), for the *cosmic covariance matrix* $s_{jk}$. In Sec. IV C, we show that $s_{jk}$ is diagonal when expressed in terms of products of Legendre polynomials, (4.15), and calculate its inverse, $s_{jk}^{-1}$. This inverse is used later on, in Sec. VII C, where we characterize the (fractional) variance of cross-correlation estimators $\hat{h}^2$ of the squared GW strain, (7.1). Finally, in Sec. IV D, we obtain the *cosmic variance* computed in [11] by simply setting the bins' indices, $j$ and $k$, of the cosmic covariance matrix equal to one another. In this many-pulsar-pair, narrow-bin limit, $\mathbb{1}$ becomes an eigenvector of the diagonal block $C_{jj}$ of the covariance matrix $\mathbb{C}$, corresponding to the (narrow) angular-separation bin at $\gamma_j$. Its eigenvalue is simply related to the average value $s_{jj}$ of the entries of that block. In Sec. IV E, we numerically examine the narrow-bin limit, demonstrating that as pulsar pairs are added, the variance of the optimal estimator decreases smoothly from the (single-pair) total variance to the (infinite-pair) cosmic variance. This approach to deriving the cosmic variance differs from that given in [11], but the results are in exact agreement.

Current PTAs do not have enough pulsar pairs to be well described by the narrow-bin limit. In Sec. V, we study the variance for this case. Since the bins have nonzero width in $\gamma$, some angular resolution is lost. But in Sec. V A we show that there is a compensating benefit: the variance of the optimal estimator can be *smaller* than the cosmic variance. In Sec. V B, we apply our formalism to the sets of pulsars currently monitored by three active PTAs. With plots, we show how these sets reduce pulsar variance, illustrating (for the ideal noiseless case) how closely the variance approaches the cosmic variance.

In Sec. VI, for pedagogic purposes, we introduce hypothetical "non-PTA" detectors, whose response to GWs differ from that of PTAs. (Among the infinite set of possible detector responses, standard PTAs are simply a particular special case.) This is helpful in understanding the statistical properties of estimators of the squared GW strain $h^2$ (discussed in Sec. VII). Non-PTA detectors include, for example, hypothetical "one-arm LIGO-like" detectors, whose response to an isotropic GW background is described in Sec. VI A. Expressing the correlation matrix for these hypothetical detectors in terms of its eigenvectors and eigenvalues (Sec VI B) allows us to construct its inverse or pseudoinverse (Sec. VI C), from which we can calculate quantities of interest.

In Sec. VII, we study the statistical properties of estimators $\hat{h}^2$ of the squared GW strain $h^2$. We assume that the shape of the GW spectrum $H(f)$ is known, but that the overall scale $h^2$ must be estimated from the observed pulsar-pair correlation data. The estimators are defined to be weighted sums of the observed correlations that are unbiased and have minimum variance. We consider three different correlation sets for constructing these estimators: cross-correlations only, auto-correlations only, or auto- and cross-correlations (which we denote "auto+cross").

In all cases, the variance $\sigma_{\hat{h}^2}^2$ of these estimators decreases as the number of pulsars $N_{pul}$ increases. But, as shown in Fig. 10, the rate of falloff depends upon which types of correlation measurements are included in the squared strain estimator. Section VII A presents the (mathematically) simplest case, where both auto- and cross-correlations are included in the estimator. A direct calculation shows that as the number of pulsars grows, the variance vanishes as $1/N_{pul}$. In Sec. VII B, we examine the variance estimator formed by using only auto-correlations, showing that it has a nonzero limit as $N_{pul} \to \infty$. Finally, in Sec. VII C, we examine the estimator formed from only pulsar-pair cross-correlations. For this case, as shown in Fig. 13, the variance vanishes as $N_{pul}^{-1/2}$.

Since the Hellings and Downs correlation of the Gaussian ensemble has nonzero cosmic variance, we found it surprising that (if pulsar-pair cross-correlations are employed) there is no cosmic variance for the $h^2$ estimator. To verify this, in Sec. VII C, we prove that $\sigma_{\hat{h}^2}^2 \to 0$ for an infinite number of pulsars distributed uniformly on the sky, using the analytic expression of the inverse $s_{jk}^{-1}$ of the cosmic covariance matrix $s_{jk}$ derived in Sec. IV C. Then, we show that this limiting behavior of $\sigma_{\hat{h}^2}^2$ has a good physical explanation. It is a consequence of the way that pulsar redshifts and timing residuals respond to a GW background, which [as shown by (6.4)] has contributions from an infinite number of (harmonic decomposition) modes. This contrasts with conventional detectors, such as the LIGO interferometers. Because the arms of LIGO-like detectors are small compared to the GW wavelength [16] they only respond to the five $l = 2$ quadrupole modes [see the discussion in Sec. VI following (6.4)].

In Sec. VII D, we calculate the fractional uncertainty $\sigma_{\hat{h}^2}/\langle \hat{h}^2 \rangle$ in the squared-strain estimator for the sets of pulsars currently monitored by three active PTA collaborations. These "best-case" lower limits (pulsar noise is neglected) are given in Table I. We also demonstrate a graphical method for constructing a "self-consistent interval" for $h^2$, based on the observed value of the estimator $\hat{h}^2$. Some care is required, because the variance $\sigma_{\hat{h}^2}^2$ depends upon $h^2$, whose true value we are trying to estimate. We illustrate this for a $1\sigma$ interval, but the construction may be carried out at any desired level of confidence.

In Sec. VIII, we turn our attention to $\chi^2$ goodness-of-fit tests. To interpret observational claims, the scientific community must assess if observed deviations away from the expected Hellings and Downs curve are consistent with expectations. Assuming that the correlation data have been binned and optimally combined, we propose several $\chi^2$ tests to carry out that assessment. The $\chi^2$ statistics, described in Secs. VIII A and VIII B, can either use "external" estimates of the squared GW strain $h^2$ (e.g., from other analysis or other PTAs), or they can construct their own estimates of this scale, by minimizing $\chi^2$. The latter leads to the same estimators $\hat{h}^2$ that we derived and characterized in Sec. VII.

The $\chi^2$ tests are meant to replace "chi by eye" for comparing observational results with theoretical expectations. Those observational results are typically presented as Hellings and Downs curves with a small number of angular bins, so our tests are designed for those binned correlations. However, the most stringent $\chi^2$ tests are the ones with the largest possible number of (nonempty) angular bins. These are obtained by putting each pulsar pair into its own angular separation bin so that $N_{bins} = N_{pairs}$.

In Sec. VIII C, we investigate the variance of the $\chi^2$ statistics for this extreme choice ($N_{bins} = N_{pairs}$) of binning. We consider the three possible correlation sets (auto+cross correlations, auto-only, and cross-only) described in Sec. VII. The variance $\sigma_{\chi^2}^2$ depends upon the fourth-order moment of the measured correlations $\rho_{ab}$, which is evaluated in Appendix F. In all cases, the variance is dominated by a non-Gaussian term $\mathbb{E}$, which is given by (F8) in the noise-free case. This term arises because, even though the GW background is Gaussian, the correlations $\rho_{ab}$ are not Gaussian.

For the auto + cross and cross-correlations-only cases, analytic calculations (Sec. VIII C 1) and numerical investigations (Sec. VIII C 3) show that as $N_{\text{pul}} \to \infty$, the fractional uncertainty $\sigma_{\chi^2} / \langle \chi^2 \rangle$ tends to zero $\propto N_{\text{bins}}^{-1/4}$. (Note: this is slower than the $\propto N_{\text{bins}}^{-1/2}$ behavior that would arise if the correlations $\rho_{ab}$ were Gaussian.) In contrast, for auto-correlations only (Sec. VIII C 2) it appears that the limit is positive, meaning that there is a (nonzero) cosmic variance for $\chi^2$. While we have not been able to prove this, some partial results may be found in Appendix G.

In Sec. VIII C 4, we evaluate the (best-case, noise-free) fractional uncertainty in $\chi^2$ for the collections of pulsars currently monitored by three active PTA collaborations. Table II gives $\sigma_{\chi^2} / \langle \chi^2 \rangle$ for the unprojected statistics; the fractional uncertainties for the projected $\chi^2$ statistics are very similar. We also present a graphical method for constructing "self-consistent one-sigma $\chi^2$ acceptance regions" for the projected $\chi^2$ statistics, starting from observed pulsar-pair correlations. Again, some care is required, because the projected $\chi^2$ statistics depend on the squared strain $h^2$, whose value must also be estimated from the observations.

Section IX examines the effects of instrumental and pulsar timing noise, assuming that the noise in different pulsars is uncorrelated. We show how to modify quantities previously computed assuming noise-free measurements. These are (1) the optimal estimator of the Hellings and Downs correlation, (2) the optimal estimator of the squared GW strain $h^2$, and (3) the various $\chi^2$ statistics. The modifications are straightforward, because the noise enters our calculations via the first four moments of the pulsar-pair correlations $\rho_{ab}$. The noise contributions to the first moment $\langle \rho_{ab} \rangle$ are obtained in Secs. IX A and IX B, to the second moment (covariance) $\mathbb{C}_{ab,cd}$ in Sec. IX C, and to the third and fourth moments (cumulants) $\mathbb{D}_{ab,cd,ef}$ and $\mathbb{E}_{ab,cd,ef,gh}$ in Sec. IX E.

In Sec. IX D, we examine how (2) is affected. In contrast to the noise-free case, the optimal estimator $\hat{h}^2$ for the squared GW strain acquires a constant (independent of $\rho_{ab}$) term. Without this, $\hat{h}^2$ cannot be an unbiased estimator if auto-correlation terms are included in the estimate. An additional complication is that the estimator $\hat{h}^2$ then depends upon the (unknown) value of the true strain. We address this with a graphical technique that constructs a self-consistent interval for $h^2$.

Lastly, in Sec. IX F, we show that for large numbers of pulsar pairs, the noise has no effect on the variance of the Hellings and Downs correlation. This relies on a detailed argument presented in Appendix C.

Our definition of the Hellings and Downs correlation, and our approach for computing the weights and corresponding variance, could be applied to any source of GWs. However, for much of the paper, we assume that the GW background arises from a Gaussian ensemble, which implies a special form for the covariance matrix. In Sec. X, we explain the consequence: our predictions for the cosmic variance create a consistency test. If the Hellings and Downs correlation (estimated according to the recipe we present) matches the Hellings and Downs prediction much more (or much less) closely than the variance we compute, then it is very unlikely that our Universe has a GW background which is described by the Gaussian ensemble. As also argued in [11], this shows that the Hellings and Downs variance is an observable quantity: it could (in principle or practice) be used to falsify the widely accepted "Gaussian ensemble" model of the PTA GW background.

This is followed by a short conclusion in Sec. XI, and a number of technical appendices, Appendix A–Appendix H. A summary of their contents may be found prior to the appendices.

Throughout the paper we use units in which the speed of light $c = 1$, meaning that distances are measured in units of time. For example, the distance $\mathcal{L}$ to a typical PTA pulsar is hundreds to thousands of years.

## II. STATISTICS OF PULSAR-PAIR REDSHIFT CORRELATIONS

Consider a PTA built from a collection of $N_{\text{pul}}$ distinct pulsars. We label the pulsars with indices $a$, $b$, $c$, and $d$, which take values in the range $1, 2, \ldots, N_{\text{pul}}$. Associated with each pulsar is a pulse redshift measurement $Z_a(t)$ which is a function of time.

The main objects of our analysis are the correlations between two pulsars. The corresponding *pulsar pair* is labeled with two letters $ab$, $cd$, $ef$, ..., which identify the individual pulsars. There are $N_{\text{pairs}} \equiv N_{\text{pul}}(N_{\text{pul}} - 1)/2$ distinct pairs; we denote the angle between the sky directions to pulsars $a$ and $b$ by $\gamma_{ab}$.

The time-averaged redshift correlation between the pulses arriving from pulsars $a$ and $b$ (that is to say, of the pulsar pair $ab$) is

$$\rho_{ab} \equiv \overline{Z_a Z_b}. \qquad (2.1)$$

Here, the overline denotes an average over the time interval $-T/2 < t < T/2$, where $T$ is the total observation period [17]. For simplicity we take $T$ to be the same for all pulsars.

Note that (a) our calculations, formalism, results, and plots also apply to pulsar timing-residual correlations (as well as to redshifts) as explained in Appendix A. (b) Here, and in most of the paper, we assume noise-free pulsar redshift measurements and thus noise-free pulsar-pair correlations $\rho_{ab}$. However, as shown in Sec. IX, our analyses can be extended to include measurement noise and intrinsic pulsar noise.

In Secs. II A and II B, we compute the first moment $\langle \rho_{ab} \rangle$ of the correlation $\rho_{ab}$, and the covariance matrix of $\rho_{ab}$. The entries of this matrix are defined by the combination of first and second moments

$$\mathbb{C}_{ab,cd} \equiv \langle \rho_{ab} \rho_{cd} \rangle - \langle \rho_{ab} \rangle \langle \rho_{cd} \rangle. \tag{2.2}$$

Note that the pairs $ab$ and $cd$ may not be the same, meaning that $\langle \rho_{ab} \rho_{cd} \rangle$ is the *correlation of the correlations*. To avoid such awkward language, we write "second moment" even when $ab$ and $cd$ denote different pairs.

The angle brackets denote the average over a Gaussian ensemble. As shown in Appendix C of [11], this ensemble corresponds to an infinite collection of weak, unpolarized, time-stationary GW sources, distributed uniformly in space [12–14]. In any narrow frequency band, there are an infinite number of sources radiating at indistinguishably close GW frequencies, but with different (random) phases. This generates "confusion noise," giving rise via the central-limit theorem to a stationary and Gaussian stochastic process.

As explained in Sec. I B, our definition of the Hellings and Downs correlation is based on angular bins. For this reason, it is helpful to arrange the entries of $\mathbb{C}_{ab,cd}$ into "bin order" to form the covariance matrix $\mathbb{C}$. We do this in Sec. II E. First, the pulsar pairs $ab$ are ordered by increasing angular separation $\gamma_{ab}$. Then, the index for those pairs is divided into ranges corresponding to the different angular bins. This puts $\mathbb{C}$ into a useful "block-matrix" form.

### A. First moment

For the unpolarized isotropic stationary Gaussian ensemble, Eq. (C15) in [11] gives the first moment of the correlation as

$$\langle \rho_{ab} \rangle = h^2 \mu_{ab}, \tag{2.3}$$

where

$$\mu_{ab} \equiv \mu_{\mathrm{u}}(\gamma_{ab}) + \delta_{ab} \mu_{\mathrm{u}}(0). \tag{2.4}$$

Here, $\delta_{ab}$ is the Kronecker delta, which vanishes if pulsars $a$ and $b$ are distinct, and is unity if they are identical. The function $\mu_{\mathrm{u}}(\gamma_{ab})$ is the Hellings and Downs curve given by (1.1). Its fundamental underlying definition is [3]

$$\mu_{\mathrm{u}}(\gamma_{ab}) \equiv \frac{1}{4\pi} \int d\hat{\Omega} \sum_A F_a^A(\hat{\Omega}) F_b^A(\hat{\Omega}). \tag{2.5}$$

This is an average of the products of antenna pattern functions $F_a^A(\hat{\Omega})$ and $F_b^A(\hat{\Omega})$ over GW directions $\hat{\Omega}$, where $\hat{\Omega}$ denotes a unit vector on the two-sphere and $A = +, \times$ labels two orthogonal polarization states of the GWs. Definitions of the antenna pattern functions and a full derivation of (1.1) starting from (2.5) are given in Appendix D in [11].

The function $\mu_{\mathrm{u}}(\gamma_{ab})$ is the sky-averaged correlation for *distinct* pulsars $a$ and $b$ separated by angle $\gamma_{ab}$, taking only Earth terms into account. The Kronecker delta which appears in (2.4) for time-stationary GW signals is discussed in detail in Appendix C 2 of [11]. It arises because the expected correlation for a pair of *identical* pulsars $a = b$ is twice the expected correlation for a pair of *distinct* pulsars $a$ and $b$ which lie along the same line of sight, but are at different distances [18] from Earth. If pulsars $a$ and $b$ are identical and the GW sources are (statistically) stationary in time, then the pulsar-pulsar term in the correlation has the same expectation value as the Earth-Earth term, producing a factor of two.

The overall scale of the first moment $\langle \rho_{ab} \rangle$ is determined by the constant

$$h^2 \equiv 4\pi \int_{-\infty}^{\infty} df \, (2\pi f)^{-2\alpha} H(f). \tag{2.6}$$

Set $\alpha = 0$ for redshift correlations, or set $\alpha = 1$ for timing-residual correlations, as explained in Appendix A. For redshifts, $h$ is a dimensionless strain, whereas for timing residuals, $h$ has units of time × (dimensionless) strain.

The real two-sided squared-strain spectral density $H(f)$ is related to the one-sided GW energy-density spectrum $\Omega_{\mathrm{gw}}(f)$ and to the characteristic strain spectrum $h_c^2(f)$ via

$$H(f) = \frac{3 H_0^2}{32\pi^3} \frac{1}{|f|^3} \Omega_{\mathrm{gw}}(|f|) = \frac{1}{16\pi} \frac{1}{|f|} h_c^2(|f|), \tag{2.7}$$

with $H(f) \geq 0$ and $H(f) = H(-f)$, as shown in [11–13].

Note that the derivation of (2.3) assumes that the spectral function $H(f)$ has a coherence time/length which is much less than the typical pulsar-pulsar and Earth-pulsar distances, so that the Earth-pulsar terms which appear in the correlation can be neglected. This should be the case for the expected PTA sources.

### B. Second moment and covariance matrix

For a Gaussian ensemble, the second moment of the pulsar-pair correlation can be found following the methods in Appendix C 3 in [11]. There, the ensemble average $\langle \rho_{12} \rho_{12} \rangle$ is evaluated, where 1 and 2 refer to distinct pulsars. A similar calculation using four arbitrary pulsars $a$, $b$, $c$, and $d$ (which might or might not be the same) gives three terms. These arise from Isserlis's theorem [19], which expresses the four-point function for zero-mean Gaussian variables as a sum of three products of two-point functions, giving

$$\begin{aligned} \langle \rho_{ab} \rho_{cd} \rangle &= \left( h^2 \mu_{ab} \right) \left( h^2 \mu_{cd} \right) + \hbar^4 \mu_{ac} \mu_{bd} + \hbar^4 \mu_{ad} \mu_{bc} \\ &= \langle \rho_{ab} \rangle \langle \rho_{cd} \rangle + \hbar^4 (\mu_{ac} \mu_{bd} + \mu_{ad} \mu_{bc}). \end{aligned} \tag{2.8}$$

The second line of (2.8) is obtained from the first line by using (2.3).

The factor which appears on the rhs of (2.8) sets the scale of the covariance, and is

$$\hbar^4 \equiv (4\pi)^2 \int_{-\infty}^{\infty} \mathrm{d}f \int_{-\infty}^{\infty} \mathrm{d}f' \operatorname{sinc}^2\big(\pi(f - f')T\big) \big(4\pi^2 f f'\big)^{-2\alpha} H(f) H(f'), \qquad (2.9)$$

where $\operatorname{sinc} x \equiv \sin(x)/x$ [20] and $T$ is the total observation time. The constant $\alpha$ is explained in Appendix A. For redshift correlations, $\alpha = 0$ and $\hbar$ is a dimensionless strain, whereas for timing-residual correlations, $\alpha = 1$ and $\hbar$ has units of time × (dimensionless) strain.

To compare the variance (proportional to $\hbar^4$) to the squared mean (proportional to $h^4$) we must have a value for their ratio $\hbar^4/h^4$. This ratio depends upon the spectrum of the GW sources, and is discussed in Appendix C 3 in [11]. Note that since $H(f) \geq 0$ and $0 \leq \operatorname{sinc}^2 x \leq 1$, it follows immediately from (2.6) and (2.9) that $\hbar^4 \leq h^4$. We evaluate their ratio for several models in Appendix B. For the simplest Gaussian ensemble of binary-inspiral sources, $\hbar^2/h^2$ is shown in Fig. 21.

The covariance matrix is defined by (2.2). For the Gaussian ensemble, making use of (2.8) we obtain

$$\mathbb{C}_{ab,cd} = \hbar^4 \big(\mu_{ac}\mu_{bd} + \mu_{ad}\mu_{bc}\big). \qquad (2.10)$$

In this expression, the pulsars $a$, $b$, $c$, and $d$ can be freely chosen: they may have arbitrary angular separations, and any or all of them may be identical or distinct. Thus, the pairs $ab$ and $cd$ may lie in different bins or in the same bin, and may even denote the same pair.

Expression (2.10) for the entries of the covariance matrix has an interesting structure. Although the lhs is (at least conceptually) a function of the two pulsar pairs $ab$ and $cd$, the rhs depends upon the angles between the pulsars in the pairs formed from the *other* possible partners $ac$, $ad$, $bc$, and $bd$. In general, those four pairs have angular separations which are very different than the angular separations of the two pairs $ab$ and $cd$.

Expression (2.10) also shows that the covariance matrix is not diagonal. A diagonal covariance matrix could be obtained by replacing the $\mu$'s on the rhs with (quantities proportional to) Kronecker deltas. But since $\mu_{ef}$ is nonzero for $e$ different than $f$, the covariance matrix has nonzero off-diagonal terms.

As a sanity check, consider a diagonal element of the covariance matrix $\mathbb{C}_{ab,ab}$, and assume that the $ab$ pulsar pair is composed of distinct pulsars, so that $a \neq b$. For this case, after setting $c = a$ and $d = b$, the rhs of (2.10) can be simplified using (2.4). This gives the total (pulsar plus cosmic) variance of $\rho_{ab}$ as

$$\mathbb{C}_{ab,ab} = \sigma_{\text{tot}}^2 = \hbar^4 \big(\mu_u^2(\gamma_{ab}) + 4\mu_u^2(0)\big). \qquad (2.11)$$

This diagonal element agrees exactly with the result found in Eq. (C28) of [11], which corresponds to (2.10) with $a = 1$, $b = 2$, $c = 1$, and $d = 2$.

### C. Third and fourth moments

The third- and fourth-order moments of the pulsar-pair correlation $\rho_{ab}$ can be calculated in a manner similar to that presented above for $\mathbb{C}_{ab,cd}$: we use Isserlis's theorem to evaluate expectation values of products of six or eight pulsar redshift measurements $Z_a, Z_b, ..., Z_h$. Since the resulting equations are messy, and are only needed to compute the variances of the $\chi^2$ statistics in Sec. VIII, we put these calculations in Appendix F. For the remaining analyses of this paper, the central object of interest is the covariance matrix $\mathbb{C}_{ab,cd}$.

### D. Useful identities

For later use, we provide two useful identities involving the covariance matrix $\mathbb{C}_{ab,cd}$. The first follows by using (2.4) to expand the rhs of (2.10) for $\mathbb{C}_{ab,cd}$, giving

$$\mu_{ac}\mu_{bd} + \mu_{ad}\mu_{bc} = \mu_u(\gamma_{ac})\mu_u(\gamma_{bd}) + \mu_u(\gamma_{ad})\mu_u(\gamma_{bc}) + \mu_u^2(0)\Big[\delta_{ac}\delta_{bd} + \delta_{ad}\delta_{bc}\Big]$$
$$+ \mu_u(0)\Big[\mu(\gamma_{bd})\delta_{ac} + \mu_u(\gamma_{ac})\delta_{bd} + \mu_u(\gamma_{bc})\delta_{ad} + \mu_u(\gamma_{ad})\delta_{bc}\Big]. \qquad (2.12)$$

The second identity provides an alternative form for the first two terms on the rhs of (2.12). Making use of (2.5), the products of Hellings and Downs curves may be written in terms of antenna pattern functions as

$$\mu_u(\gamma_{ac})\mu_u(\gamma_{bd}) + \mu_u(\gamma_{ad})\mu_u(\gamma_{bc})$$
$$= \sum_A \sum_{A'} \int \frac{\mathrm{d}\hat{\Omega}}{4\pi} \int \frac{\mathrm{d}\hat{\Omega}'}{4\pi} \Big( F_a^A(\hat{\Omega}) F_c^A(\hat{\Omega}) F_b^{A'}(\hat{\Omega}') F_d^{A'}(\hat{\Omega}') + F_a^A(\hat{\Omega}) F_d^A(\hat{\Omega}) F_b^{A'}(\hat{\Omega}') F_c^{A'}(\hat{\Omega}') \Big). \qquad (2.13)$$

We use these identities in Sec. IV to calculate the average value of a row of $C_{jk}$ in the narrow-bin limit when there are many pulsar pairs distributed uniformly on the sky. [As described in (2.14), $j$ and $k$ label angular bins.] In this limit, we demonstrate that the variance of the optimal estimator converges to the cosmic variance first calculated in [11].

### E. Block-matrix form of the covariance matrix

To form the matrix $\mathbb{C}$ from the entries $\mathbb{C}_{ab,cd}$, it is helpful to index/order pulsar pairs by increasing angular separation $\gamma_{ab}$. As explained following (2.2), each angular bin provides a lower and upper bound for this index, making it easy to restrict attention to pulsar pairs within the bin around angular separation $\gamma$.

To explicitly illustrate this indexing scheme, suppose that we have $N_{pul} = 101$ pulsars at specific sky locations, so that the number of distinct pulsar pairs is $N_{pairs} = 5050$. Order the pairs by $\cos \gamma_{ab} \in [-1, 1]$ and divide them up into $\kappa = 50$ bins of width $d(\cos \gamma) = 0.04$. *On average*, each bin will contain $n_{pairs} = 101$ pairs, with values of $\cos \gamma$ differing by at most $\pm 0.04$. (Throughout this paper, we use $n_{pairs}$ to denote the number of distinct pulsar pairs in an angular separation bin, whereas $N_{pairs}$ denotes the *total* number of distinct pulsar pairs. If necessary, we add a subscript such as $n_{pairs,j}$ to identify the angular bin.) Label the bins by $j = 1, 2, \ldots, 50$, and let $\gamma_j$ denote the average value of $\gamma_{ab}$ in bin $j$ [21].

With this angular-separation binning scheme, the covariance matrix $\mathbb{C}$, with entries given by (2.10), takes the following "block-matrix form":

$$\mathbb{C} = \begin{pmatrix} C_{11} & C_{12} & \ldots & C_{1\kappa} \\ C_{21} & C_{22} & \ldots & C_{2\kappa} \\ \vdots & \vdots & \ddots & \vdots \\ C_{\kappa 1} & C_{\kappa 2} & \ldots & C_{\kappa\kappa} \end{pmatrix}, \quad (2.14)$$

with $\kappa = 50$. We call these smaller submatrices "blocks" or "matrix blocks," and denote them by $C_{jk}$, where $j, k \in 1, 2, \ldots, \kappa$ label the bins. To distinguish the blocks from the full covariance matrix $\mathbb{C}$, we use a different symbol $C$ for them.

In this example, each block $C_{jk}$ is a matrix with approximately $101 \times 101$ entries. Note that the indices $j$ and $k$ label the blocks, and not the entries within a block. The block rows $ab$ correspond to pulsar pairs separated by angles $\gamma_{ab} \approx \gamma_j$, and its columns $cd$ to pairs separated by $\gamma_{cd} \approx \gamma_k$. In computing and discussing the statistical properties of a single bin $j$, we often use the symbol $C$ to denote the corresponding (square) block $C_{jj}$, which is located along the diagonal of (2.14). These diagonal blocks inherit the most important properties of $\mathbb{C}$: they are real, symmetric, and positive definite [22].

If $j$ and $k$ are narrow bins containing very many uniformly distributed pulsar pairs, then the block $C_{jk}$ acquires an additional important property. For this case, Sec. IV A shows that the sum of the entries along any row of the block is the same as the sum of the entries along any other row of the block. Said differently, the sum of any row of $C_{jk}$ is *independent* of the row. In this case, the only thing we will need for our analysis is the average value $s_{jk}$ of the entries of $C_{jk}$. For the diagonal blocks, we show in Sec. IV D that $s_{jj}$ is the cosmic variance $\sigma^2_{cos}(\gamma_j)$, and for the off-diagonal blocks, that $s_{jk}$ is the cosmic covariance computed in Sec. IV B.

## III. OPTIMAL ESTIMATION OF THE HELLINGS AND DOWNS CORRELATION

In this section, we define the Hellings and Downs correlation for a set of pulsars, as described in Sec. I B. For a given angular bin (in $\gamma$) it is a weighted average of the correlations for the pulsar pairs that lie in that bin. The weights are determined by the conditions that the average be *unbiased* and have *minimum variance*.

### A. Derivation of the optimal estimator and its variance

We now derive the optimal way to combine pulsar-pair correlation measurements $\rho_{ab}$, where the separation angles $\gamma_{ab}$ lie in the $j$th angular bin. To simplify the equations, we use vector/matrix notation, where the dimension of the vectors and (square) matrices is the number $n_{pairs}$ of distinct pulsar pairs in the angular correlation bin. For example, $\rho \equiv \rho_{ab}$ denotes the column vector of redshift correlation measurements in bin $j$, and the block $C \equiv C_{jj}$ is a real, positive-definite, symmetric matrix, whose elements are the covariances between the correlation measurements $\rho_{ab}$ and $\rho_{cd}$, both lying in that bin.

The optimal estimator is a linear combination

$$\rho_{opt} = w^\top \rho = \sum_{ab=1}^{n_{pairs}} w_{ab} \rho_{ab} \quad (3.1)$$

of the correlations in the bin, where $w$ is a column vector of dimensionless weights $w_{ab}$, and $w^\top$ denotes the transpose of $w$. The second equality gives an explicit expression for $\rho_{opt}$ in terms of the vector components, where the sum is restricted to $n_{pairs}$ pulsar pairs $ab$ that lie in the $j$th bin. For subsequent formulas, we only give the more compact vector/matrix equations.

Note that (3.1) is proportional to $\rho_{ab}$, and does not include a constant (i.e., independent of $\rho_{ab}$) term. If only cross-correlations ($a \neq b$) are used, then (3.1) is sufficient to construct an optimal (*unbiased* and *minimum-variance*) estimator of the Hellings and Downs correlation. Likewise, if the sum (3.1) includes auto-correlations ($a = b$) but there is no noise, then no constant term is needed. However, if (i) auto-correlations are included and (ii) there is pulsar or measurement noise, then (3.1) is not general enough to form an unbiased estimator. For that, an additional constant term must be subtracted from the rhs of (3.1). In what follows, we concentrate on the ideal noise-free case, where the constant term is zero. Later, when we consider the effects of noise in Sec. IX, this constant term must be nonzero to obtain an unbiased minimum-variance estimator.

We start by calculating the variance of the optimal estimator:

$$\sigma_{\text{opt}}^2 \equiv \langle \rho_{\text{opt}}^2 \rangle - \langle \rho_{\text{opt}} \rangle^2, \qquad (3.2)$$

where angle brackets denote the expectation value for the Gaussian ensemble used in, e.g., (2.3) and (2.8). Substituting (3.1) into (3.2), we obtain

$$\sigma_{\text{opt}}^2 = w^\top \Big( \langle \rho \rho^\top \rangle - \langle \rho \rangle \langle \rho^\top \rangle \Big) w = w^\top C w = (Cw, Cw), \qquad (3.3)$$

where the second equality follows from the definition (2.2) of the covariance matrix (restricted to block $C \equiv C_{jj}$ for bin $j$). For the third equality we introduce an inner product of column vectors $A$ and $B$ via

$$(A, B) = A^\top C^{-1} B, \qquad (3.4)$$

and use the property that the diagonal block $C$ is a symmetric matrix, so that $C^\top = C$.

To completely determine the weights, we need a normalization constraint. The expected value of the estimator follows immediately from (2.3) and (3.1), as

$$\langle \rho_{\text{opt}} \rangle = h^2 w^\top \mu = h^2 \mu^\top w = h^2 (\mu, Cw), \qquad (3.5)$$

where $\mu$ denotes the column vector with components $\mu_{ab} = \mu_u(\gamma_{ab})$, assuming $a \neq b$. Now, we must make an explicit and somewhat arbitrary choice: what value should we pick for the expected value of the Hellings and Downs correlation estimator $\langle \rho_{\text{opt}} \rangle$ for this particular angular bin? This choice defines what it means for the optimal estimator to be *unbiased*.

Using the subscript "bin" to denote this somewhat arbitrary choice for the normalization constraint, the mean value of our estimator should correspond to the expected Hellings and Downs correlation

$$\langle \rho_{\text{opt}} \rangle \equiv \rho_{\text{bin}} \equiv h^2 \mu_{\text{bin}} \qquad (3.6)$$

for *some* choice of $\mu_{\text{bin}}$ which is representative of the expected correlation in bin $j$. But what choice is most sensible? Three reasonable options are the following:

(a) $\mu_{\text{bin}} = \mu_u(\gamma_{\text{bin}})$, where $\gamma_{\text{bin}}$ is the central angle of the bin. In practice, such bin centers are typically set in advance of any analysis, as evenly spaced "round numbers."

(b) $\mu_{\text{bin}} = \mu^\top \mathbb{1}/n_{\text{pairs}}$, which is a uniform average of the expected Hellings and Downs correlation values for the different pulsar pairs in the bin. Here, $\mathbb{1}$ denotes a column vector of dimension $n_{\text{pairs}}$ containing all ones:

$$\mathbb{1} \equiv \Big( 1, 1, \ldots, 1 \Big)^\top. \qquad (3.7)$$

(c) $\mu_{\text{bin}} = \mu_u(\gamma_{\text{bin}})$, where $\gamma_{\text{bin}} = \gamma^\top \mathbb{1}/n_{\text{pairs}}$ is the mean angular separation of the pulsar pairs that lie in the bin. Here, $\gamma$ denotes a column vector of the values $\gamma_{ab}$.

We will see that this arbitrary choice only enters the mean and the (square root of the) variance as an overall scale. So, while there may be other justifiable choices which are not listed above, we will see that the choice has no effect on the most important quantity: the fractional uncertainty $\sigma_{\text{opt}}/\rho_{\text{opt}}$.

For any of these arbitrary normalization choices, we can now solve for the weights that define the optimal (minimum-variance) estimator. Since the expected value $\langle \rho_{\text{opt}} \rangle \equiv \rho_{\text{bin}} \equiv h^2 \mu_{\text{bin}}$ is independent of the weights $w$, we can divide $\sigma_{\text{opt}}^2$ by the square of this quantity before minimization over $w$. Hence, we select $w$ to minimize

$$\frac{\sigma_{\text{opt}}^2}{\rho_{\text{bin}}^2} = \frac{\sigma_{\text{opt}}^2}{\langle \rho_{\text{opt}} \rangle^2} = \frac{(Cw, Cw)}{h^4 (\mu, Cw)^2}, \qquad (3.8)$$

where we have used (3.3) and (3.5).

The inner product defined by (3.4) is positive definite and obeys the Schwarz inequality. Hence, the denominator of (3.8) is maximum when the vectors $\mu$ and $Cw$ are parallel to (and hence proportional to) one another. This implies that $Cw = q\mu$ for some $q$, or equivalently, that $w = qC^{-1}\mu$. Substituting this into the normalization condition (3.5) then lets us solve for $q$ and completely determines the optimal weights:

$$h^2(\mu, q\mu) = h^2 \mu_{\text{bin}} \implies q = \frac{\mu_{\text{bin}}}{\mu^\top C^{-1} \mu} \implies w = \frac{\mu_{\text{bin}}}{\mu^\top C^{-1} \mu} C^{-1} \mu. \qquad (3.9)$$

From the optimal weights given in (3.9), the optimal estimator and its variance are then given by (3.1) and (3.3) as

$$\rho_{\text{opt}} = \mu_{\text{bin}} \frac{\mu^\top C^{-1} \rho}{\mu^\top C^{-1} \mu} \quad \text{and} \qquad (3.10)$$

$$\sigma_{\text{opt}}^2 = \frac{\mu_{\text{bin}}^2}{\mu^\top C^{-1} \mu}. \qquad (3.11)$$

As discussed after (3.7), the ratio $\rho_{\rm opt}/\sigma_{\rm opt}$ is independent of the arbitrary choice $\mu_{\rm bin}$.

While this paper is specifically focused on GW backgrounds that are described by a Gaussian ensemble, the results of this section apply to any GW ensemble. This is because any ensemble has a covariance matrix, from which one can define the (best estimate of the) pulsar-pair correlation (3.10) and the corresponding variance (3.11). So, this way of defining the Hellings and Downs correlation for some set of pulsar pairs, and computing its variance, is quite general.

### B. Covariance between optimal estimates

In Sec. I B and Fig. 2, we explained the importance of correlations (in the Hellings and Downs correlation) between different pairs of pulsars. This correlation extends to the optimal estimators which we have constructed for different angular bins. For example, if the optimal estimator in one angular bin lies somewhat above the Hellings and Downs curve, then it is very likely that the optimal estimator for an adjacent angular bin also lies above the curve. This is because pulsar pairs are correlated, even if they do not lie in the same (arbitrarily defined) angular bin.

The correlation between deviations from the Hellings and Downs curve in the $j$th and $k$th angular bins is the covariance of the optimal estimators for those two bins, and is straightforward to compute. The optimal estimators for the two bins are given by (3.1) as

$$\rho_{{\rm opt},j} = w_j^\top \rho_j \quad \text{and} \quad \rho_{{\rm opt},k} = w_k^\top \rho_k, \tag{3.12}$$

with optimal weights given by (3.9):

$$w_j = \frac{\mu_{{\rm bin},j}}{\mu_j^\top C_{jj}^{-1} \mu_j} C_{jj}^{-1} \mu_j \quad \text{and} \quad w_k = \frac{\mu_{{\rm bin},k}}{\mu_k^\top C_{kk}^{-1} \mu_k} C_{kk}^{-1} \mu_k. \tag{3.13}$$

The covariance between $\rho_{{\rm opt},j}$ and $\rho_{{\rm opt},k}$ is defined by

$$B_{jk} \equiv \left\langle \left( \rho_{{\rm opt},j} - \langle \rho_{{\rm opt},j} \rangle \right) \left( \rho_{{\rm opt},k} - \langle \rho_{{\rm opt},k} \rangle \right) \right\rangle = \langle \rho_{{\rm opt},j} \rho_{{\rm opt},k} \rangle - \langle \rho_{{\rm opt},j} \rangle \langle \rho_{{\rm opt},k} \rangle. \tag{3.14}$$

Expanding the rhs of (3.14) using (3.12), this becomes

$$\begin{aligned} B_{jk} &= w_j^\top \left( \langle \rho_j \rho_k^\top \rangle - \langle \rho_j \rangle \langle \rho_k^\top \rangle \right) w_k \\ &= w_j^\top C_{jk} w_k \\ &= \frac{\mu_{{\rm bin},j} \mu_{{\rm bin},k}}{(\mu_j^\top C_{jj}^{-1} \mu_j)(\mu_k^\top C_{kk}^{-1} \mu_k)} \mu_j^\top C_{jj}^{-1} C_{jk} C_{kk}^{-1} \mu_k, \end{aligned} \tag{3.15}$$

where on the second line we use the notation of (2.14) for the $jk$th block of the covariance matrix, and to obtain the third line we use the weights (3.13).

Note that for the case $j = k$, the covariance $B_{jj}$ reduces to the optimal variance (3.11). The lack of the index $j$ in the latter expression is because in previous sections we dropped the bin index to simplify the notation: there was no need to distinguish one angular separation bin from another.

We will return to this topic in Sec. IV B, where we will compute this interbin covariance for the special case where both bins are narrow and contain enough pulsar pairs to faithfully cover the sky.

### C. Simplifications for narrow-bin correlation measurements

The formulas above simplify for narrow angular bins. For these, the pulsar pairs $ab$ all lie in a narrow angular bin at $\gamma$, and thus effectively have the same angular separation $\gamma = \gamma_{ab}$. For this case, the correlation measurements $\rho_{ab}$ all have the same expected value, $\langle \rho_{ab} \rangle = h^2 \mu_{ab} = h^2 \mu_{\rm u}(\gamma_{ab})$, which in vector notation reads

$$\langle \rho \rangle = h^2 \mu = h^2 \mu_{\rm u}(\gamma) \mathbb{1}, \tag{3.16}$$

where $\mathbb{1}$ is defined by (3.7). This implies that the three "different" reasonable choices for the normalization condition described in Sec. III A all correspond to the same condition:

$$\mu_{\rm bin} = \mu_{\rm u}(\gamma) \iff \mu = \mu_{\rm bin} \mathbb{1} \iff w^\top \mathbb{1} = 1. \tag{3.17}$$

For this narrow-bin case, the optimal weights (3.9), optimal estimator (3.10), and variance (3.11) are

$$w = \frac{C^{-1} \mathbb{1}}{\mathbb{1}^\top C^{-1} \mathbb{1}}, \tag{3.18}$$

$$\rho_{\rm opt} = \frac{\mathbb{1}^\top C^{-1} \rho}{\mathbb{1}^\top C^{-1} \mathbb{1}}, \quad \text{and} \tag{3.19}$$

$$\sigma_{\rm opt}^2 = \frac{1}{\mathbb{1}^\top C^{-1} \mathbb{1}}. \tag{3.20}$$

The denominators of (3.18)–(3.20) are the inverse of the covariance matrix block, summed over all of its elements. This is called the "grand sum" of $C^{-1}$. (Equations (3.18)–(3.20) are standard expressions for the optimal combination of correlated measurements that have the same expectation value, and may be found in, e.g., [10,23,24]).

An important special case is that of a single pulsar pair $ab$. For this, the optimal estimator of the correlation is $\rho_{ab}$, in agreement with the Hellings and Downs definition. The variance for this single pair is $\sigma_{opt}^2 = \mathbb{C}_{ab,ab}$, which is given in (2.11). As anticipated, this is precisely the (one-pulsar-pair) total variance $\sigma^2(\gamma)$ derived in Eq. (C28) of [11] for the Gaussian ensemble.

### D. Simplifications for narrow-bin correlations when $\mathbb{1}$ is an eigenvector of $C$

We obtain a further simplification for the narrow-bin case if the sum of every row of the corresponding covariance matrix block has the same value $\lambda$ as the sum of any other row. The condition may be written in matrix notation as

$$C\mathbb{1} = \lambda\mathbb{1}, \qquad (3.21)$$

meaning that $\mathbb{1}$ is an eigenvector of $C$ with eigenvalue $\lambda$. For this case, the variance of the optimal estimator, (3.20), can be expressed in terms of the average value of the entries of $C$.

To see this, note that since $C$ is real, symmetric, and positive definite, the eigenvalue $\lambda$ must be real and positive. Now multiply (3.21) from the left by $\lambda^{-1}C^{-1}$ to obtain

$$C^{-1}\mathbb{1} = \lambda^{-1}\mathbb{1}. \qquad (3.22)$$

This means that $\mathbb{1}$ is also an eigenvector of $C^{-1}$ with a real positive eigenvalue $\lambda^{-1}$. So, we can take the dot product of (3.22) on the left with $\mathbb{1}^\top$, leading to

$$\mathbb{1}^\top C^{-1}\mathbb{1} = \mathbb{1}^\top \lambda^{-1}\mathbb{1} = n_{pairs}/\lambda, \qquad (3.23)$$

where the number of pairs has appeared because it is the dimension of the vector $\mathbb{1}$.

So, with these assumptions (3.20) simplifies to

$$\sigma_{opt}^2 = \lambda/n_{pairs} \equiv s, \qquad (3.24)$$

where $s$ is the average value of the entries of $C$. In addition, the properly normalized optimal weights of (3.18) are then given by

$$w = \mathbb{1}/n_{pairs}. \qquad (3.25)$$

In this case, the optimal weights are all equal, and (3.19) becomes

$$\rho_{opt} = \mathbb{1}^\top \rho/n_{pairs}. \qquad (3.26)$$

This is the uniform average of the correlation measurements $\rho_{ab}$.

To summarize, if the sum of each row of the covariance matrix $C$ for narrow-bin correlations gives the same number $\lambda$, then: (i) the variance of the optimal estimator is simply the average value, $s \equiv \lambda/n_{pairs}$, of the entries of $C$, and (ii) the optimal weights are all equal.

We will show in Sec. IV that, in the limit of an infinite number of pulsar pairs distributed uniformly on the sky, each row of a given narrow-bin covariance matrix block $C$ sums to the same value. Thus, the variance of the optimal estimator may be found from the average value of the entries of $C$, and is the cosmic variance. This follows from rotational symmetry: with an infinite number of uniformly distributed pairs, any pair at angle $\gamma$ is equivalent to any other pair at the same angle. The considerations of Fig. 2 do not apply to this limiting case.

### E. Simple example: An infinite set of measurements with nonzero variance

We would normally expect that as more and more noise-free measurements are included in some estimate, the uncertainty in the estimate decreases to zero. However, as we have explained, the Hellings and Downs correlation does not behave this way. In the limit of many measurements, the uncertainty decreases to a nonzero value, which is the cosmic variance.

Here, we provide a simple example to illustrate how correlations between the different measurements (described by off-diagonal terms in the covariance matrix) are responsible for this behavior. Consider the following $n_{pairs} \times n_{pairs}$ covariance matrix block (from along the diagonal of the full covariance matrix, as described in Sec. II E), where the terms proportional to $a$ arise from correlations between the different measurements:

$$C = \begin{pmatrix} 1 & a & \dots & a \\ a & 1 & \dots & a \\ \vdots & \vdots & \ddots & \vdots \\ a & a & \dots & 1 \end{pmatrix}. \qquad (3.27)$$

The parameter $a$ must lie in the range $-1/(n_{pairs} - 1) < a < 1$ in order for $C$ to be a positive-definite matrix [25].

We find the optimal estimator and its variance using the results of Sec. III D. It is easy to verify that $\mathbb{1}$ is an eigenvector of $C$ with eigenvalue

$$\lambda = 1 + (n_{pairs} - 1)a. \qquad (3.28)$$

Thus, the optimal estimator is

$$\rho_{opt} = \mathbb{1}^\top \rho/n_{pairs}, \qquad (3.29)$$

and its variance is

$$\sigma_{\rm opt}^2 = \frac{\lambda}{n_{\rm pairs}} = \frac{1 + (n_{\rm pairs} - 1)a}{n_{\rm pairs}}. \qquad (3.30)$$

Now, one might think that since $\rho_{\rm opt}$ is a uniform average of the correlation measurements $\rho_{ab}$, then the variance of the optimal estimator should scale like $1/n_{\rm pairs}$, and hence vanish in the limit $n_{\rm pairs} \to \infty$. But it follows immediately from (3.30) that in the limit of large numbers of measurements the variance approaches

$$\lim_{n_{\rm pairs} \to \infty} \sigma_{\rm opt}^2 = a, \qquad (3.31)$$

which is nonzero for $a > 0$.

In the next section, we will see that the cosmic variance exhibits the same behavior. From this example, we conclude that it is the nonzero off-diagonal elements of the covariance matrix that are responsible for this. (It does not arise from unequal weighting of the correlation measurements because in the limit of many pulsar pairs, uniformly distributed on the sky, all of the weights are equal.)

## IV. COSMIC COVARIANCE AND COSMIC VARIANCE

This section considers a pair of narrow Hellings and Downs correlation bins at two different angles, each containing a large number of pulsar pairs, uniformly distributed on the sky. For each bin, we construct the optimal estimator of the Hellings and Downs correlation in that bin. We then compute the covariance of those two optimal estimators, as defined in Sec. III B, obtaining a simple analytic expression for the cosmic covariance. If the two bins are the same, we prove that this reduces to the cosmic variance found in [11].

At the end of the section, we consider the more realistic case, where the number of pulsars is finite, and show how the cosmic variance is approached as the number of pulsars grows.

### A. Narrow-bin pulsar averaging

Our approach to deriving the cosmic covariance follows the formulation of Secs. III B, III C, and III D. Since each angular bin is narrow, the quantities of interest are the average values of the entries in the inverse covariance matrix for that bin, as shown in Sec. III C. The averages are computed along each row of the matrix. If each angular bin contains a large number of pulsar pairs, uniformly distributed on the sky, then we can prove that every row of the covariance matrix block has the same average value. This allows further simplifications, as shown in Sec. III D.

Throughout this section we consider a pair of narrow angular-separation bins $j$ and $k$ centered at $\gamma_j$ and $\gamma_k$, and let $ab$ and $cd$ denote the (many, uniformly distributed) pulsar pairs lying in those respective bins. Equation (2.10) and identities (2.12) and (2.13) provide the following expression for the entries $\mathbb{C}_{ab,cd}$ of the covariance matrix $C_{jk}$ of the pulsar pairs in those two bins:

$$\begin{aligned}
\mathbb{C}_{ab,cd} &= \hbar^4 \big( \mu_{ac}\mu_{bd} + \mu_{ad}\mu_{bc} \big) \\
&= \hbar^4 \sum_A \sum_{A'} \int \frac{d\hat{\Omega}}{4\pi} \int \frac{d\hat{\Omega}'}{4\pi} \left( F_a^A(\hat{\Omega}) F_c^A(\hat{\Omega}) F_b^{A'}(\hat{\Omega}') F_d^{A'}(\hat{\Omega}') + F_a^A(\hat{\Omega}) F_d^A(\hat{\Omega}) F_b^{A'}(\hat{\Omega}') F_c^{A'}(\hat{\Omega}') \right) \\
&\quad + \hbar^4 \mu_{\rm u}(0) \big[ \mu_{\rm u}(0)(\delta_{ac}\delta_{bd} + \delta_{ad}\delta_{bc}) + \mu_{\rm u}(\gamma_{bd})\delta_{ac} + \mu_{\rm u}(\gamma_{ac})\delta_{bd} + \mu_{\rm u}(\gamma_{bc})\delta_{ad} + \mu_{\rm u}(\gamma_{ad})\delta_{bc} \big]. \qquad (4.1)
\end{aligned}$$

The second equality can be written more compactly using "symmetrization notation" [26] for indices,

$$Q_{(cd)} \equiv \frac{1}{2} \big( Q_{cd} + Q_{dc} \big). \qquad (4.2)$$

This yields

$$\begin{aligned}
\mathbb{C}_{ab,cd} = 2\hbar^4 \bigg[ &\sum_A \sum_{A'} \int \frac{d\hat{\Omega}}{4\pi} \int \frac{d\hat{\Omega}'}{4\pi} F_a^A(\hat{\Omega}) F_{(c}^A(\hat{\Omega}) F_{d)}^{A'}(\hat{\Omega}') F_b^{A'}(\hat{\Omega}') \\
&+ \mu_{\rm u}^2(0)\delta_{a(c}\delta_{d)b} + \mu_{\rm u}(0) \Big( \mu_{\rm u}(\gamma_{a(c})\delta_{d)b} + \mu_{\rm u}(\gamma_{b(c})\delta_{d)a} \Big) \bigg]. \qquad (4.3)
\end{aligned}$$

Recall [see the discussion after (2.10)] that the above expression for $\mathbb{C}_{ab,cd}$ is valid for all possible values of $a$, $b$, $c$, and $d$ even though we will often restrict attention to distinct pulsar pairs for which $a < b$ and $c < d$.

We now compute the average value of the entries $\mathbb{C}_{ab,cd}$ in (4.3), where the average is computed along the $ab$ row of $C_{jk}$. We will see that this average value is independent of the row $ab$. This means that the matrix $C_{jk}$, when multiplied by the column vector $\mathbb{1}_k$ of dimension $n_{{\rm pairs},k}$, yields a vector proportional to $\mathbb{1}_j$, which has (a possibly different) dimension $n_{{\rm pairs},j}$. This is sufficient to carry out the simplifications described in Sec. III D.

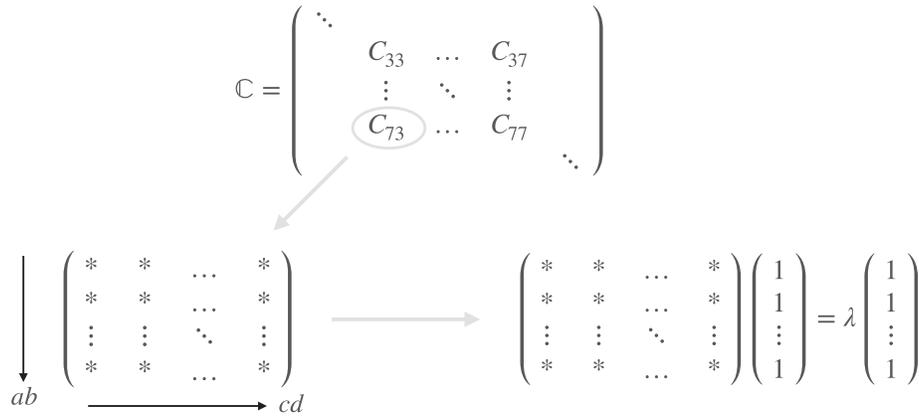

FIG. 4. Four sub-blocks of the full covariance matrix $\mathbb{C}$ are shown. Each corresponds to a narrow bin containing many uniformly distributed pulsar pairs $ab$ and $cd$. For each sub-block $C_{jk}$, we compute the average value $s_{jk}$ of its entries, which is simply related to the eigenvalue $\lambda$ of the $\mathbb{1}$ vector (see text).

Figure 4 provides a pictorial guide to clarify our notation. The top shows the full covariance matrix $\mathbb{C}$, which contains four submatrices $C_{jk}$ corresponding to narrow bins labeled by $j$ and $k$. For the illustration, these are arbitrarily selected to be the third and seventh bins. The two submatrices along the diagonal are square, but may have different dimensions. Both dimensions are large, because both bins contain many pulsar pairs. The two submatrices off the diagonal are (for the general case) nonsquare.

The bottom left of Fig. 4 shows the rectangular submatrix $C_{73}$, whose many entries are denoted by $*$'s. Its rows are indexed by pulsar pairs $ab$ that lie in the narrow bin at angle $\gamma_7$, while its columns are indexed by pulsar pairs $cd$ that lie in the narrow bin at angle $\gamma_3$.

The bottom right illustrates the implications of the statement "the average values of the submatrix entries, averaged along a row, are row-independent" [27]. This is true if and only if each of the four submatrix blocks satisfies an equation of the form $C_{jk}\mathbb{1} = \lambda_{jk}\mathbb{1}$. Pictorially, this is illustrated by the matrix equation at the bottom right of Fig. 4, keeping in mind that the $\mathbb{1}$ vectors which appear on the two sides of that matrix equation may have different dimensions. Indeed, the only property of the four matrices which we will need is the average value of their entries, denoted by $s_{jk}$. For the pictorial example, these average values are $s_{33}$, $s_{37} = s_{73}$, and $s_{77}$.

In what follows, we derive a formula for $s_{jk}$. This holds for any indices $j$ and $k$ that correspond to narrow bins containing many uniformly distributed pulsar pairs. In parsing the equations, the reader should find it helpful to keep Fig. 4 in mind.

To compute the average value $s_{jk}$ of a matrix element within the block, we fix the row index $ab$, and average the entries along the columns $cd$ within the block. The pairs labeled by $cd$ all have (approximately) the same angular separation $\gamma_k$, and by assumption are uniformly distributed on the sky. Thus, denoting the average by $(s_{jk})_{ab}$, we have

$$(s_{jk})_{ab} \equiv \langle \mathbb{C}_{ab,cd} \rangle_{cd \in \gamma_k}, \tag{4.4}$$

where $\langle \rangle_{cd \in \gamma_k}$ denotes the average over all pulsar pairs $cd$ separated by angle $\gamma_k$. In [11], this is called a "pulsar average" and is explicitly defined in Appendix A of [11] as a normalized integral over three variables.

From (4.3), one can see that there are three terms that contribute to the average (4.4). The first explicitly involves the antenna pattern functions, and the second and third involve either a single Kronecker delta function or a product of two Kronecker deltas. We show in Appendix C that, in the limit of many pulsar pairs, the latter two terms make negligible contributions to the average: they correspond to a set of measure zero.

Thus, for large numbers of pulsar pairs, the only term that contributes to the average value of the entries along the row labeled by $ab$ is

$$(s_{jk})_{ab} = 2\hbar^4 \sum_A \sum_{A'} \int \frac{d\hat{\Omega}}{4\pi} \int \frac{d\hat{\Omega}'}{4\pi} F_a^A(\hat{\Omega}) F_b^{A'}(\hat{\Omega}') \left\langle F_{(c}^A(\hat{\Omega}) F_{d)}^{A'}(\hat{\Omega}') \right\rangle_{cd \in \gamma_k}. \tag{4.5}$$

(The symmetrization on $c$ and $d$ has no effect, because we have assumed that the pulsar pairs in the bin at angle $\gamma_k$ are uniformly distributed on the sky.) Pulsar averages of this type are introduced in Appendix G in [11], where they are called *two-point* functions, and the specific average that appears in (4.5) is evaluated. The uniform sky average over all pulsar pairs $cd$ separated by angle $\gamma_k$ yields

$$\langle F_c^A(\hat{\Omega}) F_d^{A'}(\hat{\Omega}') \rangle_{cd \in \gamma_k} = \mu_{AA'}(\gamma_k, \beta), \qquad (4.6)$$

where the functions $\mu_{AA'}(\gamma, \beta)$ are given explicitly in Eqs. (G5) and (G9) in [11]; these functions vanish if $A \neq A'$. Here, $\beta$ is the angle between $\hat{\Omega}$ and $\hat{\Omega}'$, so $\cos\beta = \hat{\Omega} \cdot \hat{\Omega}'$. Substituting (4.6) into (4.5) and summing over polarizations $A'$, we obtain

$$(s_{jk})_{ab} = 2\hbar^4 \sum_A \int \frac{d\hat{\Omega}}{4\pi} \int \frac{d\hat{\Omega}'}{4\pi} F_a^A(\hat{\Omega}) F_b^A(\hat{\Omega}') \mu_{AA}(\gamma_k, \cos^{-1}(\hat{\Omega} \cdot \hat{\Omega}')). \qquad (4.7)$$

It is easy to show that this expression is independent of the pulsar pair (row index) $ab$, because $a$ only enters through $F_a^A(\hat{\Omega})$ and $b$ only enters through $F_b^A(\hat{\Omega}')$.

To see this, pick coordinates for the $\hat{\Omega}$ integration in which pulsar $a$ points along the $z$ axis. Next, pick coordinates for the $\hat{\Omega}'$ integration in which pulsar $b$ lies in the $x, z$ plane. Then, since all $ab$ pairs have the same separation angle $\gamma_{ab} = \gamma_k$, the average value is independent of the specific choice of the $ab$ pair.

Since the value of $(s_{jk})_{ab}$ does not depend upon the row $ab$, we can write $(s_{jk})_{ab} = s_{jk}$, meaning that the average of the entries along any row of the covariance matrix depends only upon the corresponding bin, and not upon which row (pulsar pair) is selected. To compute this average value, use (4.6) again (with pulsar labels $cd$ replaced by $ab$ and $\gamma_k$ replaced by $\gamma_j$) to average (4.7) over $ab$ pairs separated by angle $\gamma_j$. Since the entries have the same average value along each row, the additional averaging does not change that value, and we obtain

$$s_{jk} = 2\hbar^4 \sum_A \int \frac{d\hat{\Omega}}{4\pi} \int \frac{d\hat{\Omega}'}{4\pi} \mu_{AA}(\gamma_j, \cos^{-1}(\hat{\Omega} \cdot \hat{\Omega}')) \mu_{AA}(\gamma_k, \cos^{-1}(\hat{\Omega} \cdot \hat{\Omega}'))$$

$$= \hbar^4 \int_0^\pi d\beta \sin\beta \Big( \mu_{++}(\gamma_j, \beta) \mu_{++}(\gamma_k, \beta) + \mu_{\times\times}(\gamma_j, \beta) \mu_{\times\times}(\gamma_k, \beta) \Big). \qquad (4.8)$$

To obtain the final line of (4.8), we use

$$\int \frac{d\hat{\Omega}}{4\pi} \int \frac{d\hat{\Omega}'}{4\pi} Q(\hat{\Omega} \cdot \hat{\Omega}') = \frac{1}{2} \int_0^\pi d\beta \sin\beta \, Q(\cos\beta) \qquad (4.9)$$

for the spherical average of any function $Q(x)$ of a single variable. The normalization can be checked by setting $Q(x) = 1$.

The quantity $s_{jk}$ only depends on the angles $\gamma_j$ and $\gamma_k$ that define two angular bins. When both bins are narrow, and both bins contain many uniformly distributed pulsar pairs, then $s_{jk}$ is the average value of any row or of any column of the corresponding correlation matrix block $C_{jk}$. Thus, the vector $\mathbb{1}_k$ is an "eigenvector" of this (possibly rectangular) matrix $C_{jk}$, which permits the simplifications of Sec. III D.

## B. Cosmic covariance and correlation between optimal estimators for two narrow bins

In Sec. IV A, we obtained a simple formula for the average value $s_{jk}$ of the entries in the covariance matrix block corresponding to angular separation bins labeled by $j$ and $k$. We assumed that each of the bins is (a) narrow and (b) contains many pulsar pairs that uniformly cover the sky.

The covariance between the optimal estimators of the Hellings and Downs correlation for the two bins is given in (3.15). Since the bins are narrow, the quantities $\mu^\top C^{-1}\mu$ simplify as shown in Sec. III C and Fig. 4 because the column vector $\mu$ is proportional to $\mathbb{1}$. Furthermore, because there are large numbers of pulsar pairs, the pulsar averaging methods of Sec. IV A may be applied, which permit the simplifications of Sec. III D. Under these conditions, (3.15) simplifies to

$$B_{jk} = w_j^\top C_{jk} w_k = \frac{1}{n_{\text{pairs},j} n_{\text{pairs},k}} \mathbb{1}_j^\top C_{jk} \mathbb{1}_k = \langle \mathbb{C}_{ab,cd} \rangle_{ab \in \gamma_j; cd \in \gamma_k} = s_{jk}, \qquad (4.10)$$

where $s_{jk}$ is defined by (4.8). This *cosmic covariance* $s_{jk}$ is a function of two angles $\gamma_j$ and $\gamma_k$, which are the separation angles of the (many) pulsar pairs in the $j$th and $k$th angular bins. It quantifies the amount by which deviations in the Hellings and Downs correlation (away from the Hellings and Downs curve) are correlated between different pulsar separation angles.

We evaluate the cosmic covariance by carrying out the integral in (4.8), using the functions $\mu_{AA'}(\gamma,\beta)$ which are explicitly given in Eqs. (G5) and (G9) in [11]. To evaluate the integrals, it is convenient to set $\gamma_< = \min(\gamma_j, \gamma_k)$ and $\gamma_> = \max(\gamma_j, \gamma_k)$. Because $s_{jk}$ is a symmetric function of the two angles, it can be written as a function of $\gamma_<$ and $\gamma_>$. The integral over $\beta$ is evaluated separately in the three intervals $\beta \in [0, \gamma_<]$, $\beta \in [\gamma_<, \gamma_>]$, and $\beta \in [\gamma_>, \pi]$. Adding these three contributions together gives

$$
\begin{aligned}
\frac{s_{jk}}{2\hbar^4} = &\frac{1}{12}\left(\cos\gamma_< \cos\gamma_> - \cos\gamma_< - \cos\gamma_> - 3\right)\ln\frac{1+\cos\gamma_<}{2} + \frac{49}{432}\cos\gamma_< \cos\gamma_> \\
&+ \frac{1}{12}\left(\cos\gamma_< \cos\gamma_> + \cos\gamma_< + \cos\gamma_> - 3\right)\ln\frac{1-\cos\gamma_>}{2} + \frac{1}{4}\left(\cos\gamma_< - \cos\gamma_>\right) - \frac{5}{48} \\
&+ \frac{1}{12}\left(\cos\gamma_< \cos\gamma_> + 3\right)\left(\mathrm{Li}_2\frac{1-\cos\gamma_<}{2} - \mathrm{Li}_2\frac{1-\cos\gamma_>}{2} - \ln\frac{1-\cos\gamma_>}{2}\ln\frac{(1+\cos\gamma_<)(1+\cos\gamma_>)}{4}\right). \quad (4.11)
\end{aligned}
$$

This is explicitly symmetric $s_{jk} = s_{kj}$, since interchanging $\gamma_j$ and $\gamma_k$ does not change the values of $\gamma_<$ or $\gamma_>$.

The special function $\mathrm{Li}_2$, which appears in the cosmic covariance, is the dilogarithm (or second polylogarithm) [28–30]. This remarkable function [31] is defined by the sum

$$
\mathrm{Li}_2(z) = \sum_{n=1}^{\infty} \frac{z^n}{n^2}, \quad (4.12)
$$

which converges on and inside the unit circle $|z|^2 \le 1$ in the complex plane. While it may be analytically continued outside of the circle, the sum in (4.12) is sufficient for our purposes, because the arguments in (4.11) lie on the real axis between $-1$ and 1.

The structure of the cosmic covariance is displayed in the left panel of Fig. 5. This shows that the angular range between 0 and 180° consists of three regions, from 0 to ≈50°, from ≈50° to ≈130°, and from ≈130° to 180°. The fluctuations in each of these three regions are strongly correlated, and adjacent regions are anticorrelated.

These three regions correspond to the ranges in which the Hellings and Downs curve has one or the other sign. It indicates that the variance in the Hellings and Downs correlation also has a strongly quadrupolar nature. A second way to illustrate this point is via the *coherence function* $s_{jk}/\sqrt{s_{jj}s_{kk}}$, shown in the right panel of Fig. 5, which is normalized to take values between $-1$ and $+1$.

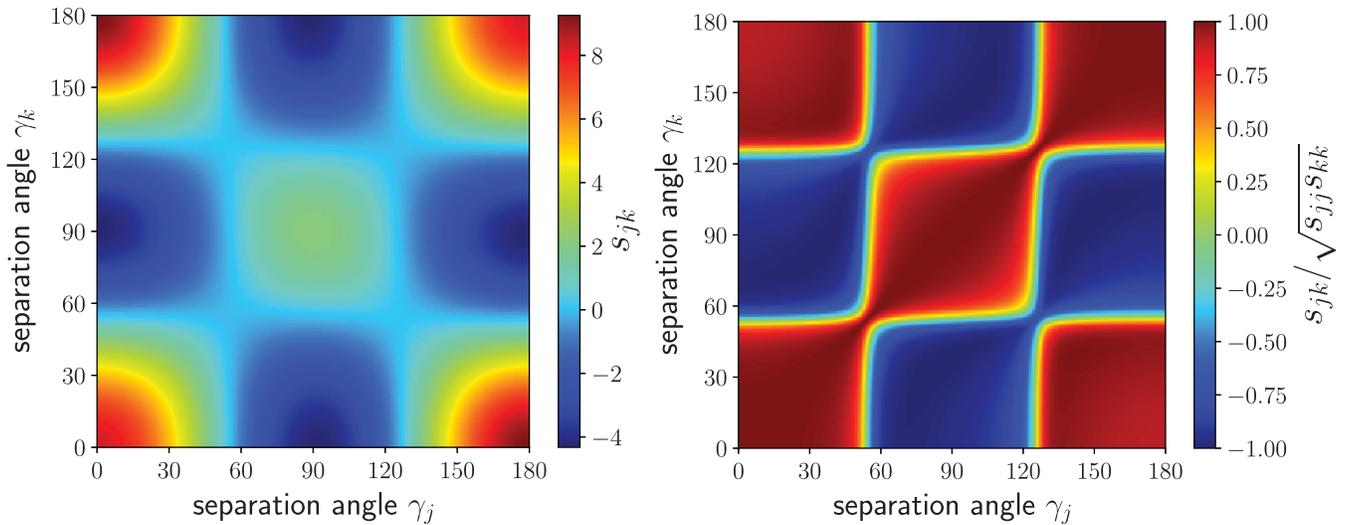

FIG. 5. Left panel: the cosmic covariance $s_{jk}$ is a function of the angles $\gamma_j$ and $\gamma_k$. Its shape along the diagonal is the cosmic variance. The map shows the rhs of (4.11) scaled by a factor of one thousand, which is equivalent to setting $2\hbar^4 = 1000$. Right panel: the cosmic coherence (see text).

This is also called the Pearson correlation coefficient or "Pearson's r." Here, the values are close to $+1$ or to $-1$, indicating that the expected deviations away from the Hellings and Downs curve are tightly correlated and thus fluctuate together in the three angular regions.

### C. Diagonalizing and inverting the cosmic covariance

The inverse of the covariance matrix appears in (the weights and variances associated with) the optimal estimator of the Hellings and Downs correlation, for example in (3.18)–(3.20). We want to characterize these in the cosmic variance limit, when there are many pulsar pairs uniformly distributed on the sky. In this limit, the average value of the covariance matrix approaches the cosmic covariance $s_{jk}$ computed in Sec. IV B. In a similar fashion, the average value of the inverse covariance matrix approaches the average value of the inverse of $s_{jk}$. So, in this section, we diagonalize and invert $s_{jk}$. Later, we will use that result to obtain cosmic variance limits for the squared GW strain.

To carry out this diagonalization, we expand the Hellings and Downs correlation in harmonic functions. This approach was pioneered in [32,33] and then further developed in [34,35]. We follow the treatment given in Appendix C 6 in [11]. For example, the Hellings and Downs curve (1.1) may be expressed ([11], Eq. (C49)) as

$$\mu_u(\gamma) = \sum_{l=0}^{\infty} (2l+1) C_l P_l(x), \qquad (4.13)$$

where $P_l$ is a Legendre polynomial of order $l$, $x = \cos \gamma$, and the coefficients are

$$C_l = \begin{cases} 0 & \text{if } l < 2 \\ (l-2)!/(l+2)! = 1/(l+2)(l+1)l(l-1) & \text{if } l \geq 2. \end{cases} \qquad (4.14)$$

The $C_l$ arise as expectation values in the Gaussian ensemble, and for our purposes may be treated as definite quantities.

The cosmic covariance $s_{jk}$ is a real symmetric matrix. To write it in harmonic form, we follow the same method as used in Appendix C 6 in [11] to obtain

$$s_{jk} = 4\hbar^4 \sum_{l=2}^{\infty} \frac{2l+1}{2} C_l^2 P_l(\cos \gamma_j) P_l(\cos \gamma_k). \qquad (4.15)$$

As a simple check, we can set $\gamma_j = \gamma_k$, which correctly recovers the harmonic form of the cosmic variance, first given in Eq. (C53) in [11]. A numerical check (it is sufficient to sum $l = 2, 3, \ldots, 10$) shows excellent agreement between (4.11) and (4.15).

To invert $s_{jk}$, consider first that the inverse $M^{-1}$ of a matrix $M$ is defined by $MM^{-1} = I$ where $I$ is the identity matrix. For finite-dimensional matrices, the components satisfy

$$\sum_k M_{jk} M_{km}^{-1} = \delta_{jm}. \qquad (4.16)$$

For the case of interest, the matrix $M$ is $s_{jk}$. In the limit of an infinite number of pulsar pairs distributed uniformly on the sky, this matrix has continuous row and column labels and can be written as

$$s(x, x') \equiv 4\hbar^4 \sum_{l=2}^{\infty} \frac{2l+1}{2} C_l^2 P_l(x) P_l(x'), \qquad (4.17)$$

where $x = \cos \gamma$ and $x' = \cos \gamma'$ lie in the interval $[-1, 1]$. By analogy with (4.16), if $M(x, x')$ is a continuous matrix, then we define its inverse $M^{-1}(x, x')$ as a "solution" [but see remark after (4.25)] to

$$\int_{-1}^{1} dx' M(x, x') M^{-1}(x', x'') = \delta(x - x''), \qquad (4.18)$$

where $\delta(x)$ is the Dirac delta function. Note that the integration measure $dx' = \sin \gamma' d\gamma'$ corresponds to uniformly distributed points on the sphere.

The Dirac delta function on the rhs of (4.18) may be expressed as a sum of Legendre polynomials. The orthogonality relation for Legendre polynomials,

$$\int_{-1}^{1} dx P_l(x) P_{l'}(x) = \frac{2}{2l+1} \delta_{ll'}, \qquad (4.19)$$

implies that for $x$ and $x'$ in the interval $[-1, 1]$ one has

$$\delta(x - x') = \sum_l \frac{2l+1}{2} P_l(x) P_l(x'). \qquad (4.20)$$

This allows us to rewrite the rhs of (4.18).

We now demonstrate a convenient way to invert $M$, if it is diagonal in a Legendre function basis. Suppose that $M$ and its inverse are written as a sum of products of Legendre polynomials

$$M(x, x') = \sum_l \frac{2l+1}{2} Q_l P_l(x) P_l(x'), \quad \text{and} \qquad (4.21)$$

$$M^{-1}(x, x') = \sum_l \frac{2l+1}{2} q_l P_l(x) P_l(x'). \qquad (4.22)$$

Substituting these into the lhs of (4.18), using (4.19) to carry out the integral over $x'$, and replacing the rhs of (4.18) with (4.20), we obtain

$$\sum_l \frac{2l+1}{2} Q_l q_l P_l(x) P_l(x'') = \sum_l \frac{2l+1}{2} P_l(x) P_l(x''). \qquad (4.23)$$

Identifying the Legendre expansion coefficients on both sides implies

$$Q_l q_l = 1. \qquad (4.24)$$

Thus, to invert $M$, we simply invert the diagonal values, setting $q_l = 1/Q_l$.

We now use (4.24) to invert the continuous matrix $s(x, x')$, by letting $M \equiv s$. Identifying (4.17) and (4.21) gives $Q_l = 4\hbar^4 C_l^2$, so (4.24) implies $q_l = 1/4\hbar^4 C_l^2$. The continuous matrix inverse is then given by (4.22) as

$$s^{-1}(x, x') = (4\hbar^4)^{-1} \sum_{l=2}^{\infty} \frac{2l+1}{2C_l^2} P_l(x) P_l(x'). \qquad (4.25)$$

Note that the $l = 0$ and $l = 1$ terms are absent from this sum. This means that (4.17) and (4.25) cannot satisfy (4.18). Rather, they satisfy this equation if the $l = 0$ and $l = 1$ terms are dropped from the Dirac delta function on the rhs of (4.20). This is because $s(x, x')$ has a two-dimensional null space formed from all linear combinations of $P_0(x)$ and $P_1(x)$. The inverse operator $s^{-1}(x, x')$ is only defined on the complement of that null space.

Later, we will need to evaluate sums over uniformly distributed pulsars, when the number of pulsar pairs goes to infinity. These sums, which typically involve the vector $\mu_{ab} \equiv \mu_j$, and the matrices $s_{jk}$ and $s_{jk}^{-1}$, can be converted to Riemann sums and hence to integrals as follows.

To construct vectors of dimension $N_{bins}$ and matrices of size $N_{bins} \times N_{bins}$, first select $N_{bins}$ values of $x = \cos \gamma$ which are uniformly spaced in the interval $(-1, 1)$:

$$x_j = (2j + 1 - N_{bins})/N_{bins} \quad \text{for } j = 0, 1, \dots N_{bins} - 1. \qquad (4.26)$$

The vector $\mu_j$ is obtained by evaluating the function $\mu_u$, given in (1.1) or (4.13), at these discrete values: $\mu_j = \mu_u(x_j)$. (For notational convenience we use both $x$ and $\gamma$ as arguments.) The matrix $s_{jk}$ is defined by (4.15), and (4.17) shows that its entries are obtained by evaluating the continuous function at the grid of points, hence

$$s_{jk} \equiv s(x_j, x_k). \qquad (4.27)$$

While the continuous matrix inverse $s^{-1}(x, x')$ defined by (4.25) may also be evaluated on the discrete grid of points, this would not yield the matrix inverse of (4.27): it has the wrong normalization to satisfy

$$\sum_k s_{jk} s_{km}^{-1} = \delta_{jm}, \qquad (4.28)$$

even in the limit of large $N_{bins}$.

To obtain an inverse matrix $s_{jk}^{-1}$ satisfying (4.28), in the limit of many pairs, one must multiply $s^{-1}(x_j, x_k)$ by two factors of $\Delta x \equiv 2/N_{bins}$. From (4.25) this gives

$$s_{jk}^{-1} = \left(\frac{2}{N_{bins}}\right)^2 (4\hbar^4)^{-1} \sum_{l=2}^{\infty} \frac{2l+1}{2C_l^2} P_l(\cos \gamma_j) P_l(\cos \gamma_k). \qquad (4.29)$$

Note that the equality in (4.29) holds only in the limit of large $N_{bins}$. In that limit, (4.28) may be demonstrated by converting the sum to a Riemann sum and then to an integral [36]. We make use of $s_{jk}^{-1}$ and $s^{-1}(x, x')$ in Secs. VII C and VIII C 3.

The failure of $s_{jk}^{-1}$ to exactly equal $s^{-1}(x_j, x_k)$ for a finite number of pulsar pairs has a close analog in Fourier analysis (where the inverse operation is the Fourier transform). Suppose that we are handed a function of time $y(t)$ and its Fourier inverse $\tilde{y}(f)$, which is a function of frequency. Now, we want to construct a discrete time sequence and its discrete Fourier inverse. The discrete time sequence is trivial to obtain: sample $y(t)$ at uniformly spaced times $t = k\Delta t$, where $\Delta t$ is the time step and $k = 0, 1, \dots, N - 1$. In contrast, the discrete inverse (a sequence in frequency $f$) is *not* obtained by simply sampling the continuous inverse $\tilde{y}(f)$ at the corresponding discrete frequencies. However, for a band-limited function, they approach one another in the large-$N$ limit with $N\Delta t$ held constant. A more detailed discussion may be found in Appendix D of Ref. [14].

### D. Relating the variance of the narrow-bin optimal estimator to the cosmic variance

Using (4.8) and the results of Sec. III D, an expression for the variance of the optimal estimator in the narrow-bin limit follows immediately from setting $\gamma_j = \gamma_k$. For the blocks along the diagonal of $\mathbb{C}$, this is true to a good approximation. So, setting $j = k$ and $\gamma_j = \gamma_k \equiv \gamma$ in (4.8), we see that $\mathbb{1}$ is an eigenvector of the matrix block $C_{jj}$. Thus, the average value of an entry on one of these diagonal blocks is

$$s = 2\hbar^4 \frac{1}{2} \int_0^{\pi} d\beta \sin \beta \left( \mu_{++}^2(\gamma, \beta) + \mu_{\times\times}^2(\gamma, \beta) \right) = 2\hbar^4 \widetilde{\mu}^2(\gamma), \qquad (4.30)$$

where we dropped the indices from $s_{jk} = s_{jj}$ on the lhs of the above equation.

The (cosmic variance) function $\widetilde{\mu^2}(\gamma)$ is computed in Appendix G in [11] and may also be obtained directly from the diagonal of the cosmic covariance (4.11) by setting $\gamma_j = \gamma_k = \gamma$. We find

$$\frac{\widetilde{\mu^2}(\gamma)}{\mu_u^2(0)} = \frac{49}{48}\cos^2\gamma - \frac{15}{16} - \frac{3}{2}(\cos^2\gamma + 3)\ln\left(\frac{1-\cos\gamma}{2}\right)\ln\left(\frac{1+\cos\gamma}{2}\right)$$
$$+ \frac{3}{4}(\cos\gamma - 1)(\cos\gamma + 3)\ln\left(\frac{1-\cos\gamma}{2}\right) + \frac{3}{4}(\cos\gamma + 1)(\cos\gamma - 3)\ln\left(\frac{1+\cos\gamma}{2}\right), \quad (4.31)$$

which agrees with Eq. (G11) in [11]. We have divided $\widetilde{\mu^2}(\gamma)$ by $\mu_u^2(0)$ to make (4.31) independent of the choice of normalization of the Hellings and Downs correlation.

The variance of the optimal estimator of the Hellings and Downs correlation is given by (3.24) as the average value $s$ of the entries of the covariance matrix, given by (4.30). Hence,

$$\sigma_{opt}^2(\gamma) = s = 2\hbar^4\widetilde{\mu^2}(\gamma) = \sigma_{cos}^2(\gamma), \quad (4.32)$$

where $\sigma_{cos}^2(\gamma)$ is exactly the cosmic variance found by different arguments in Appendix C5 in [11]. This provides a simple and alternative way to derive the cosmic variance.

### E. Numerical illustration showing the approach to the cosmic variance

There are two limiting cases for which we can give analytic expressions for the variance. If there is only a single pulsar pair at each angular separation, then the variance is the total variance, given by $\sigma_{tot}^2(\gamma)$ in (2.11). If there are a large number of uniformly distributed pulsar pairs, then the variance is the cosmic variance $\sigma_{cos}^2(\gamma)$, given by (4.32). Here, we study the transition between these two limits, when the number of pulsar pairs is finite.

In the left panel of Fig. 6, we show the transition from $\sigma_{tot}^2(\gamma)$ to $\sigma_{cos}^2(\gamma)$ as we average the correlations of more and more pulsar pairs with angular separation $\gamma$. For these plots, we simulated 10 different realizations of $n_{pairs} = 1$, 2, 5, 10, 25, 50, 100, and 200 pulsar pairs, all separated by the same angle $\gamma$ and distributed uniformly on the sky. We then calculated $\sigma_{opt}^2(\gamma)$ for each value of $n_{pairs}$ and each value of $\gamma$. (To reduce the fluctuations associated with the random placement of the pulsar pairs on the sky, we averaged together the results of the 10 different realizations to get the final set of transition plots.)

In the right panel of Fig. 6, we plot the minimum number of pulsar pairs needed for the optimal variance $\sigma_{opt}^2(\gamma)$ to reach a value that lies within $1/e \approx 0.37$ of the cosmic variance $\sigma_{cos}^2(\gamma)$. In terms of the standard deviation $\sigma_{opt}(\gamma)$, this corresponds to reaching a value of approximately $1.17\sigma_{cos}(\gamma)$ for each value of $\gamma$. Near the minima of $\sigma_{cos}(\gamma)$, in the vicinity of $\gamma \approx 54°$ and $\gamma \approx 126°$, about 6000 pulsar pairs are required.

## V. VARIANCE OF THE OPTIMAL BINNED ESTIMATOR

### A. Beating the cosmic variance

One interesting consequence of doing the binned estimation is that for some ranges of angular separation $\gamma$,

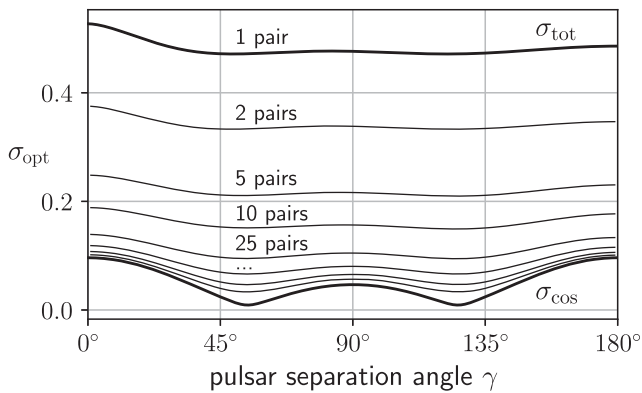
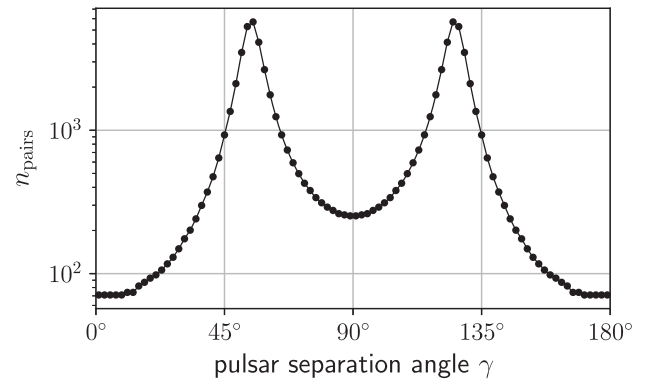

FIG. 6. Left panel: the variance of the optimal estimator as the number of pulsar pairs at each angular separation $\gamma$ is increased. Shown in the plot, from top to bottom, is $\sigma_{opt}(\gamma)$ calculated for 1, 2, 5, 10, 25, 50, 100, and 200 pairs. (For this plot we have set $\hbar^4 = 1/2$.) Right panel: the minimum number of pulsar pairs needed for $\sigma_{opt}^2(\gamma)$ to reach a value that is $1/e \approx 37\%$ larger than the cosmic variance. For this number of pulsar pairs, $\sigma_{opt}^2(\gamma) \approx (1 + 1/e)\sigma_{cos}^2(\gamma)$.

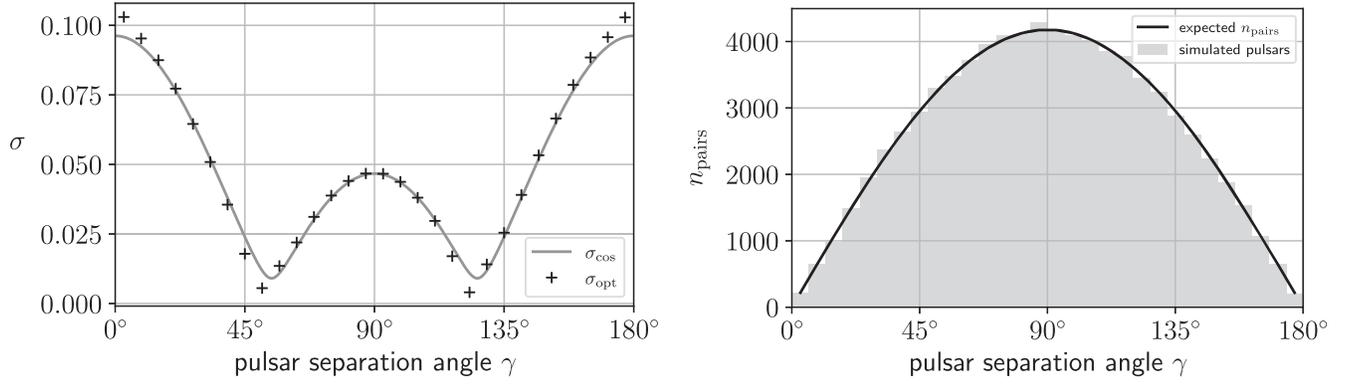

FIG. 7. Left panel: the variance of the optimal estimator for a simulation with 400 pulsars distributed uniformly on the sky, with 6° angular bins. This is compared to the (narrow-bin) cosmic variance. The variance of the optimal estimator dips below the cosmic variance for sufficiently large numbers of pulsar pairs per bin. This is possible because angular resolution has been sacrificed, see Appendix E. (For this plot we have set $\hbar^4 = 1/2$.) Right panel: the number of pulsar pairs in each angular-separation bin for the left-hand-panel case. The expected number is proportional to $\sin \gamma$.

the loss of angular resolution allows one to "beat" the cosmic variance for a sufficiently large number of pulsar pairs in the bin. This is illustrated numerically in Fig. 7 for the case of a simulation involving 400 pulsars distributed uniformly on the sky and 6° angular-separation bins. For this calculation, we used normalization condition III A, which corresponds to $\mu_{bin} \equiv \mu^{\top} \mathbb{1} / n_{pairs}$. An analytic demonstration of this result, for a simple "two-component" angular-separation bin is given in Appendix E.

### B. Examples

Here, we apply the results of Sec. III A to make plots showing the total variance $\sigma_{tot}^2(\gamma)$, cosmic variance $\sigma_{cos}^2(\gamma)$, and variance $\sigma_{opt}^2(\gamma)$ of the optimal estimator for a finite number of pulsars distributed nonuniformly on the sky, with 6° angular-separation bins. We use the current sky locations of the pulsars monitored by three active PTA collaborations: European Pulsar Timing Array (EPTA), North American Nanohertz Observatory for Gravitational Waves (NANOGrav), and Parkes Pulsar Timing Array (PPTA). We also construct an International Pulsar Timing Array (IPTA) by forming the union of the pulsars monitored by the individual PTAs. See Table IV in Appendix H for the names and angular coordinates of the pulsars.

The individual PTA collaborations are currently monitoring 42, 66, and 26 pulsars respectively, with a total of 88 distinct pulsars for the IPTA [37]. A skymap of the pulsars is shown in Fig. 8, which is a Mollweide projection in equatorial coordinates. Note that the pulsars are clustered in the direction of the galactic center, which has equatorial coordinates (ra, dec) = (17h46m, −29°); this is indicated by black dots in Fig. 8. For reference, the center of the sky maps is (ra, dec) = (12h, 0°).

Plots showing the expected Hellings and Downs correlation plus/minus the uncertainties associated with the total variance, cosmic variance, and variance of the optimal binned estimator are given in Fig. 9.

For these plots, we use thirty 6°-wide angular-separation bins, equally spaced between 0 and 180°. To model the relative amplitude of the expected correlations and their uncertainties, we set $\hbar^2 = 0.5622h^2$ and $h^2 = 1$. This corresponds to timing-residual measurements for a Gaussian ensemble of binary-inspiral sources, as described in Appendix B. Finally, for the optimal binning, we have chosen to normalize the weights according to condition

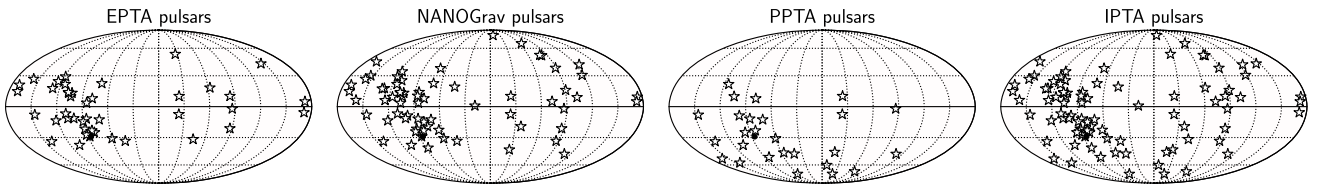

FIG. 8. Sky locations of the pulsars employed by the EPTA, NANOGrav, PPTA, and IPTA collaborations. The black dot indicates the direction to the Galactic center. Table IV in Appendix H lists the pulsar names and sky locations.

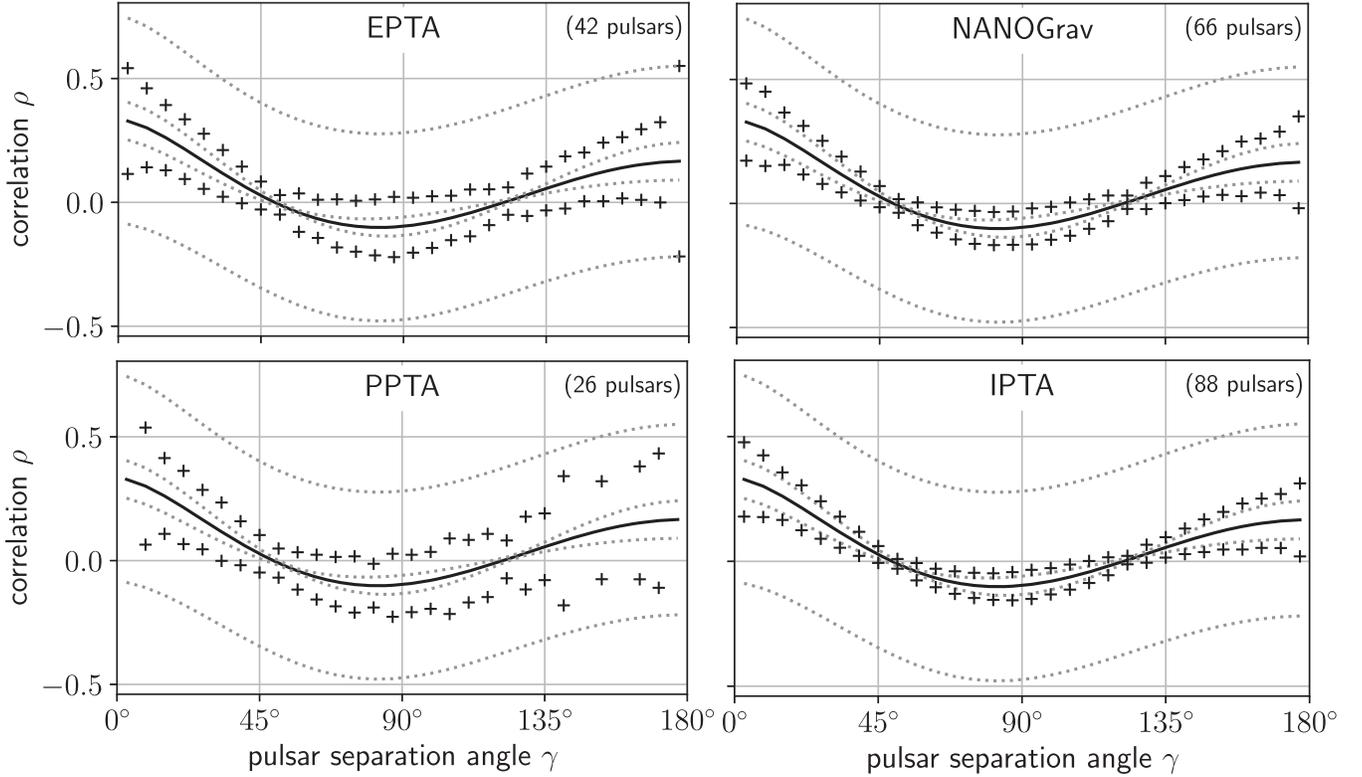

FIG. 9. The mean Hellings and Downs timing-residual correlations (solid black curve) and the $\pm\sigma_{\mathrm{opt}}$ uncertainty of the optimal estimator with $30 \times 6°$ angular-separation bins ("+" symbols). This illustrates the expected match for data which has no measurement noise or intrinsic pulsar noise. We take the pulsars currently monitored by the EPTA, NANOGrav, and PPTA collaborations, and merge these for the IPTA. (See Fig. 3 for a closeup of the IPTA variances.) We assume the binary-inspiral GW background described in Appendix B, which has $\hbar^2/h^2 \approx 0.5622$ (timing-residual correlations, $\alpha = 1$) and normalize the mean with $h^2 = 1$. Two other curves are shown for comparison. The dotted lines furthest from the solid black curve are the single pulsar pair variance: $\pm\sigma_{\mathrm{tot}}$. The dotted lines closest to the solid black curve are the cosmic variance corresponding to an infinite number of pulsar pairs: $\pm\sigma_{\mathrm{cos}}$. The PPTA plot has no "+"-symbols at $3°$, $147°$, $159°$, and $177°$ since $n_{\mathrm{pairs}} = 0$ in those bins.

(c) in Sec. III A, that is, $\mu_{\mathrm{bin}} \equiv \mu_{\mathrm{u}}(\gamma_{\mathrm{bin}})$, where $\gamma_{\mathrm{bin}} \equiv \gamma^{\top}\mathbb{1}/n_{\mathrm{pairs}}$ is the average angular separation of the pulsar pairs in a bin.

These plots do not include any contributions from measurement noise or intrinsic pulsar noise. As such, they show the fundamental limits on the recovery of the Hellings and Downs correlation, based on using optimal estimators for the given set of pulsars and a specific choice of binning.

Since the PTAs are improving their results by adding additional pulsars, these plots will improve with time. They may be easily constructed as follows:

(i) starting with a set of pulsar sky directions, form all pulsar pairs;

(ii) divide the pulsar pairs among a set of angular-separation bins;

(iii) for each bin, compute the average angular separation $\gamma_{\mathrm{bin}}$ of the $n_{\mathrm{pairs}}$ pairs in that bin;

(iv) for each bin, choose a normalization $\mu_{\mathrm{bin}}$ for the optimal estimator as described in Sec. III A, for example, $\mu_{\mathrm{bin}} \equiv \mu_{\mathrm{u}}(\gamma_{\mathrm{bin}})$;

(v) for each bin, calculate the $n_{\mathrm{pairs}}$-dimensional vector $\mu_{ab}$ and $\mathbb{C}_{ab,cd}$ for all pulsar pairs $ab$ and $cd$ in the bin, using (2.4) and (2.10); the $n_{\mathrm{pairs}} \times n_{\mathrm{pairs}}$-dimensional covariance matrix $C$ is defined by that block of $\mathbb{C}$;

(vi) for each bin, calculate the variance $\sigma_{\mathrm{opt}}^2$ of the optimal estimator using (3.11) for each bin.

Following this simple procedure, interested readers can produce similar plots for different arrays of pulsars and different choices of angular bins.

## VI. NON-PTA DETECTORS

The focus of this paper is the Hellings and Downs correlation (1.1) induced by GWs in PTAs. However, to aid in our understanding, it is useful to consider "non-PTA" detectors, which have a different response to GWs than a PTA. This is particularly helpful in understanding the properties of estimators of the squared GW strain $h^2$ [38]. These estimators, which are linear combinations of pulsar-pair correlation measurements $\rho_{ab}$, are described in Sec. VII.

### A. A more general form for the detector response and its expected correlation

The most familiar non-PTA detector type is the "short-arm interferometer" GW detector, described in Appendix D. 4 of [32]. To construct this, place one arm of a LIGO-like detector along the line of sight $\hat{p}$ to a pulsar. Along that path, a laser is used to measure the separation between a pair of freely falling mirrors located near Earth. We assume that the distance between the mirrors (or the effective distance, if there are multiple bounces [16]) is small compared to the GW wavelength. In practice, at typical PTA frequencies this detector would be overwhelmed by low-frequency noise, but in our idealized thought experiment we replace pulsar redshift with the fractional change in length between the mirrors.

Let us contrast the response ("antenna pattern") of this LIGO-like detector with one arm along direction $\hat{p}$, to the response of a PTA pulsar with sky direction $\hat{p}$. For a unit-amplitude circularly polarized plane GW traveling in direction $\hat{\Omega}$, the redshift responses are

$$\text{one-arm LIGO:} \quad F(\hat{\Omega}) = \frac{1}{2}(2\pi i f \mathcal{L})\hat{p}^\mu \hat{p}^\nu e_{\mu\nu}(\hat{\Omega}) \quad \text{and} \tag{6.1}$$

$$\text{PTA pulsar:} \quad F(\hat{\Omega}) = \frac{1}{2}\frac{\hat{p}^\mu \hat{p}^\nu}{1 + \hat{\Omega}\cdot\hat{p}}e_{\mu\nu}(\hat{\Omega})\left[1 - \eta\, e^{-2\pi i f \mathcal{L}(1+\hat{\Omega}\cdot\hat{p})}\right]. \tag{6.2}$$

Here, the Einstein summation convention is in effect for the spatial indices $\mu$ and $\nu$, $e_{\mu\nu} \equiv e_{\mu\nu}^+ - i e_{\mu\nu}^\times$ is the (complex) circular polarization tensor formed from the traditional linear GW basis (Eq. (D6) in [11]), and $f\mathcal{L}$ is the dimensionless distance to the pulsar, measured in GW wavelengths.

The factor of $2\pi i f$ in (6.1) arises because $F(\hat{\Omega})$ is the redshift (rather than the timing residual) response. This factor is discussed in Appendix A. It can be obtained starting from the conventional (one-way) formula for GW strain [14], which is the fractional variation $h = \Delta\mathcal{L}/\mathcal{L} = \frac{1}{2}\hat{p}^\mu \hat{p}^\nu e_{\mu\nu}(\hat{\Omega})$, where $\mathcal{L}$ is the detector arm length. The redshift is the time derivative of $\Delta\mathcal{L}$, introducing a factor of $2\pi i f$, from which (6.1) follows.

In (6.2), the unity term in square brackets gives the Earth term, and the second term in square brackets gives the pulsar term. If desired, a similar pulsar term could be incorporated into the one-arm LIGO-like detector [39]. The real constant $\eta$ allows us to correctly incorporate the pulsar term by setting $\eta = 1$ [40], or to turn off the pulsar term by setting $\eta = 0$. "Flipping this switch" makes it easier to separate and understand the effects of the pulsar term.

The response (6.1) of a one-arm LIGO-like detector may be obtained as the low-frequency ($\equiv$ short-arm) limit of the PTA response (6.2) [41,42]. This is no coincidence. If our fictional one-arm LIGO-like detector could also operate at wavelengths much *smaller* than its arm length, then its response would be described by the same expression [(6.2) with $\eta = 1$] as a PTA pulsar [43]. If we start with (6.2), and take the limit $f\mathcal{L} \to 0$, we obtain (6.1). This is because the pulsar term in (6.2) may be approximated using $e^x \approx 1 + x$ and when subtracted from the Earth term, cancels the denominator $1 + \hat{\Omega}\cdot\hat{p}$.

This short-arm limit is valid for the currently operating advanced LIGO detectors in their most sensitive frequency band around $f \approx 100$ Hz where the GW wavelength $\approx 3000$ km is much larger than the 4 km arm length [16]. By assumption/construction, the short-arm limit is valid for our hypothetical one-arm LIGO-like detectors.

Equations (6.1) and (6.2) for the detector response can be recast in harmonic space by decomposing the GW perturbations into a sum of (gradients and curls of) spherical harmonic functions $Y_{lm}(\theta, \phi)$, as described in [32]. Any GW field may be expressed in this form, whose value is determined by a countably infinite set of amplitudes, labeled by $l = 2, 3, \ldots$ and $m = -l, -l+1, \ldots, l$. For the Gaussian ensemble, each of these amplitudes is an independent zero-mean Gaussian random variable, with a variance that depends only upon $l$. The response of the above two types of detectors to the $l$, $m$ mode is (respectively) given by the first term of Eq. (D4) from Ref. [32], and by Eq. (92) (including the pulsar term) from that same reference. These are

$$\text{one-arm LIGO:} \; F_{lm}(\hat{p}) \propto \delta_{l2}Y_{lm}(\hat{p}) \; \text{and} \tag{6.3}$$

$$\text{PTA pulsar:} \; F_{lm}(\hat{p}) \propto C_l^{1/2}\left[1 - \eta\, e^{-2\pi i f \mathcal{L}(1+\hat{\Omega}\cdot\hat{p})}\right]Y_{lm}(\hat{p}), \tag{6.4}$$

where the $C_l$ are given by (4.14), and $l$-independent numerical factors are omitted. The key point is that the one-arm LIGO-like detector responds to *only the $l = 2$ modes*, whereas the PTA pulsar detector responds to *all modes with $l \geq 2$*. Here, $Y_{lm}(\hat{p})$ is the spherical harmonic function $Y_{lm}(\theta_p, \phi_p)$, where $\theta_p, \phi_p$ are the spherical coordinates of the pulsar direction vector $\hat{p}$.

Our claim about the distinct patterns of response may appear surprising. Fundamentally, it is a statement about the differing antenna patterns of the detectors, and can be understood by comparing (6.1) and (6.2). Said succinctly, a LIGO-like detector has a purely quadrupolar antenna pattern and response, whereas a PTA pulsar detector responds to all modes.

To determine the output at time $t$ of an idealized (one- or two-arm) LIGO-like detector, it is sufficient to know the five independent spatial components of the metric strain $h_{\mu\nu}(t)$ at the location of the instrument. In contrast, this knowledge is *insufficient* to predict the redshift of a pulsar [44] at time $t$, because of the $1 + \hat{\Omega} \cdot \hat{p}$ term in the denominator of (6.2). Because the denominator depends upon $\hat{\Omega}$, to predict the redshift one must also know the specific contribution to $h_{\mu\nu}(t)$ from each different sky direction. This requires (or, given observations, provides) information that is not local to the neighborhood of the detector.

As shown in (6.1), a LIGO detector with one arm in direction $\hat{p}^\mu$ has an antenna pattern $\hat{p}^\mu \hat{p}^\nu$. A two-arm differential LIGO detector (arm directions $\hat{p}_a^\mu$, $\hat{p}_b^\mu$) has an antenna pattern $\hat{p}_a^\mu \hat{p}_a^\nu - \hat{p}_b^\mu \hat{p}_b^\nu$. Both are *quadratic* functions of the three orthogonal Cartesian directions $\hat{p}^x$, $\hat{p}^y$, $\hat{p}^z$, restricted to a two-sphere. Thus, their expression in terms of spherical harmonics contains only $l = 2$ terms. In fact, this connection between the order of the polynomial (here two) and the order $l$ of the harmonic is the origin of the name "spherical harmonics." In contrast, the factor of $1 + \hat{\Omega} \cdot \hat{p}$ in the denominator of (6.1) gives the PTA pulsar detector an antenna pattern that also responds to modes with $l > 2$.

Note that even for a (one- or two-arm) LIGO-like detector, the five components of $h_{\mu\nu}$ are functions of time, and thus have (in principle) infinite information content. The corresponding $l = 2$ amplitudes may be thought of as functions of time, although it is simpler to regard them as functions of frequency as in Eq. (14) of [32]. Thus, there are five of these per frequency bin; the restriction to $l = 2$ only reduces the amount of spatial information.

Returning back to (6.3) and (6.4), we can use these response functions to calculate the expected value of the correlation between detector responses $Z_a(t)$ and $Z_b(t)$ for the standard Gaussian ensemble. The expected correlation is obtained by summing products of the response functions for two detectors over both $l$ and $m$ [12,13,32]. Let $\hat{p}_a$ and $\hat{p}_b$ denote the directions to pulsars $a$ and $b$, or the arm directions of a pair of one-arm LIGO-like detectors. Then

$$\langle \rho_{ab} \rangle = h^2 \mu_{ab} = h^2 \sum_l \sum_{m=-l}^{l} F_{lm}(\hat{p}_a) F_{lm}^*(\hat{p}_b), \quad (6.5)$$

where $\mu_{ab}$ is the overlap reduction function for the two detectors.

To evaluate the expected correlation for either of the two different GW detector types, substitute (6.3) or (6.4)

into (6.5) and use the *addition theorem* for spherical harmonics,

$$\sum_{m=-l}^{l} Y_{lm}(\hat{p}_a) Y_{lm}^*(\hat{p}_b) = \frac{2l+1}{4\pi} P_l(\cos\gamma_{ab}), \quad (6.6)$$

to evaluate the sum over $m$, where $\cos\gamma_{ab} \equiv \hat{p}_a \cdot \hat{p}_b$. One obtains

one-arm LIGO: $\mu_{ab} = P_2(\cos\gamma_{ab})$ and $\quad (6.7)$

PTA pulsar: $\mu_{ab} = \sum_l (2l+1) C_l P_l(\cos\gamma_{ab}) [1 + \eta^2 \delta_{ab}],$

$(6.8)$

where for (6.8) we assume (i) that the Earth-to-pulsar distances are many GW wavelengths $1 \ll f\mathcal{L}$, and (ii) that the Earth-to-pulsar distances $\mathcal{L}_a$ and $\mathcal{L}_b$ differ by many GW wavelengths $1 \ll |f(\mathcal{L}_a - \mathcal{L}_b)|$. This means that the Earth-pulsar cross-terms average to zero, as discussed in detail in [11]. Thus, (6.8) contains only Earth-Earth and pulsar-pulsar terms. For pairs of distinct pulsars $a \neq b$, the Kronecker delta term vanishes, and we recover the Hellings and Downs correlation $\mu_u(\gamma_{ab})$ as written in (4.13) and (4.14).

It should now be clear to the reader that the two examples we have just given are members of an infinite collection of possible detector response functions. To define these, we replace $(2l+1)C_l$ in (6.8) by an arbitrary set of (positive) coefficients $Q_l$. A one-arm LIGO-like detector has $Q_l = 0$ for $l \neq 2$, whereas a PTA has a specific set of coefficients which are nonzero for all $l \geq 2$. By analogy, one can construct (at least, in our imagination) detectors for any (positive) choice of these coefficients, with corresponding correlation matrices $\mu_{ab}$.

The generalized detector response functions are defined in analogy with (6.8). They have a correlation matrix

$$\mu_{ab} = (1 + \eta^2 \delta_{ab}) U_{ab}, \quad (6.9)$$

where $U$ is a real symmetric matrix of dimension $N_{pul} \times N_{pul}$, having entries

$$U_{ab} = \sum_l Q_l P_l(\cos\gamma_{ab}). \quad (6.10)$$

The non-negative quantities $Q_l \geq 0$ which define $U_{ab}$ determine which Legendre polynomials appear. The non-negative constant $\eta^2 \geq 0$ which appears in (6.9) allows us to control the pulsar term. Depending upon the context, $N_{pul}$ denotes either the number of pulsars in a PTA, or the number of non-PTA detectors.

A PTA has $\eta = 1$ and $Q_l = (2l+1)C_l$, where the $C_l$ are defined by (4.14). Alternative detectors have different values for the $Q_l$, but always with at least one of these values nonzero. For example, we can have $Q_l$ nonzero for a single value $l = L$, or for some finite range, say $2 \le l \le L$. The one-arm LIGO-like detector corresponds to $Q_l = \delta_{l2}$ and $\eta = 0$.

### B. Eigenvectors and eigenvalues of the correlation matrix $\mu_{ab}$

We now investigate the properties of the correlation matrix $\mu_{ab}$ defined by (6.9) and (6.10). We start by finding its eigenvalues and eigenvectors. For this, it is helpful to first examine the matrix $U_{ab}$.

The eigenvalues of $U_{ab}$ must be non-negative, since we assume $Q_l \ge 0$ in (6.10). To see this, use the addition theorem (6.6) to write $U_{ab}$ as

$$U_{ab} = \sum_l \frac{4\pi}{2l+1} Q_l \sum_{m=-l}^{l} Y_{lm}(\hat{p}_a) Y_{lm}^*(\hat{p}_b), \qquad (6.11)$$

where $\hat{p}_a$ and $\hat{p}_b$ are the sky directions to the two pulsars. Now, consider the real number $v^\top U v = \sum_{a,b} v^a U_{ab} v^b$, where $v^a$ is any real $N_{\text{pul}}$-component vector. The matrix $U$ is positive definite if $v^\top U v > 0$ for all $v$, or positive semidefinite if $v^\top U v \ge 0$ for all $v$. We have

$$\sum_{a,b} v^a U_{ab} v^b = \sum_l \frac{4\pi}{2l+1} Q_l \sum_{m=-l}^{l} |y_{lm}|^2, \qquad (6.12)$$

where $y_{lm} \equiv \sum_a v^a Y_{lm}(\hat{\Omega}_a)$. Since we have assumed $Q_l \ge 0$, the rhs of (6.12) is non-negative. This establishes that $U$ is positive semidefinite, meaning that its eigenvalues are either zero or positive.

To count the vanishing eigenvalues of $U_{ab}$, first consider a simple case: suppose that the $Q_l$ are nonzero only for $l = L$. Since $Q_L$ is positive, from (6.12) an eigenvector $v^a$ with vanishing eigenvalue must satisfy

$\sum_a v^a Y_{Lm}(\hat{p}_a) = 0$, for $m = -L, …, L$. This is a set of $2L + 1$ (real) homogeneous linear equations in $N_{\text{pul}}$ (real) variables, which are the components of $v^a$. If the number of these components satisfies $N_{\text{pul}} \le 2L + 1$, then for generic sky locations, the system of linear equations is overdetermined. Since the equations are homogeneous, the unique solution is the trivial one, $v^a = 0$. Thus, for $N_{\text{pul}} \le 2L + 1$ there are no eigenvectors with eigenvalue zero, so that all $N_{\text{pul}}$ eigenvalues are positive. Alternatively, if $N_{\text{pul}} > 2L + 1$, then the equations $\sum_a v^a Y_{Lm}(\hat{p}_a) = 0$ have $N_{\text{pul}} - (2L + 1)$ linearly independent nonvanishing solutions $v^a$, corresponding to $N_{\text{pul}} - (2L + 1)$ vanishing eigenvalues. The remaining $2L + 1$ eigenvalues are positive.

The result generalizes immediately to the case where more than one $Q_l$ is nonzero. Thus, define

$$N_{\text{DOF}} \equiv \sum_{\{l | Q_l \ne 0\}} (2l + 1), \qquad (6.13)$$

which is the number of real degrees of freedom (DOF) associated with the nonzero $Q_l$. If $N_{\text{pul}} \le N_{\text{DOF}}$, then all of the eigenvalues of $U$ are positive. Otherwise, $U$ has $N_{\text{DOF}}$ positive eigenvalues and $N_{\text{pul}} - N_{\text{DOF}}$ vanishing eigenvalues.

For notational clarity, for any square matrix $M$, we define the integer

$$N_M^+ \equiv \text{number of nonzero eigenvalues of } M. \qquad (6.14)$$

For example, $N_U^+$ is the number of positive eigenvalues of $U_{ab}$. (Here, the matrices $M$ of interest are symmetric and hence can be diagonalized. For such matrices, $N_M^+$ is also the rank of $M$.)

With this notation, the number of positive and zero eigenvalues of $U_{ab}$ are therefore given by

$$\text{number of positive eigenvalues of } U = N_U^+ = \min(N_{\text{DOF}}, N_{\text{pul}}), \quad \text{and}$$
$$\text{number of zero eigenvalues of } U = N_{\text{pul}} - N_U^+ = \max(N_{\text{pul}} - N_{\text{DOF}}, 0). \qquad (6.15)$$

Note that a real PTA has $Q_l \ne 0$ for $l \ge 2$, so $N_U^+ = N_{\text{pul}}$. Thus, for a real PTA, $U_{ab}$ has no zero eigenvalues, regardless of the number of pulsars.

It is now straightforward to count the number of vanishing eigenvalues of the correlation matrix $\mu_{ab}$. If the pulsar term is absent, so that $\eta = 0$, then from (6.9) one has $\mu_{ab} = U_{ab}$, and (6.15) also describes the eigenvalues of $\mu_{ab}$. On the other hand, if $\eta \ne 0$, it is easy to show that $\mu_{ab}$ is positive definite and has no zero eigenvalues for any GW response function.

To see this, note from (6.10) that the diagonal terms of $U_{ab}$ are all positive and equal to one another, because (no summation convention) $P_l(\cos \gamma_{aa}) = P_l(\cos 0) = P_l(1) = 1$, and $Q_l \ge 0$. Denote these diagonal elements by

$$U(0) \equiv U_{aa} = \sum_l Q_l > 0. \qquad (6.16)$$

Since they are equal, from (6.9) we can write

$$\mu_{ab} = U_{ab} + \eta^2 U(0) \delta_{ab}, \qquad (6.17)$$

where $\delta_{ab}$ is the identity matrix. Thus, if $v^a$, $\lambda$ are an eigenvector and corresponding eigenvalue of $U$, then $v^a$, $\lambda + \eta^2 U(0)$ are the eigenvector and corresponding eigenvalue for $\mu_{ab}$. Since $\lambda \geq 0$, and $\eta^2$ and $U(0)$ are both positive, it follows that all the eigenvalues of $\mu_{ab}$ are positive. Thus, for any GW detector response, if the pulsar term is present, so that $\eta \neq 0$, then all of the eigenvalues of $\mu_{ab}$ are positive.

### C. Inverse of the correlation matrix $\mu_{ab}$

Having understood the properties of the eigenvalues of the correlation matrix $\mu_{ab}$, we now turn to the definition of its inverse. This is needed for the calculations in Sec. VII A.

For a PTA, or for any GW detector response which includes a pulsar term, the matrix $\mu_{ab}$ has only positive eigenvalues, and hence its inverse $\mu_{ab}^{-1}$ is defined. However, if (i) the pulsar term is absent (so $\eta = 0$), (ii) the number of degrees of freedom $N_{\mathrm{DOF}}$ is finite, and (iii) the number of pulsars is greater than $N_{\mathrm{DOF}}$, then the matrix $\mu_{ab}$ has vanishing eigenvalues, and the conventional matrix inverse is not defined.

In such cases, it is helpful to define a modified inverse, which we denote by $\mu_{ab}^{+}$. This is

$$\mu_{ab}^{+} \equiv \sum_{\{n|\lambda_n > 0\}} \lambda_n^{-1} v_n^a v_n^b, \qquad (6.18)$$

where $\lambda_n$, $v_n^a$ are the eigenvalues and corresponding normalized eigenvectors of $\mu_{ab}$, and the sum includes only those terms for which the eigenvalues do not vanish. This is called the *Moore-Penrose pseudoinverse* of the matrix: it was found by Moore [45,46] and rediscovered three decades later by Penrose [47]; see [48] for an account. If all of the eigenvalues of $\mu_{ab}$ are positive, then $\mu_{ab}^{+}$ is the ordinary matrix inverse $\mu_{ab}^{-1}$, so that $\sum_b \mu_{ab}^{+} \mu_{bc} = \sum_b \mu_{ab} \mu_{bc}^{+} = \delta_{ac}$.

If some of the eigenvalues of $\mu_{ab}$ are zero, then $\mu_{ab}^{+}$ is still well defined, but has some properties that are different than the normal matrix inverse. In particular, $\sum_b \mu_{ab}^{+} \mu_{bc} = \sum_b \mu_{ab} \mu_{bc}^{+} = \pi_{ac}$, where $\pi_{ab}$ is a projector onto the subspace spanned by the eigenvectors corresponding to the nonvanishing eigenvalues. Thus,

$$\pi_{ab} \equiv \sum_{\{n|\lambda_n > 0\}} v_n^a v_n^b. \qquad (6.19)$$

In a basis formed from the eigenvectors $v_n^a$, (6.19) shows that the matrix $\pi_{ab}$ vanishes off the diagonal, and has $N_\pi^+ = N_\mu^+$ ones and $N_{\mathrm{pul}} - N_\mu^+$ zeroes along the diagonal. The trace of $\pi_{ab}$, which is basis independent, is $\sum_a \pi_{aa} = N_\mu^+$. These follow immediately from the trace of (6.19), which is the number of nonzero eigenvalues of $\mu_{ab}$.

## VII. ESTIMATING THE SQUARED GW STRAIN $h^2$

PTAs are constructed to detect low-frequency GWs; the most important quantity that they should determine is the squared amplitude $h^2$ of those waves. In this section, we investigate how PTA correlation data can best be used to estimate or constrain $h^2$; the estimator is denoted $\hat{h}^2$.

If the pulsar measurements are free of noise, or if their noise is uncorrelated between different pulsars, then our expectation is that including more pulsars in the PTA will reduce the uncertainty in the strain estimate, because it adds more information. This is indeed the case. However, we will see that the rate at which that uncertainty decreases, and whether it tends to zero for large numbers of pulsars, depends upon the details.

In this section, we consider three cases, starting with data from $N_{\mathrm{pul}}$ pulsars. (i) All of the possible $N_{\mathrm{pul}}(N_{\mathrm{pul}} + 1)/2$ pulsar-pair correlation measurements, including both auto- and cross-correlations, are used. (ii) Only $N_{\mathrm{pul}}$ auto-correlation measurements are used. (iii) Only the $N_{\mathrm{pul}}(N_{\mathrm{pul}} - 1)/2$ cross-correlation measurements from distinct pairs of pulsars are used. We will show that the typical uncertainties in the $h^2$ estimates have different scaling behavior for these different cases, as the number of pulsars grows. To set the stage, the three different scaling behaviors are illustrated in Fig. 10, which shows the (inverse) variance of the estimators. Later in this section, we derive these results, and discuss and compare them in detail.

In addition, to gain some insight, we will also consider GW responses which differ from those of PTA pulsars. As described in Sec. VI, these could correspond, for example, to fictitious LIGO-like detectors, floating in space near

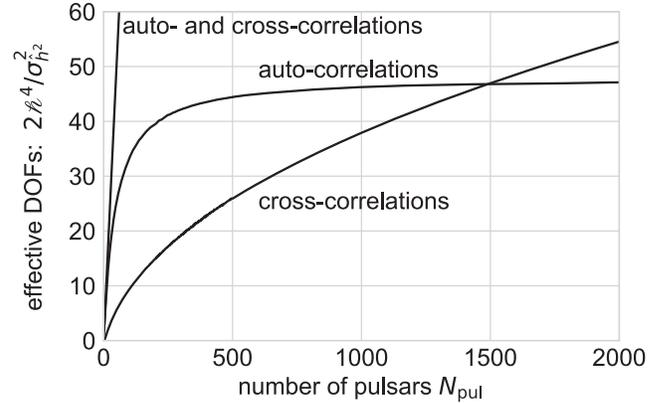

FIG. 10. The inverse variance of three different estimators of the squared GW strain, plotted as a function of the number of uniformly distributed pulsars. For large numbers of pulsars, the estimator based only on auto-correlations has a finite limit corresponding to 48 degrees of freedom. The estimator which employs only cross-correlations, and the estimator which employs both auto- and cross-correlations, have variances that vanish as $N_{\mathrm{pul}} \to \infty$. However, the latter vanishes much faster.

Earth, but sensitive to the same low frequencies as current PTAs.

In all three cases (i)–(iii) above, we use $\hat{h}^2$ to denote the optimal estimator of the squared strain GW $h^2$. This estimator is a linear combination of (measured or calculated) pulsar-pair correlations $\rho_{ab}$ with weights $w_{ab}$, and is defined by

$$\hat{h}^2 \equiv \sum_{ab \in \mathcal{S}} w_{ab} \rho_{ab}. \quad (7.1)$$

The only difference between cases (i)–(iii) is the choice of the set $\mathcal{S}$, which determines *which* pulsar pairs $ab$ are used in the summation. For case (i) we have $\mathcal{S} = \{ab | a \leq b\}$, for case (ii) we have $\mathcal{S} = \{ab | a = b\}$, and for case (iii) we have $\mathcal{S} = \{ab | a < b\}$. In what follows, we will indicate these limits on the sums using the shorthand forms $a \leq b$, $a = b$, and $a < b$ respectively.

Note that (7.1) is proportional to $\rho_{ab}$, and does not include a constant (i.e., independent of $\rho_{ab}$) term. The constant term is also absent from (3.1), which gives the optimal estimator of the Hellings and Downs correlation for a particular angular-separation bin. The situation here is the same. A constant term in (7.1) is only needed if (i) auto-correlations ($\rho_{aa}$) are included in the sum and (ii) there is pulsar or measurement noise. The reasoning given after (3.1) also applies here. If both (i) and (ii) hold, then (7.1) is not general enough to form an unbiased estimator for $h^2$: an additional constant term must be subtracted from the rhs of (7.1). Since we primarily focus on the ideal noise-free case, we do not include a constant term in (7.1) or in what follows. We will return to this issue later, in Sec. IX, when we consider the effects of noise.

To find the optimal (unbiased, minimum-variance) estimator for the squared strain $h^2$, we determine the weights in (7.1) with exactly the same technique that was used in Sec. III A. The resulting estimator $\hat{h}^2$ is a linear combination of the pulsar-pulsar correlations $\rho_{ab}$, given by

$$\hat{h}^2 = \left( \sum_{ab,cd \in \mathcal{S}} \mu_{ab} \mathbb{C}^{-1}_{ab,cd} \mu_{cd} \right)^{-1} \sum_{ef,gh \in \mathcal{S}} \mu_{ef} \mathbb{C}^{-1}_{ef,gh} \rho_{gh}. \quad (7.2)$$

The expected value of $\hat{h}^2$ is $h^2$, and the variance of $\hat{h}^2$ is

$$\sigma^2_{\hat{h}^2} = \left( \sum_{ab,cd \in \mathcal{S}} \mu_{ab} \mathbb{C}^{-1}_{ab,cd} \mu_{cd} \right)^{-1}. \quad (7.3)$$

Note that $\hat{h}^2$ is *not* Gaussian distributed: it is a sum of products of multivariate-Gaussian random variables. Thus, it is described by a *generalized chi-squared distribution* [49].

The pairs $ab$, $cd$, etc. which appear in (7.2) and (7.3) differ for cases (i)-(iii), exactly as described after (7.1).

In the following three subsections, we investigate the behavior of the variance for those three different cases.

Expressions (7.2) and (7.3) for $\hat{h}^2$ and its variance $\sigma^2_{\hat{h}^2}$ are similar in form to the standard cross-correlation estimator given in [42,50,51]; in the literature this is called the "optimal statistic." Our expression differs in two ways. First, the standard cross-correlation estimator corresponds to choice (iii) above, and only includes cross-correlations. With choices (i) and (ii) our definition can also include auto-correlations. Second, the standard optimal statistic assumes that the covariance matrix is dominated by instrumental and pulsar noise. Our expressions are general enough to cover the entire range from noise-free to noise-dominated, although we primarily focus on the noise-free case.

In the absence of noise, the optimal estimator $\hat{h}^2$ is determined by the (observed or calculated) correlations $\rho_{ab}$ and by the geometry of the pulsar sky positions. The details of the GW spectral shape and its overall scale do not matter. This is because the (nongeometrical) factors of $h^4$, which appear in the expression for the covariance matrices $\mathbb{C}$ [see (2.10)] cancel out from the numerator and denominator of (7.2). In contrast, the variance $\sigma^2_{\hat{h}^2}$ is proportional to $h^4$ and hence depends on $h^2$. This is the true *but unknown* value of the squared GW strain that we are trying to estimate.

If these estimators are applied to observational data, one cannot avoid the question: "what value of $h^2$ should be used to evaluate $\sigma^2_{\hat{h}^2} = \sigma^2_{\hat{h}^2}(h^2)$?" In this case, since the true value of $h^2$ is unknown, we show how to construct a "self-consistent interval" for the $h^2$ estimate. We obtain this by evaluating $h^2 \pm \sigma_{\hat{h}^2}(h^2)$ for many different (assumed) values of $h^2$, and finding the interval of $h^2$ values consistent within $\pm \sigma_{\hat{h}^2}(h^2)$ of the observed value of $\hat{h}^2$. We illustrate this method graphically at the end of Sec. VII D. A more general discussion of this self-consistent-interval method, allowing for the presence of measurement and pulsar noise, may be found in Sec. IX D.

## A. Case (i): Using auto-correlations and cross-correlations

We first consider the optimal estimator (7.1) for the squared strain $h^2$ which fully employs *both* the cross-correlations *and* the auto-correlations. While this is defined by the index set $a \leq b$, in this subsection, it is convenient to let $a$ and $b$ run independently over the range $1, ..., N_{pul}$, so that (7.1) includes $N^2_{pul}$ terms. In addition to including the auto-correlation terms $a = b$, this also includes each possible cross-correlation term twice. These two possible terms, for example $\rho_{37}$ and $\rho_{73}$, are each assigned identical weights $w_{37} = w_{73}$. This means that (by doubling the cross-correlation weights and keeping the auto-correlation weights the same) the double sum over all $N^2_{pul}$ pulsar pairs could be replaced by an equivalent sum over $a \le b$. However, for the calculations in this subsection, it is helpful to keep all $N_{\text{pul}}^2$ terms in (7.1), so that

$$\hat{h}^2 \equiv \sum_{a=1}^{N_{\text{pul}}} \sum_{b=1}^{N_{\text{pul}}} w_{ab} \rho_{ab} \equiv \sum_{a,b} w_{ab} \rho_{ab}, \qquad (7.4)$$

where the final summation is shorthand notation for the previous two sums. The same extension of the summations apply to (7.2) and (7.3).

Here, and elsewhere in this paper, we use the following convention for commas that occur under summation signs. A comma between indices under a summation sign, for example the "$a$, $b$" in (7.4), means that within the sum, each index ranges separately over $1, \dots, N_{\text{pul}}$. If there is no comma, for example "$ab$", then the sum over pairs $ab$ might not be uniformly indexed, but instead is described by some set $\mathcal{S}$, which is described in the text. Note that commas that are not under a summation sign, for example in objects like $\mathbb{C}_{ab,cd}$, $\mathbb{D}_{ab,cd,ef}$, and $\mathbb{E}_{ab,cd,ef,gh}$, have no such meaning. Instead, they serve to group the arguments into pulsar pairs.

The variance of $\hat{h}^2$ is given by (7.3), where each index $a$, $b$, $c$, and $d$ is independently summed from 1 to $N_{\text{pul}}$. We can evaluate (7.3) in closed form, by explicitly inverting the covariance matrix. [With these index conventions, the inverse $\mathbb{C}^{-1}$ is formally undefined. However, as we show below, its action on symmetric matrices such as $\mu_{ab}$ is well defined, so that (7.3) is also well defined.]

With the index labeling conventions described in the previous two paragraphs, the inverse of $\mathbb{C}_{ab,cd}$ (restricted to the space of symmetric matrices) is

$$\mathbb{C}_{ab,cd}^{-1} = \frac{1}{4} \hbar^{-4} \left( \mu_{ac}^{-1} \mu_{bd}^{-1} + \mu_{ad}^{-1} \mu_{bc}^{-1} \right). \qquad (7.5)$$

To understand the restriction, consider the original definition of $\mathbb{C}$ as given in (2.10), but with the indexing conventions of this subsection: each of the four indices runs from 1 to $N_{\text{pul}}$. $\mathbb{C}$ is a linear map from the space of $N_{\text{pul}} \times N_{\text{pul}}$ square matrices to itself. Such a linear map may be expressed in terms of the eigenvalues $\Lambda_n$ and normalized eigenvectors $W_n^{ab}$ of $\mathbb{C}$, defined by

$$\sum_{c,d} \mathbb{C}_{ab,cd} W_n^{cd} = \Lambda_n W_n^{ab} \quad \text{and} \quad \sum_{a,b} W_n^{ab} W_m^{ab} = \delta_{nm}. \quad (7.6)$$

In terms of these, $\mathbb{C}$ may be expressed as

$$\mathbb{C}_{ab,cd} = \sum_n \Lambda_n W_n^{ab} W_n^{cd}, \qquad (7.7)$$

which can be checked by confirming the action of $\mathbb{C}$ on each of the eigenvectors in turn, using (7.6).

Since the number of eigenvectors is the dimension of the space of linearly independent $N_{\text{pul}} \times N_{\text{pul}}$ matrices, the range of $n$ is $n = 1, 2, \dots, N_{\text{pul}}^2$. From the definition of $\mathbb{C}$ given in (2.10), one can see that it is symmetric in the indices $c$ and $d$. Thus, it maps any antisymmetric matrix to zero, and therefore must have at least $N_{\text{pul}}(N_{\text{pul}} - 1)/2$ vanishing eigenvalues. The remaining $N_{\text{pul}}(N_{\text{pul}} + 1)/2$ eigenvalues are positive if $\mu_{ab}$ is positive definite.

By the inverse of $\mathbb{C}$, we mean the matrix

$$\mathbb{C}_{ab,cd}^{-1} = \sum_{\{n | \Lambda_n > 0\}} \Lambda_n^{-1} W_n^{ab} W_n^{cd}, \qquad (7.8)$$

which excludes the eigenvectors corresponding to vanishing eigenvalues. Strictly speaking, $\mathbb{C}^{-1}$ as defined in (7.8) is the Moore-Penrose pseudoinverse $\mathbb{C}^+$ as discussed in Sec. VI B. That is to say, if we multiply $\mathbb{C}^{-1}$ with $\mathbb{C}$, we obtain the projection operator

$$\sum_{c,d} \mathbb{C}_{ab,cd} \mathbb{C}_{cd,ef}^{-1} = \sum_{\{n | \Lambda_n \neq 0\}} W_n^{ab} W_n^{ef}. \qquad (7.9)$$

Because the zero eigenvectors are missing from the sum, the rhs of (7.9) is not the identity matrix $\delta_{ae} \delta_{bf}$. Instead, it is a projector onto the space of symmetric matrices.

Making use of (2.10) and (7.5) we obtain

$$\sum_{c,d} \mathbb{C}_{ab,cd} \mathbb{C}_{cd,ef}^{-1}$$
$$= \frac{1}{4} \sum_{c,d} \left( \mu_{ac} \mu_{bd} + \mu_{ad} \mu_{bc} \right) \left( \mu_{ce}^{-1} \mu_{df}^{-1} + \mu_{cf}^{-1} \mu_{de}^{-1} \right)$$
$$= \frac{1}{2} (\delta_{ae} \delta_{bf} + \delta_{af} \delta_{be}). \qquad (7.10)$$

The rhs is precisely (7.9). It is easy to show that this is the linear map of rank $N_{\text{pul}}(N_{\text{pul}} + 1)/2$ which maps a matrix $M$ onto $(M + M^\top)/2$. Thus, it preserves a symmetric matrix, annihilates an antisymmetric matrix, and extracts the symmetric part of an arbitrary matrix.

Please be warned that (7.5) does not give the inverse of $\mathbb{C}$, as it is used in the other sections of this paper. First, the indexing structure is different, and second, for a binned quantity we are typically only interested in a submatrix block of the full $\mathbb{C}$. The inverse of a submatrix block is not the corresponding submatrix block of the inverse.

We now evaluate the (inverse) variance $\sigma_{\hat{h}^2}^2$ given in (7.3). Making use of (7.5) we find

$$(\sigma_{\hat{h}^2}^2)^{-1} = \sum_{a,b,c,d} \mu_{ab} \mathbb{C}_{ab,cd}^{-1} \mu_{cd}$$
$$= \frac{1}{4} \hbar^{-4} \sum_{a,b,c,d} \mu_{ab} \left( \mu_{ac}^{-1} \mu_{bd}^{-1} + \mu_{ad}^{-1} \mu_{bc}^{-1} \right) \mu_{cd}$$
$$= \frac{1}{2} \hbar^{-4} \sum_b \delta_{bb}$$
$$= \frac{1}{2} \hbar^{-4} N_{\text{pul}}. \qquad (7.11)$$

Thus, the fractional variance for the squared GW strain is

$$\frac{\sigma_{\hat{h}^2}^2}{\langle \hat{h}^2 \rangle^2} = \frac{2}{N_{\text{pul}}} \frac{\hslash^4}{h^4}, \tag{7.12}$$

where we have inverted (7.11) and used $\langle \hat{h}^2 \rangle = h^2$. This is the result that holds for a physical PTA. It demonstrates that if both the auto-correlation and cross-correlation information are fully exploited, then the squared GW strain can be determined with arbitrary precision, given a sufficient number of low-noise pulsars. Each additional pulsar provides an additional degree of freedom, as seen in Fig. 11.

Initially, we found this result surprising for two reasons. First, in the many-pulsar limit, the optimal estimator of the Hellings and Downs correlation has a nonzero (cosmic) variance. Why does the variance of the squared-strain estimate vanish in that same limit? Second, previous work (e.g., Eq. (C40) in Sec. C 4 of Ref. [11]) shows that an apparently similar measure of the squared strain (the time average of $s \equiv h_{\mu\nu} h^{\mu\nu}$) has fractional variance 2/5. How can the estimator $\hat{h}^2$ defined by (7.4) provide more precise information?

To answer these two questions, consider GW detector responses which differ with those of a PTA, as examined in

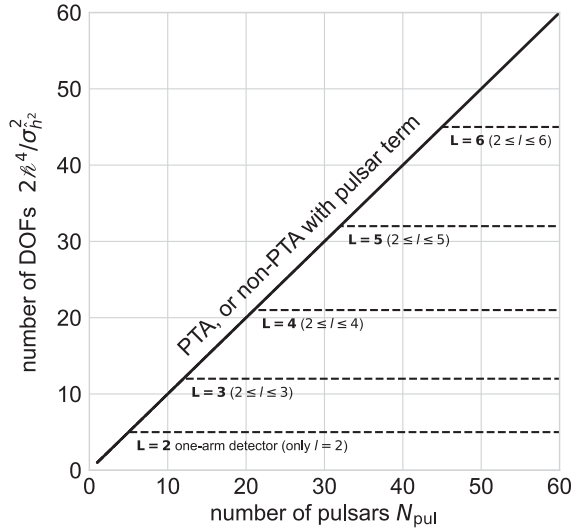

FIG. 11. The inverse variance (7.14) of the squared-strain estimator which optimally exploits both auto- and cross-correlations, as a function of the number of pulsars employed. Diagonal line: for any type of detector with a pulsar term, such as PTAs, $\sigma_{\hat{h}^2}^{-2}$ is a linear function of the number of pulsars, corresponding to one degree of freedom per pulsar. Hence, given a sufficient number of pulsars, the uncertainty in the $h^2$ estimate can be made arbitrarily small. Dashed curves: if there is no pulsar term *and* the detector is sensitive to only a finite set of modes, then the inverse variance follows the diagonal line until it reaches the corresponding number of degrees of freedom. Then it saturates and follows the corresponding dashed curve, so $\sigma_{\hat{h}^2}^2$ does not decrease further as more pulsars are added.

Sec. VI A. If there are a sufficient number of pulsars, then for some responses, the correlation matrix $\mu_{ab}$ has vanishing eigenvalues. This means that $\mu_{ab}$ is not invertible. In such cases, the $\mu_{ab}^{-1}$ which appear on the rhs of (7.5) change to $\mu_{ab}^{+}$, which is the Moore-Penrose pseudoinverse of $\mu_{ab}$ defined by (6.18). Then the inverse variance becomes

$$\begin{aligned}
(\sigma_{\hat{h}^2}^2)^{-1} &= \frac{1}{4} \hslash^{-4} \sum_{a,b,c,d} \mu_{ab} \left( \mu_{ac}^{+} \mu_{bd}^{+} + \mu_{ad}^{+} \mu_{bc}^{+} \right) \mu_{cd} \\
&= \frac{1}{4} \hslash^{-4} \left( \sum_{b,c} \pi_{bc} \pi_{bc} + \sum_{b,d} \pi_{bd} \pi_{bd} \right) \\
&= \frac{1}{2} \hslash^{-4} \sum_{b} \pi_{bb} \\
&= \frac{1}{2} \hslash^{-4} N_{\mu}^{+}. \tag{7.13}
\end{aligned}$$

Here, the projection operator $\pi_{ab}$ is defined in (6.19), from which it follows that its trace is $N_{\mu}^{+}$, which is the number of positive eigenvalues of the correlation matrix $\mu_{ab}$.

There are two possibilities for the variance. Looking at the rhs of (7.13):

(i) If the pulsar term is present, meaning that $\eta > 0$ in (6.9), then $\mu_{ab}$ has no vanishing eigenvalues. In this case, $N_{\mu}^{+} = N_{\text{pul}}$ and the inverse variance of (7.13) agrees with that of (7.11).

(ii) If the pulsar term is absent, corresponding to $\eta = 0$ in (6.9), then $\mu_{ab} = U_{ab}$. In this case, (6.15) shows that $N_{\mu}^{+} = N_{U}^{+} = \min(N_{\text{pul}}, N_{\text{DOF}})$. Hence,

$$(\sigma_{\hat{h}^2}^2)^{-1} = \frac{1}{2} \hslash^{-4} \ \min(N_{\text{pul}}, N_{\text{DOF}}). \tag{7.14}$$

Here, as the number of pulsars grows, the behavior depends upon the number of DOF to which the detector responds. If $N_{\text{DOF}}$ is infinite, then the behavior is identical to case (i). However, if the number of DOF is finite, then the variance decreases to a minimum value and does not decrease further as more pulsars are added. This is shown by the dashed curves in Fig. 11. It demonstrates that for the squared GW strain, "cosmic variance" is a property of the detector rather than the Universe.

This provides a clear answer to the second question. Consider an array of one-arm LIGO-like ($L = 2$) detectors, each of which is sensitive to some linear combination of $N_{\text{DOF}} = 2L + 1 = 5$ modes of the GW field. If the array contains five or more detectors (i.e., $N_{\text{pul}} \geq 5$), then from (7.14) the variance is

$$\sigma_{\hat{h}^2}^2 = \frac{2}{5} \hslash^4. \tag{7.15}$$

But five one-arm LIGO-like detectors are enough to measure all five independent components of the GW strain

$h_{\mu\nu}$ at Earth, since $h_{\mu\nu}$ is a traceless symmetric $3 \times 3$ matrix. Adding additional LIGO-like detectors only adds redundant information, because the strain is completely determined: there is no additional DOF to measure. The corresponding fractional uncertainty $\sigma_{\hat{h}^2}/\langle\hat{h}^2\rangle$ agrees exactly with $\sigma_s/\langle s\rangle$ from Eq. (C40) in Ref. [11] where the cosmic variance of $s \equiv$ time average$(h_{\mu\nu}h^{\mu\nu})$ is computed as an instructive exercise. This makes sense, because $s$ is determined by exactly the same five DOF.

We can also clearly see why a PTA detector provides vanishing variance as $N_{\rm pul} \to \infty$, which answers the first question above. In contrast to a LIGO-like detector, which is only sensitive to the $l = 2$ mode, a PTA detects all modes of the GW field. This is due to the divisor $1 + \hat{\Omega} \cdot \hat{p}$ that appears in (6.2). This factor of $(1 + x)^{-1} = 1 - x + x^2 - x^3 + \cdots$, where $x = \hat{\Omega} \cdot \hat{p}$, provides additional powers of $\hat{p}$, contributing the $l > 2$ spherical harmonics $Y_{lm}(\hat{p})$ in (6.4). Each independent mode provides additional information, and with enough pulsars, all of those modes can be accessed. Hence, by employing a large enough number of pulsars, (7.12) shows that the uncertainty in the squared GW strain can be reduced as much as desired.

While the calculations of this section and the previous section require careful attention to detail, the results themselves have an appealingly simple and intuitive physical interpretation. The optimal combination of auto- and cross-correlation terms extracts as much information as is possible from the GWs. Each additional pulsar added to the array provides another such degree of freedom, reducing the variance in proportion. In the absence of the pulsar term, this continues until the number of pulsars reaches the number of degrees of freedom probed by the detector. Then the variance saturates, because the detector cannot extract further information from the local measurements. On the other hand, if the pulsar term is present, then each additional pulsar brings access to a new set of degrees of freedom: the GW fluctuations in the space-time region surrounding that new pulsar. In this case, the variance continues to decrease without bound as new pulsars are added to the PTA.

### B. Case (ii): Using only auto-correlations

In this section, we construct an optimal estimator $\hat{h}^2 = \sum_a w_{aa}\rho_{aa}$ of $h^2$, which uses only pulsar-pair auto-correlations $\rho_{aa}$, where $a = 1, 2, \ldots, N_{\rm pul}$. Because it has access to much less information than the estimator of the previous section, we expect this estimator to have a larger variance. Indeed, unlike the estimators of $h^2$ that use only cross-correlations or both cross- and auto-correlations, we will see that this estimator has a (nonzero) cosmic variance. Even with noise-free measurements from an infinite number of pulsars distributed uniformly on the sky, this auto-correlation-only estimator cannot estimate $h^2$ with arbitrary precision.

We evaluate the inverse variance of the estimator by starting with the general expression (7.3) and including only the auto-correlation terms in the sums. Thus,

$$\left(\sigma_{\hat{h}^2}^2\right)^{-1} = \sum_{a,b}\mu_{aa}\mathbb{C}_{ab}^{-1}\mu_{bb}, \tag{7.16}$$

where $\sum_{a,b}$ is same shorthand notation used in (7.4), and $\mathbb{C}_{ab}^{-1} \equiv \mathbb{C}_{aa,bb}^{-1}$ is the inverse of the $N_{\rm pul} \times N_{\rm pul}$ covariance matrix [52]

$$\begin{aligned}
\mathbb{C}_{ab} &\equiv \mathbb{C}_{aa,bb} \\
&= 2\hbar^4\mu_{ab}^2 \\
&= 2\hbar^4 U_{ab}^2(1 + \eta^2\delta_{ab})^2 \\
&= 2\hbar^4\left(U_{ab}^2 + (2\eta^2 + \eta^4)U^2(0)\delta_{ab}\right). \tag{7.17}
\end{aligned}$$

The second line follows from (2.10), the third line follows from (6.9), and the final line follows from (6.16), because all of the diagonal elements of $U_{ab}$ are equal to $U(0)$. Note that $U_{ab}^2$ denotes the matrix whose elements are the squares of the elements of the matrix $U_{ab}$. It does not denote the matrix product of $U$ with itself.

To evaluate the inverse variance, use (6.17) to write the diagonal elements of $\mu_{ab}$ as $\mu_{aa} = (1 + \eta^2)U(0)$. Hence, the inverse variance (7.16) is

$$\left(\sigma_{\hat{h}^2}^2\right)^{-1} = (1 + \eta^2)^2U^2(0)\sum_{a,b}\mathbb{C}_{ab}^{-1}, \tag{7.18}$$

which is proportional to the grand sum of $\mathbb{C}_{ab}^{-1}$. This is no surprise, because Sec. III C shows that narrow-band estimators give rise to a grand sum. Indeed, this estimator is as narrow band as possible: it only has contributions from the bin at zero separation angle. Later, Sec. III D shows that if the vector $\mathbb{1}$ (here, a vector of dimension $N_{\rm pul}$ containing all ones) is an eigenvector of $\mathbb{C}_{ab}$, then (7.18) can be computed from its eigenvalue. We now employ the same argument, showing that if there are many pulsars uniformly distributed on the sky, then $\mathbb{1}$ is an eigenvector of $\mathbb{C}_{ab}$ and hence of $\mathbb{C}_{ab}^{-1}$.

It is easy to see that $\mathbb{1}$ is an eigenvector of $\mathbb{C}_{ab}$ in the large pulsar limit. Focus attention on the $U_{ab}^2$ term in (7.17), since $\mathbb{1}$ is trivially an eigenvector of the other term, which is proportional to the identity matrix $\delta_{ab}$. Equation (6.10) shows that $U_{ab}^2$ only depends on the angle $\gamma_{ab}$ between the directions $\hat{p}_a$ and $\hat{p}_b$ to pulsars $a$ and $b$. It follows from symmetry that averaging uniformly over $\hat{p}_b$ (for fixed $\hat{p}_a$) yields a result which is independent of $\hat{p}_a$.

To compute the action of $U_{ab}^2$ and $\delta_{ab}$ on $\mathbb{1}/N_{\rm pul}$, we use (6.10) to write $U_{ab}^2$ in terms of Legendre polynomials, then average over the directions $\hat{p}_b$. When there are many pulsars uniformly distributed on the sky, this gives

$$
\begin{aligned}
\frac{1}{N_{\text{pul}}} \sum_b U_{ab}^2 &\simeq \frac{1}{4\pi} \int d\hat{p}_b \, U_{ab}^2 \\
&= \frac{1}{4\pi} \int d\hat{p}_b \sum_{l,l'} Q_l Q_{l'} P_l(\cos\gamma_{ab}) P_{l'}(\cos\gamma_{ab}) \\
&= \frac{1}{2} \int_{-1}^{1} dx \sum_{l,l'} Q_l Q_{l'} P_l(x) P_{l'}(x) \\
&= \sum_{l,l'} Q_l Q_{l'} \frac{\delta_{ll'}}{2l+1} \\
&= \sum_l \frac{Q_l^2}{2l+1}.
\end{aligned}
\tag{7.19}
$$

In the first line, the approximation sign is a reminder that we have averaged over a large number of pulsars uniformly distributed on the sky. The second line follows from the definition of $U_{ab}$ in (6.10). The third line follows by setting $x = \cos\gamma_{ab}$, also explicitly demonstrating that the average is independent of the direction $\hat{p}_a$ of pulsar $a$. The final two lines follow from the orthogonality of the Legendre polynomials (4.19). Now, we return to (7.17) and consider the term proportional to $\delta_{ab}$. The action of this term on $\mathbb{1}/N_{\text{pul}}$ is trivial, since

$$
\frac{1}{N_{\text{pul}}} \sum_b \delta_{ab} = \frac{1}{N_{\text{pul}}}.
\tag{7.20}
$$

So, combining (7.19) and (7.20) with (7.17), it follows that

$$
\frac{1}{N_{\text{pul}}} \sum_b \mathbb{C}_{ab} \simeq 2\hbar^4 \left( \sum_l \frac{Q_l^2}{2l+1} + \frac{1}{N_{\text{pul}}} (2\eta^2 + \eta^4) U^2(0) \right) \equiv \frac{\lambda}{N_{\text{pul}}}.
\tag{7.21}
$$

Thus, $\mathbb{1}$ is an eigenvector of $\mathbb{C} \equiv \mathbb{C}_{ab}$, with the eigenvalue $\lambda$ defined above.

Using this last result together with the results of Sec. III D, we can now write down an expression for the variance of the auto-correlation-only estimator of $h^2$ in the large pulsar limit. From (7.18), we immediately have

$$
\sigma_{\hat{h}^2}^2 = \frac{1}{(1+\eta^2)^2 U^2(0)} \frac{1}{\sum_{a,b} \mathbb{C}_{ab}^{-1}}.
\tag{7.22}
$$

Then, using (3.23), we can replace $\left(\sum_{a,b} \mathbb{C}_{ab}^{-1}\right)^{-1}$ with $\lambda/N_{\text{pul}}$. This gives

$$
\sigma_{\hat{h}^2}^2 \simeq \frac{2\hbar^4}{(1+\eta^2)^2 U^2(0)} \sum_l \frac{Q_l^2}{2l+1} + \frac{2\hbar^4}{N_{\text{pul}}} \frac{(2\eta^2 + \eta^4)}{(1+\eta^2)^2}.
\tag{7.23}
$$

The variance (7.23) is nonzero in the limit $N_{\text{pul}} \to \infty$. In contrast, the variances of the $h^2$ estimators constructed from both auto- and cross-correlations (as we saw in Sec. VII A) and from cross-correlations only (as we shall see in Sec. VII C) vanish in this limit.

We now evaluate the variance (7.23) of the squared-strain estimator which employs only PTA auto-correlations, in the many-pulsar limit. Set $Q_l = (2l+1)C_l$ with the $C_l$ given by (4.14). Then

$$
\frac{1}{U^2(0)} \sum_l \frac{Q_l^2}{2l+1} = \frac{1}{U^2(0)} \sum_l (2l+1) C_l^2 = \frac{1}{2\hbar^4} \frac{\sigma_{\cos}^2(0)}{U^2(0)} = \frac{1}{12},
\tag{7.24}
$$

where we used (4.15) with $\gamma = \gamma' = 0$ to get the second equality, and where $\sigma_{\cos}^2(0)/U^2(0) = 2\hbar^4 \mu^2(0)/\mu_u^2(0) = \hbar^4/6$ follows from (4.31) and (4.32). Then, setting $\eta = 1$ in (7.23) to properly incorporate the pulsar term, and making use of (7.24) to evaluate the first term, we obtain

$$
\sigma_{\hat{h}^2}^2 \simeq \frac{\hbar^4}{24} \left( 1 + \frac{36}{N_{\text{pul}}} \right)
\tag{7.25}
$$

for the variance of the auto-correlation-only PTA estimator of $h^2$. This approximation, and numerically determined values for $\sigma_{\hat{h}^2}^{-2}$, are shown in Fig. 12. As $N_{\text{pul}} \to \infty$, these asymptote to $\hbar^4/24$, corresponding to 48 effective DOF. Without the pulsar term, also shown in the figure, the convergence is more rapid, to 12 effective DOF.

It is also instructive to evaluate the variance of this $h^2$ estimator for an array of one-arm LIGO-like detectors in the $N_{\text{pul}} \to \infty$ limit. Here, we set $Q_l = \delta_{l2}$, so that $U(0) = 1$ and set $\eta = 0$ to turn off the pulsar term. The variance (7.23) is then

$$
\sigma_{\hat{h}^2}^2 \simeq \frac{2}{5} \hbar^4,
\tag{7.26}
$$

which agrees exactly with our expectations from (7.15) for the estimator of $h^2$ that uses both auto- and cross-correlations. This variance, and the corresponding variance for an $l = 3$ only detector (meaning, $Q_l = \delta_{l3}$) are also shown in Fig. 12.

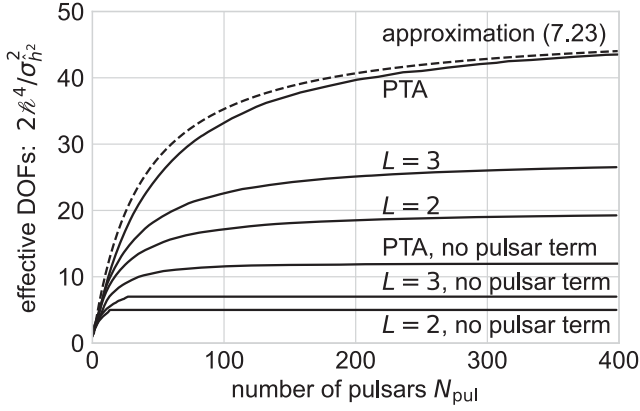

FIG. 12. Inverse variance of strain estimators formed entirely from auto-correlations, as a function of the number of (uniformly distributed) pulsars, for different types of GW detectors, determined numerically. All have nonvanishing cosmic variance as $N_{pul} \to \infty$. We plot $2\hbar^4/\sigma_{\hat{h}^2}^2$, which is the effective number of statistically independent degrees of freedom. The top solid curve for a PTA and the dashed approximation of (7.25) both asymptote to 48. The next two solid curves show the corresponding quantities for detectors having only a single $l$ mode, i.e., $Q_l = \delta_{lL}$ with $L = 3$ or $L = 2$. These include a pulsar term and asymptote to $4 \times 7 = 28$ and $4 \times 5 = 20$ DOF respectively. The bottom three "no pulsar term" curves have the pulsar term turned off ($\eta = 0$). As explained in the text, the $L = 2$ and $L = 3$ curves saturate at $N_{pul} = (L + 1)(2L + 1)$, corresponding to 15 and 28 pulsars respectively, with 5 and 7 DOF. A PTA without the pulsar term asymptotes to 12 effective DOF (see text).

The reader will note from Fig. 12 that if there is no pulsar term present, then the variance of the finite-mode detectors saturates if the number of pulsars is large enough. For the detector of mode $L$, meaning $Q_l = \delta_{lL}$, this occurs when the number of pulsars $N_{pul} = (L + 1)(2L + 1)$. For an array of one-arm LIGO-like detectors ($L = 2$) the variance saturates if the array contains $N_{pul} = 15$ or more detectors. Note that this is more than are needed if both cross-correlation and auto-correlation information is used. Then, as previously discussed, just five detectors are sufficient to saturate the variance.

We can determine the number of detectors needed to saturate the variance if the detectors only respond to a finite number of DOF and the pulsar term is absent. It suffices to decompose the matrix whose entries are the squares of the entries of $U_{ab}$ into Legendre polynomials. The matrix $U_{ab}^2$ then takes the form (6.10), with coefficients formed from quadratic combinations of the $Q_l$'s, weighted with Wigner $3j$ symbols [53,54]. For $Q_l = \delta_{lL}$, the $U_{ab}^2$ matrix has $(L + 1)(2L + 1)$ DOF, as defined by (6.13). If modes $l \le L_{max}$ are included, then the number of DOF is bounded by $(2L_{max} + 1)^2$. So, if the number of pulsars exceeds this, then the variance saturates for the same reason as explained in Sec. VII A.

## C. Case (iii): Using only cross-correlations

This section examines the optimal estimator $\hat{h}^2$ of the squared GW strain which is constructed using only PTA cross-correlations. In other words, the pulsar-pair correlations $\rho_{ab}$ which appear in the sum (7.2) are restricted to distinct pulsars $a < b$.

We computed the variance (7.3) of this estimator numerically, for $N_{pul}$ PTA pulsars distributed randomly on the sky. The results are plotted in Fig. 13. They show that $\sigma_{\hat{h}^2}^2$ tends to zero as $N_{pul} \to \infty$. When there are more than about 30 pulsars, the variance is well approximated (blue dashed curve) by

$$\sigma_{\hat{h}^2}^2 \approx 1.61 \frac{\sqrt{N_{pul} + 75}}{N_{pul}} \hbar^4. \qquad (7.27)$$

Thus, as the number of (randomly distributed) pulsars grows, the variance in the estimate of $\hat{h}^2$ decreases $\propto N_{pul}^{-1/2}$. Note that this decrease with increasing $N_{pul}$ is slower than that of the $h^2$ estimator that uses both auto- and cross-correlation measurements. That variance decreases $\propto 1/N_{pul}$, as shown in (7.12) and Fig. 10.

To confirm the $N_{pul} \to \infty$ behavior suggested by Fig. 13, we compute the variance $\sigma_{\hat{h}^2}^2$ for a PTA containing a large number of uniformly distributed pulsars. Imagine that the set of all pulsar pairs are ordered by increasing angular separation, and note that since $\mu_{ab}$ varies smoothly, the summand of (7.3) is effectively averaging over nearby entries in the inverse covariance matrix $\mathbb{C}^{-1}$. Since there are many pulsars uniformly distributed on the sky, we are

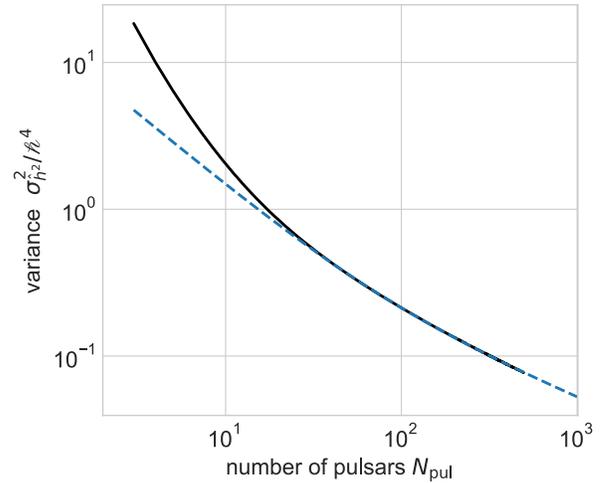

FIG. 13. The black curve shows the variance $\sigma_{\hat{h}^2}^2$ of the PTA "cross-correlations only" estimator $\hat{h}^2$ for the squared strain. This is obtained from (7.3) for $N_{pul}$ pulsar positions randomly distributed on the sky. The blue dashed curve shows the fit (7.27): for large numbers of pulsars $\sigma_{\hat{h}^2}^2 \propto N_{pul}^{-1/2}$. Extrapolation (correctly) suggests that the variance vanishes as $N_{pul} \to \infty$.

justified [55] in replacing $\mathbb{C}^{-1}$ with the inverse of the cosmic covariance matrix, $s_{jk}^{-1}$. This is given in diagonal form as a harmonic sum in (4.29). We are also justified in replacing $\mu_{ab}$ and $\mu_{cd}$ with their harmonic sum forms given in (4.13). This is because auto-correlation terms such as

$\delta_{ab}\mu_u(0)$, which appear in (2.4), vanish for distinct pulsar pairs $a < b$. Thus, substituting (4.13) and (4.29) (with $N_{\text{bins}}$ replaced by $N_{\text{pairs}}$) into (7.3), and converting the Riemann sums into integrals by taking the limit $N_{\text{pairs}} \to \infty$, we obtain

$$
\begin{aligned}
(\sigma_{\hat{h}^2}^2)^{-1} &\simeq \sum_{ab,cd} \mu_{ab} \mathbb{C}_{ab,cd}^{-1} \mu_{cd} \\
&= \int_0^\pi \sin\gamma \, d\gamma \int_0^\pi \sin\gamma' \, d\gamma' \, \mu_u(\gamma) s^{-1}(\gamma,\gamma') \mu_u(\gamma') \\
&= \int_{-1}^1 dx \int_{-1}^1 dx' \sum_{l,l',l'' \geq 2}^{L} (2l+1)(2l'+1)(2l''+1) \frac{C_l C_{l''}}{8\hbar^4 C_{l'}^2} P_l(x) P_{l'}(x) P_{l'}(x') P_{l''}(x').
\end{aligned}
\tag{7.28}
$$

Here, we have changed variables to $x \equiv \cos\gamma$ and $x' \equiv \cos\gamma'$, assuming that the pulsars are uniformly distributed on the sky.

The sums over $l$, $l'$ and $l''$ in (7.28) arise from expressions (4.13) and (4.25), which are harmonic sums over all modes $l \geq 2$. However, in order to also consider the limiting behavior $\sigma_{\hat{h}^2}^2$ for the non-PTA detectors described in Sec. VI A, we terminate the harmonic sums at $l, l', l'' = L$ where $L \geq 2$ is a finite integer. For a physical PTA, which includes all modes, $L \to \infty$.

The integrals in (7.28) may be evaluated using the orthogonality relation (4.19) for Legendre polynomials. This converts the integrals over $x$ and $x'$ into $2\delta_{ll'}/(2l+1)$ and $2\delta_{l'l''}/(2l+1)$, respectively. The sums over $l'$ and $l''$ may then be evaluated, eliminating the Kronecker deltas, and (7.28) becomes

$$
(\sigma_{\hat{h}^2}^2)^{-1} \simeq \sum_{l=2}^{L} \frac{2l+1}{2\hbar^4} = \frac{1}{2\hbar^4} \left((L+1)^2 - 4\right).
\tag{7.29}
$$

As more harmonics are included, $L$ and the rhs of (7.29) grow larger. Thus, for a PTA, $\sigma_{\hat{h}^2}^2$ vanishes as $N_{\text{pul}} \to \infty$, just as Fig. 13 suggests.

Equation (7.29) also applies to non-PTA detectors. To see this, note that $C_l$ cancels out of (7.28) and so does not affect the value of (7.29), *provided* $C_l \neq 0$. In other words, the actual numerical value of $C_l$ is irrelevant; all that matters is whether or not $C_l$ vanishes. If $C_l$ is nonzero, then the corresponding value of $l$ is included in the sum of (7.29), whereas if $C_l$ is zero, then the corresponding term is absent. Hence, (7.29) holds for the non-PTA detectors of Sec. VI A, which are defined by correlation-matrix expansion coefficients $Q_l$ for $2 \leq l \leq L$. For example, taking $L = 2$ for a one-arm LIGO-like detector, we find

$$
\sigma_{\hat{h}^2}^2 \simeq \frac{2}{5} \hbar^4.
\tag{7.30}
$$

This is the same result that we have seen before for both the auto + cross and auto-only correlation estimators, see (7.15) and (7.26). Plots of the inverse variance for the cross-correlation-only squared-strain estimator $\hat{h}^2$ are given in Fig. 14 for the different detector types discussed above.

### D. Uncertainty of the squared-strain estimators for the four PTAs

We now take the sky locations of the pulsars currently being monitored by the different PTAs, as given in

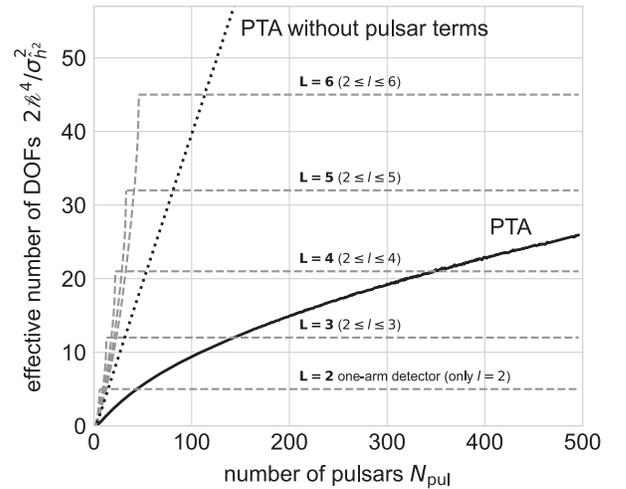

FIG. 14. Plots of $2\hbar^4/\sigma_{\hat{h}^2}^2$ for the cross-correlation-only estimator of $h^2$ for different detector types. These include a PTA with and without pulsar terms, and non-PTA detectors sensitive only to modes $2 \leq l \leq L$. For these non-PTA detectors, we have neglected pulsar terms; at saturation they measure almost one degree of freedom per pulsar. One can see that in the limit $L \to \infty$, these converge to the "PTA without pulsar terms" curve, which statistically measures 2/5 of a degree of freedom per pulsar.

TABLE I. Fractional uncertainties for the different squared-strain estimators discussed in this section, assuming noise-free measurements for each PTA's pulsars. The values in row (1) are obtained directly from (7.12), whereas those in rows (2) and (3) are obtained by numerically evaluating (7.3) for the PTA's specific pulsar sky directions. While the order of increasing uncertainty is (1), (2), (3), if there were more than $\approx 1500$ pulsars uniformly distributed on the sky, the ordering would be (1), (3), (2); see the final column and Fig. 10. The values in row (4) are the fractional error $\sigma_s/\langle s \rangle = \sqrt{2/5}\hbar^2/h^2$ from (7.15) if the pulsars are replaced by one-arm LIGO-like detectors sensitive to only the $l = 2$ quadrupole mode of the background. Numerical values of $\hbar^2/h^2$ for the binary-inspiral GW background of Appendix B are given in Table III. For strain ($\alpha = 0$) set $\hbar^2/h^2 \approx 0.3905$; for timing residuals ($\alpha = 1$) set $\hbar^2/h^2 \approx 0.5622$.

| PTA: | EPTA | NANOGrav | PPTA | IPTA | Ideal |
|---|---|---|---|---|---|
| Number of pulsars: | 42 | 66 | 26 | 88 | $\infty$ |
| (1) $\sigma_{\hat{h}^2}/\langle\hat{h}^2\rangle$ from auto+cross-correlations: | $0.2182\,\hbar^2/h^2$ | $0.1741\,\hbar^2/h^2$ | $0.2774\,\hbar^2/h^2$ | $0.1508\,\hbar^2/h^2$ | 0 |
| (2) $\sigma_{\hat{h}^2}/\langle\hat{h}^2\rangle$ from auto-correlations only: | $0.3222\,\hbar^2/h^2$ | $0.2850\,\hbar^2/h^2$ | $0.3533\,\hbar^2/h^2$ | $0.2629\,\hbar^2/h^2$ | $0.2041\,\hbar^2/h^2$ |
| (3) $\sigma_{\hat{h}^2}/\langle\hat{h}^2\rangle$ from cross-correlations only: | $0.6818\,\hbar^2/h^2$ | $0.5639\,\hbar^2/h^2$ | $0.8005\,\hbar^2/h^2$ | $0.5028\,\hbar^2/h^2$ | 0 |
| (4) $\sigma_s/\langle s \rangle$ for a set of one-arm detectors: | $0.6325\,\hbar^2/h^2$ | $0.6325\,\hbar^2/h^2$ | $0.6325\,\hbar^2/h^2$ | $0.6325\,\hbar^2/h^2$ | $0.6325\,\hbar^2/h^2$ |

Appendix H. For these, we can predict the (best-case, noise-free) fractional uncertainties, $\sigma_{\hat{h}^2}/\langle\hat{h}^2\rangle$, of the three different squared-strain estimators. These uncertainties are given in Table I. For comparison, the last row of the table gives the fractional uncertainty $\sigma_s/\langle s\rangle$ from (7.15), where $s = $ time average$(h_{\mu\nu}h^{\mu\nu})$. As discussed in Sec. VII A after (7.15), we assume that five or more of the pulsars are replaced by ideal one-arm LIGO-like detectors ($Q_l = \delta_{l2}$) and $s$ is estimated using both auto- and cross-correlations.

It is interesting to contrast the (inverses of the) fractional uncertainties in $\hat{h}^2$ with traditional SNRs. The expected SNR for a detection statistic is defined to be the expected value of that statistic in the *presence* of the signal, divided by the rms value of the statistic in the *absence* of the signal. It appears to be practically the same thing as the inverse of the fractional uncertainty, $\langle\hat{h}^2\rangle/\sigma_{\hat{h}^2}$. However, they are different: in our calculation, $\sigma_{\hat{h}^2}$ is *not* the rms value of the "detection statistic" $\hat{h}^2$ in the absence of the GW background (i.e., due purely to pulsar and instrumental noise). Rather, it reflects the variations in $\hat{h}^2$ among different universes in the Gaussian ensemble, each of which would give a different value of $\hat{h}^2$.

Thus, we can employ $\hat{h}^2$ in two different ways, depending upon which variance we select. First, we can use $\hat{h}^2$ as a detection statistic, to test the hypothesis "there is a GW signal present in the data." Alternatively, we can use it for parameter estimation, to determine the most likely value of the squared strain (and its expected fractional uncertainty) *assuming the presence of a GW signal*. For the first of these applications, the variance of $\hat{h}^2$ is calculated assuming only measurement and pulsar noise (see Sec. IX). Thus, if $\hat{h}^2$ is employed as a detection statistic, its SNR could be considerably larger (or smaller) than the inverse of the fractional uncertainties given in Table I.

There is a useful graphical way to think about the second application: estimating the squared GW strain from observational data. This parameter estimation problem is illustrated in Fig. 15, which shows the results of a single numerical simulation. The simulation generates one realization of a Gaussian ensemble with $h^2 = 1$; we then use the (simulated) observational data to construct the optimal estimator $\hat{h}^2$. This simulates "observational reality," where only a single set of measurements and correlation values is obtained. Here, the (one, randomly selected)

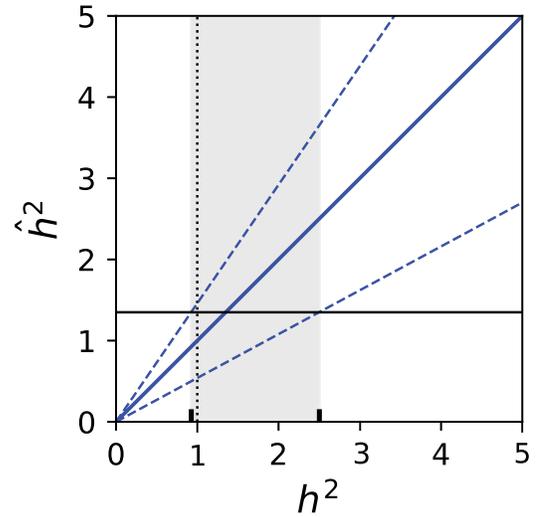

FIG. 15. A "one-sigma self-consistent" $h^2$ interval calculated for one realization of simulated noise-free cross-correlation-only data, for 40 pulsars placed at random sky locations. The blue lines show the expected value of the estimator $\hat{h}^2$ and the $\pm\sigma_{\hat{h}^2}(h^2)$ band about it. The one-sigma interval for $h^2$ is defined by the intersection of the one-sigma interval $h^2 \pm \sigma_{\hat{h}^2}(h^2)$ (blue dashed lines) about $h^2$ (solid blue line) with the observed value of $\hat{h}^2$ for many different (assumed) values of $h^2$. The vertical gray region spans the $h^2$ interval indicated by the thick black tick marks on the $h^2$ axis. The dotted black line corresponds to the injected value $h^2 = 1$.

realization gave $\hat{h}^2 = 1.35$. What uncertainty should we associate with this estimate?

As discussed earlier, there is a complication: the width of the $\pm 1\sigma$ uncertainty band depends upon the unknown *true* value $h^2$ of the squared GW strain. So, at what value of $h^2$ should this uncertainty be evaluated? Figure 15 shows our proposed "self-consistent" approach. Since we do not know the *actual* value of $h^2$, we construct the optimal estimator and its variance (vertical axis) for *all possible* values of $h^2$ (horizontal axis).

Examination of Fig. 15 shows that only some values of $h^2$ are consistent with the (simulated) observation (i.e., credible) at the $\pm 1\sigma$ level. For example, suppose that the actual value of $h^2$ were 3. In this case, we would expect the estimated $\hat{h}^2$ to lie in the interval $\hat{h}^2 \in [1.6, 4.4]$. As can be seen from the figure, this is inconsistent (at the $\pm 1\sigma$ level) with the observation. The self-consistent interval for $h^2$ lies between the tick marks on the horizontal axis, and corresponds to the gray region.

## VIII. TESTING THE CORRELATION MODEL AGAINST DATA

The most important result of the present paper and of Ref. [11] is this: even with data from many noise-free pulsars, and even with signals coming from a perfectly Gaussian GW background, the pulsar-averaged Hellings and Downs correlation in our Universe will *not* follow the Hellings and Downs curve exactly. The following question then arises: after experiments have binned and (optimally) averaged the correlations, are the deviations away from the Hellings and Downs curve consistent with expectations? A simple way to quantify and answer this question is via $\chi^2$ goodness-of-fit tests.

In the following subsections, we define two types of $\chi^2$ statistics, which we call "unprojected" and "projected" $\chi^2$. We start in Sec. VIII A with the "unprojected" statistic, which assumes that both the overall scale and spectral shape of the GW background are known exactly, *a priori*. This means that the true values of $h^2$, $\hbar^4$, etc., are "magically" given to us, with no reliance on the data used to form the $\chi^2$ statistic. While unrealistic, this simplifies the calculation of the expected value and variance of this unprojected $\chi^2$, which may then be used to test the Gaussian ensemble hypothesis of the GW background against the measured data.

In Sec. VIII B, we define a "projected" $\chi^2$ statistic. In addition to testing the match between the observed correlation data and the Hellings and Downs curve, this statistic also estimates the scale of the squared GW strain $h^2$. Hence it eliminates the need for a fictitious "oracle" who provides the correct value of $h^2$. We show that this estimate comes at low cost: for a given value of the squared strain $h^2$, the fractional differences between expected value and variance of the projected and unprojected $\chi^2$ statistics are negligible. This is because projection effectively reduces the (large) number of degrees of freedom by one. This is the same reduction in degrees of freedom that takes place when the variance of a data set is computed using the sample mean as opposed to the population mean. Since the cases of interest have many pulsars and hence many degrees of freedom, it is not surprising that the properties of the two $\chi^2$ statistics are almost identical.

The projected $\chi^2$ statistic provides an estimate of the squared GW strain, but that estimate, and the value of $\chi^2$, depend upon the (unknown) true value of $h^2$. This is the same situation that we encountered at the end Sec. VII D, as illustrated in Fig. 15. It arises because $\chi^2$ is a quadratic form defined by the (inverse of the) covariance matrix $\mathbb{C}$, which in turn depends upon $\hbar^4$, which in turn depends upon $h^2$. Our solution follows the same philosophy as in the previous section: we evaluate $\chi^2$ for *all* possible values of $h^2$. Then, we ask if those observationally derived quantities are self-consistent. This means that $\chi^2(h^2)$ should lie within the bounds $\langle \chi^2 \rangle \pm \sigma_{\chi^2}(h^2)$, where $h^2$ is restricted to an interval consistent with the observed value of the estimator $\hat{h}^2$, as discussed at the end of Sec. VII D. The combination of these two bounds creates a pass/fail "acceptance window," which we describe at the end of Sec. VIII C 4.

For the remainder of this preamble, we review the notation and some related issues. As described above, our starting point is a set of correlations which have been binned and optimally averaged. The notation reflects this: indices $j, k, \cdots$ denote angular separation bins, $\rho_{\text{opt},j}$ is the value of the optimal estimator of the correlation in the $j$th bin, and $n_{\text{pairs},j}$ is the number of distinct pulsar pairs in the $j$th bin. The average value of the angles $\gamma_{ab}$ for the pulsar pairs $ab$ within bin $j$ is denoted $\gamma_{\text{bin},j}$. The quantity $\mu_{\text{bin},j} \equiv \mu_u(\gamma_{\text{bin},j})$ is the expected value of the Hellings and Downs curve for that bin, which we typically use for normalization.

Binning is arbitrary: how many bins should be used, and where should the bin boundaries be placed? Among the possibilities are two important extreme cases: (a) combine all pulsar pairs into a single bin, and (b) put every pulsar pair into its own separate bin. We will see that (b) provides the highest discriminating power and the best statistical test. However, the intermediate cases are also of interest, particularly to convince the scientific community regarding detection claims.

In reading the remainder of this section, it may be helpful to imagine that the choice of bins has been "fixed in advance" by some external agency, say a review board or detection committee. That choice might be one of the extremes (a) or (b) described above, or lie somewhere in between.

To parse the equations, it is also helpful to keep in mind that there are different vector spaces that enter. At the top level, we have a vector space whose dimension $N_{\text{bins}}$ is equal to the number of bins, with vector and matrix components denoted by indices $j, k, \ell, m$. Then, each of those bins has its own vector space, whose dimension $n_{\text{pairs},j}$ is the number of distinct pulsar pairs that lie in that $j$th bin. In those vector spaces, the vector and matrix components are indexed by pulsar pairs. For example, if the expressions involve only cross-correlations, then these pulsar pairs are of the form $ab, cd, \ldots, gh$ with $a < b$, $c < d, \ldots, g < h$. On the other hand, if auto-correlations are also included, then we also must incorporate terms with $a = b$, $c = d$ and so on. Nevertheless, we still have some freedom in how to define the bins. One choice would be to put all of those auto-correlation $\rho_{aa}$ terms into the same bin at $\gamma = 0$. At the other extreme, we could put each pulsar's auto-correlation into a separate bin, so that there were distinct bins at vanishing angular separation. This latter case corresponds to choice (b) above.

In the remainder of this section, we construct $\chi^2$ statistics which include three different sets $\mathcal{S}$ of pulsar-pair correlations: (i) using all auto + cross correlations, $\mathcal{S} = \{ab | a \leq b\}$; (ii) using auto-correlations only, $\mathcal{S} = \{ab | a = b\}$; and (iii) using cross-correlations only, $\mathcal{S} = \{ab | a < b\}$. These are the same sets we used in Sec. VII to construct different estimators of the squared strain. Each of these choices has a corresponding unprojected and projected $\chi^2$ statistic.

Here, we restrict our attention to noise-free correlation measurements. In Sec. IX, we discuss how the $\chi^2$ analyses are modified by including the effects of pulsar and observational noise.

### A. The unprojected $\chi^2$ test

Suppose that the squared GW strain $h^2$ is known. Define a $\chi^2$ statistic by

$$\chi^2 = \sum_{j,k} (\rho_{\text{opt},j} - \langle \rho_{\text{opt},j} \rangle) B_{jk}^{-1} (\rho_{\text{opt},k} - \langle \rho_{\text{opt},k} \rangle), \quad (8.1)$$

where the expectation value $\langle \rho_{\text{opt},j} \rangle = h^2 \mu_{\text{bin},j}$ is a "known quantity." The matrix $B_{jk}^{-1}$ is the inverse of the covariance $B_{jk}$ between the Hellings and Downs correlation estimates in bins $j$ and $k$, given in (3.14) and (3.15):

$$B_{jk} = \frac{\mu_{\text{bin},j} \mu_{\text{bin},k}}{(\mu_j^\top C_{jj}^{-1} \mu_j)(\mu_k^\top C_{kk}^{-1} \mu_k)} \mu_j^\top C_{jj}^{-1} C_{jk} C_{kk}^{-1} \mu_k. \quad (8.2)$$

As we remarked above, this equation contains quantities with different (vector) dimensions. For example, the vector $\mu_{\text{bin},j}$ has a dimension equal to the number of bins: $j = 1, \ldots, N_{\text{bins}}$. In contrast, for a given value of $j$, the dimension of the square matrix $C_{jj}$ is $n_{\text{pairs},j}$, which is the number of distinct pulsar pairs that lie in the $j$th bin. This is also the dimension of the vector $\mu_j$. The rectangular matrix $C_{jk}$ has as many rows as the number of distinct pulsar pairs in bin $j$ and as many columns as there are distinct pulsar pairs in bin $k$, as indicated using the block form of $\mathbb{C}$ in (2.14).

In the remainder of this section, to simplify equations and calculations, and to gain geometric insight, we adopt standard notation for vectors and tensors in metric spaces. We exploit the fact that $B_{jk}$ is a real, square, positive-definite $N_{\text{bins}} \times N_{\text{bins}}$ matrix, where $N_{\text{bins}}$ is the number of angular separation bins that contain at least one pulsar pair. Hence, it is a positive-definite quadratic form,

$$g_{jk} \equiv B_{jk}, \quad (8.3)$$

on the vector space indexed by angular separation bins. This notation will be familiar to many readers because it is widely used for general relativity (GR) [26]. However, in contrast with GR, where $g_{jk}$ is a $4 \times 4$ matrix of signature $(-, +, +, +)$, here it is an $N_{\text{bins}} \times N_{\text{bins}}$ matrix of signature $(+, +, \cdots, +)$. With this notation, one must distinguish between covariant (down) and contravariant (up) indices of vectors and tensors. The covariant metric $g_{jk} \equiv B_{jk}$ and its contravariant inverse,

$$g^{jk} \equiv B_{jk}^{-1}, \quad (8.4)$$

are used to lower and raise indices, where $B^{-1}$ denotes the matrix inverse (and *not* the inverse of the matrix elements). We also adopt the Einstein summation convention [26] for equations. That is, if the same index label appears once in the covariant position and once in the contravariant position, then that index is summed over all angular bins.

To illustrate our conventions, we define the two contravariant vectors

$$\mu_{\text{bin}}^j \equiv g^{jk} \mu_{\text{bin},k} = \sum_k B_{jk}^{-1} \mu_{\text{bin},k} \qquad \text{and} \qquad \rho_{\text{opt}}^j \equiv g^{jk} \rho_{\text{opt},k} = \sum_k B_{jk}^{-1} \rho_{\text{opt},k}. \quad (8.5)$$

The $\chi^2$ statistic from (8.1) may then be written as

$$\chi^2 = (\rho_{\text{opt},j} - h^2 \mu_{\text{bin},j})(\rho_{\text{opt}}^j - h^2 \mu_{\text{bin}}^j). \quad (8.6)$$

Equivalently, if we define the fluctuations of the observed correlations away from the Hellings and Downs curve as

$$n_j \equiv \rho_{\text{opt},j} - \langle \rho_{\text{opt},j} \rangle = \rho_{\text{opt},j} - h^2 \mu_{\text{bin},j}, \quad (8.7)$$

then the $\chi^2$ statistic from (8.1) and (8.6) is simply the squared length

$$\chi^2 = n_j n^j. \qquad (8.8)$$

Note that the fluctuations have zero mean $\langle n_j \rangle = 0$, so their correlation and covariance matrices are equal. These are

$$\langle n_j n_k \rangle = B_{jk} = g_{jk}. \qquad (8.9)$$

These follow directly from (8.7) and (3.14).

$$\sigma_{\chi^2}^2 \equiv \langle (\chi^2)^2 \rangle - \langle \chi^2 \rangle^2 = g^{jk} g^{\ell m} \big( \langle n_j n_k n_\ell n_m \rangle - \langle n_j n_k \rangle \langle n_\ell n_m \rangle \big), \qquad (8.11)$$

where the first equality defines the variance, and the second equality follows by substitution of (8.8).

The fourth-order expectation value that appears in (8.11) is deceptive. Since the GW background is described by a Gaussian ensemble, one might suppose that Isserlis's theorem [19] could be applied to $\langle n_j n_k n_\ell n_m \rangle$. If so, one could write it as a sum of three products of two-point functions. However, this is not the case: the quantities $n_j$ are not Gaussian random variables. Instead, they are quadratic expressions in the Gaussian random GW field amplitudes.

Evaluating the fourth-order expectation value using the results of Appendix F gives

$$\langle n_j n_k n_\ell n_m \rangle = \langle n_j n_k \rangle \langle n_\ell n_m \rangle + \langle n_j n_\ell \rangle \langle n_k n_m \rangle + \langle n_j n_m \rangle \langle n_k n_\ell \rangle + \mathbb{E}_{jk\ell m}. \qquad (8.12)$$

The first three terms on the rhs are quadratic expectation values given by (8.9). If the $n_j$ were Gaussian random variables, so that Isserlis's theorem could be applied, then only these three terms would arise, and the final term would be absent.

The final term in (8.12) arises entirely from the non-Gaussian behavior of $n_j$. This non-Gaussian contribution to the variance of $\chi^2$ is

$$\mathbb{E}_{jk\ell m} = \sum_{ab,cd,ef,gh} w_{j,ab} w_{k,cd} w_{\ell,ef} w_{m,gh} \mathbb{E}_{ab,cd,ef,gh}, \qquad (8.13)$$

with $\mathbb{E}_{ab,cd,ef,gh}$ given by (F8) and weights $w_{j,ab}$ given by (3.13).

We now evaluate the variance of $\chi^2$. Starting from (8.11), we use (8.12) and (8.9) to obtain

$$\sigma_{\chi^2}^2 = g^{jk} g^{\ell m} \big( g_{j\ell} g_{km} + g_{jm} g_{k\ell} + \mathbb{E}_{jk\ell m} \big) = 2 N_{\text{bins}} + \mathbb{E}, \qquad (8.14)$$

where we have defined

$$\mathbb{E} \equiv g^{jk} g^{\ell m} \mathbb{E}_{jk\ell m}. \qquad (8.15)$$

If the $n_j$ were Gaussian random variables, then the $\mathbb{E}$ term would be absent from (8.14), and $\chi^2$ would have the statistics of a standard $\chi^2$-distributed random variable with $N_{\text{bins}}$ degrees of freedom. The mean value would be $N_{\text{bins}}$

The expected value of $\chi^2$ follows immediately by taking the expected value of (8.8) and using (8.9). We obtain

$$\langle \chi^2 \rangle = g^{jk} \langle n_j n_k \rangle = g^{jk} g_{jk} = \delta_j^{\,j} = N_{\text{bins}}, \qquad (8.10)$$

where $N_{\text{bins}}$ is the number of angular separation bins. As described above, $N_{\text{bins}}$ may be as small as 1 in case (a) or as large as the total number of pulsar pairs in case (b).

To further characterize $\chi^2$, we need to know its expected variations away from the mean $\langle \chi^2 \rangle$. The magnitude of these variations is quantified by the variance of $\chi^2$, which is and the variance would be $2 N_{\text{bins}}$. However, in Sec. VIII C we will see that the $\mathbb{E}$ term is much larger than the other terms. Thus, the statistical properties of $\chi^2$ are quite different than those of an ordinary $\chi^2$-distributed random variable with $N_{\text{bins}}$ degrees of freedom.

Note that all of these expressions simplify in the extreme case (b) described above, where every pulsar pair lies in its own bin. In that case, the weights $w_{j,ab}$ are all unity [56], and $B_{jk} \equiv g_{jk}$ becomes the full covariance matrix $\mathbb{C}_{ab,cd}$.

## B. The "projected" $\chi^2$ statistic (which also estimates the squared amplitude of the GW background)

We now define a "projected" $\chi^2$ statistic, which also uses the observed correlations to construct an estimator $\hat{h}^2$ of $h^2$. We will see that for a large number of bins and for any given value of $h^2$, the expected value and variance of the projected $\chi^2$ statistic are very similar to those of the unprojected $\chi^2$ statistic. This has two positive features. First, it means that we get an $h^2$ estimate "almost for free," and second, it means that later, in discussing their behavior, we do not need to distinguish between the projected and the unprojected $\chi^2$ statistics.

There is, however, a complication. The projected $\chi^2$ statistic depends on the unknown value of the squared GW strain $h^2$ via the dependence of the quadratic form $B_{jk}^{-1}$ on the covariance matrix $\mathbb{C}$ [see (8.2)], which in turn depends on $h^4$, which in turn depends on $h^2$. This means that to test if observationally determined correlations are consistent with expectations, we need to evaluate the projected $\chi^2$ statistic at different (assumed) values of $h^2$. We then ask if those values, $\chi^2(h^2)$, agree with its expected value to within $\pm\sigma_{\chi^2}(h^2)$, over the one-sigma range of $h^2$ values consistent with the observed value of $\hat{h}^2$. This procedure will be discussed in more detail at the end of Sec. VIII C 4.

The projected $\chi^2$ statistic is defined by

$$\chi^2(h^2) \equiv \min_{\tilde{h}^2 \geq 0} \left\{ \sum_{j,k} \left( \rho_{\text{opt},j} - \tilde{h}^2 \mu_{\text{bin},j} \right) B_{jk}^{-1}(h^2) \left( \rho_{\text{opt},k} - \tilde{h}^2 \mu_{\text{bin},k} \right) \right\}, \tag{8.16}$$

where we have explicitly indicated the $h^2$ dependence of $B_{jk}^{-1}$ and thus of $\chi^2$. We have chosen in this subsection not to use the compact "raised-and-lowered" index notation of the previous subsection since quantities with raised indices have a "hidden" $h^2$ dependence. The squared-strain estimator $\hat{h}^2$ is the value of the real quantity $\tilde{h}^2 \geq 0$ at which the rhs of (8.16) is minimized.

We can easily obtain an explicit expression for $\hat{h}^2$, in terms of the observed correlations $\rho_{\text{opt},j}$. This is found by taking the derivative with respect to $\tilde{h}^2$ of the quantity inside the curly brackets in (8.16), setting the derivative to zero, and solving the resulting linear equation to obtain $\tilde{h}^2$. Its value is

$$\hat{h}^2 = \frac{1}{\mu_{\text{bin}}^2} \sum_{j,k} \mu_{\text{bin},j} B_{jk}^{-1} \rho_{\text{opt},k}, \tag{8.17}$$

where the positive scalar quantity

$$\mu_{\text{bin}}^2 \equiv \sum_{j,k} \mu_{\text{bin},j} B_{jk}^{-1} \mu_{\text{bin},k} \tag{8.18}$$

is the squared norm of the vector $\mu_{\text{bin},j}$. Expression (8.17) for $\hat{h}^2$ is a special case of the form (7.1) which we used to find the best $h^2$ estimator in Sec. VII. Its variance is $\sigma_{\hat{h}^2}^2 = 1/\mu_{\text{bin}}^2$.

Equation (8.17) provides a squared-strain estimate $\hat{h}^2$ formed from the optimal ("average") Hellings and Downs correlations for the different angular bins. If every single pulsar pair is put into its own distinct bin, then (8.17) reduces to the "best" $h^2$ estimator given in (7.2). It is easy to see that the estimators formed from the binned averages cannot have a smaller variance than the best estimator. This is because the general form of the best estimator allows it to equal the binned estimator, simply by selecting appropriate weights.

An explicit expression for $\chi^2$ is obtained by setting $\tilde{h}^2$ on the rhs of (8.16) to the value found in (8.17). This minimizes the quantity in curly brackets in (8.16), showing that the projected $\chi^2$ statistic is

$$\chi^2(h^2) = \sum_{j,k} \left( \rho_{\text{opt},j} - \hat{h}^2 \mu_{\text{bin},j} \right) B_{jk}^{-1}(h^2) \left( \rho_{\text{opt},k} - \hat{h}^2 \mu_{\text{bin},k} \right). \tag{8.19}$$

In the absence of noise, $\rho_{\text{opt},j}$ and $\hat{h}^2$ are independent of $h^2$, because the overall factors of $\hbar^4$ in the inverse covariance matrices [see (3.12), (3.13), and (8.17), (8.18)] cancel out. In the presence of instrumental or pulsar noise, there is no such cancellation. In this case, $\rho_{\text{opt},j}$ and $\hat{h}^2$ depend on $h^2$ as described in Sec. IX D. While we do not explicitly indicate that dependence in (8.19) or in the following equations, it is not a problem. Our philosophy/approach for interpreting the $\chi^2$ test (described below) is unaffected by this additional dependence of $\chi^2(h^2)$ on $h^2$.

As before, $\chi^2$ is the squared length of a vector

$$\chi^2(h^2) = \sum_{j,k} \hat{n}_j B_{jk}^{-1}(h^2) \hat{n}_k, \tag{8.20}$$

but now this vector is the *estimated* fluctuations $\hat{n}_j$ rather than the *actual* fluctuations $n_j$. The fluctuation estimators constructed from the observational data are

$$\hat{n}_j \equiv \rho_{\text{opt},j} - \hat{h}^2 \mu_{\text{bin},j} = \sum_k P_{jk} \rho_{\text{opt},k}, \tag{8.21}$$

where we have defined the projection operator

$$P_{jk} \equiv \delta_{jk} - u_j \sum_\ell B_{k\ell}^{-1} u_\ell \quad \text{for } u_\ell \equiv \mu_{\text{bin},\ell}/\sqrt{\mu_{\text{bin}}^2}, \tag{8.22}$$

with $\mu_{\text{bin}}^2$ defined by (8.18). The unit vector $u_j$ is a normalized version of the vector $\mu_{\text{bin},j}$, and satisfies

$$\sum_{j,k} u_j B_{jk}^{-1} u_k = 1. \tag{8.23}$$

Thus, $P_{jk}$ projects onto the vector space orthogonal to $u_k$, so $\hat{n}_j = \sum_k P_{jk} n_k$ is obtained from $n_j$ by removing its component parallel to the expected value of the correlation $h^2 \mu_{\text{bin},j}$. Thus, $\chi^2$ as given by (8.20) is the squared length (with respect to the metric $B_{jk}$) of the estimate for the part of $n_j$ that differs from the Hellings and Downs expectation. Note that for the noise-free case, the projection operator $P_{jk}$ is independent of $h^2$ since $B_{kl}^{-1}$ is proportional to $\hbar^{-4}$ while each of $u_j$ and $u_\ell$ are proportional to $\hbar^2$ due to the normalization by $\sqrt{\mu_{\text{bin}}^2}$ in (8.22).

In contrast with the $n_j$ defined previously in (8.7), the $\hat{n}_j$ are *estimators* of the fluctuations away from the mean. We will see that they have one fewer degree of freedom than the $n_j$. This is because the $\hat{n}_j$ are obtained entirely from the measurements, whereas the $n_j$ may only be found by exploiting *a priori* knowledge of the true value of $h^2$. The $\hat{n}_j$'s have zero mean, i.e., $\langle \hat{n}_j \rangle = 0$, and covariance

$$\langle \hat{n}_j \hat{n}_k \rangle = B_{jk} - u_j u_k = \sum_\ell P_{j\ell} B_{\ell k}. \tag{8.24}$$

The quantity $B_{jk} - u_j u_k$ is a metric on the $(N_{\text{bins}} - 1)$-dimensional vector space orthogonal to the unit vector $u_j$.

We can calculate the ensemble mean and variance of $\chi^2$ by proceeding as in Sec. VIII A. From (8.20) and (8.24), the expected value of $\chi^2$ is

$$\langle \chi^2 \rangle = \sum_{j,k} B_{jk}^{-1} \langle \hat{n}_j \hat{n}_k \rangle = \sum_{j,k} B_{jk}^{-1} \left( B_{jk} - u_j u_k \right) = N_{\text{bins}} - 1. \tag{8.25}$$

From (8.20), the variance in $\chi^2$ is

$$\sigma_{\chi^2}^2 = \sum_{j,k,\ell,m} B_{jk}^{-1} B_{\ell m}^{-1} \left( \langle \hat{n}_j \hat{n}_k \hat{n}_\ell \hat{n}_m \rangle - \langle \hat{n}_j \hat{n}_k \rangle \langle \hat{n}_\ell \hat{n}_m \rangle \right). \tag{8.26}$$

To evaluate the fourth-order expectation value of the $\hat{n}_j$'s, we use $\hat{n}_j = \sum_k P_{jk} \rho_{\text{opt},k}$ from (8.21) and the results of Appendix F to evaluate the fourth-order expectation value of the $\rho_{\text{opt},j}$'s. This leads to

$$\langle \hat{n}_j \hat{n}_k \hat{n}_\ell \hat{n}_m \rangle = \langle \hat{n}_j \hat{n}_k \rangle \langle \hat{n}_\ell \hat{n}_m \rangle + \langle \hat{n}_j \hat{n}_\ell \rangle \langle \hat{n}_k \hat{n}_m \rangle + \langle \hat{n}_j \hat{n}_m \rangle \langle \hat{n}_k \hat{n}_\ell \rangle + \hat{\mathbb{E}}_{jklm}, \tag{8.27}$$

where the quadratic expectation values are given by (8.24), and

$$\hat{\mathbb{E}}_{jk\ell m} \equiv \sum_{j',k',\ell',m'} P_{jj'} P_{kk'} P_{\ell\ell'} P_{mm'} \mathbb{E}_{j'k'\ell'm'}. \tag{8.28}$$

Starting with (8.26) and making use of (8.27) and (8.25), we obtain

$$\sigma_{\chi^2}^2 = 2(N_{\text{bins}} - 1) + \hat{\mathbb{E}}, \tag{8.29}$$

where

$$\hat{\mathbb{E}} \equiv \sum_{j,k,\ell,m} B_{jk}^{-1} B_{\ell m}^{-1} \hat{\mathbb{E}}_{jk\ell m} \tag{8.30}$$

is the trace of the projected non-Gaussian term. Note that for the noise-free case, the expected value and variance of the projected $\chi^2$ statistic are independent of the squared GW strain $h^2$. This is because $\hat{\mathbb{E}}$ as given by (8.30) depends only on the ratio $\mathfrak{h}^8/\hbar^8$. While this ratio depends upon the type of the GW sources, it is independent of the overall scale $h^2$. For an explicit example, see Table III.

The mean and variance of the projected $\chi^2$ statistic are very similar to those of the unprojected $\chi^2$ statistic, which was discussed in Sec. VIII A. We will see that their values and distributions are also very similar. If we compare (8.25) and (8.29) to (8.10) and (8.14), we see that $N_{\text{bins}}$ is replaced by $N_{\text{bins}} - 1$ and the non-Gaussian term involving $\mathbb{E}_{jk\ell m}$ is replaced by its projected version involving $\hat{\mathbb{E}}_{jk\ell m}$, given by (8.28). This reduction by one degree of freedom takes place whenever one estimates the variance of a data set using the sample mean (here $\hat{h}^2$) as opposed to the population mean (here $h^2$). For example, the formula for the normalization of the sample variance has a factor of $1/(N-1)$ whereas the corresponding formula for the population variance has $1/N$.

### C. Variance $\sigma_{\chi^2}^2$ for the $\chi^2$ statistics

Here, we investigate the variance of the unprojected and projected $\chi^2$ statistics constructed in Secs. VIII A and VIII B. The first has variance given by (8.14)

$$\sigma_{\chi^2}^2 = 2N_{\text{bins}} + \mathbb{E}, \tag{8.31}$$

where $\mathbb{E}$ is given by (8.15) and $N_{\text{bins}}$ is the number of angular separation bins. The projected $\chi^2$ statistic has a variance given by (8.29). It may be obtained from (8.31) and (8.15) by replacing $N_{\text{bins}}$ with $N_{\text{bins}} - 1$ and $\mathbb{E}$ with the projected version $\hat{\mathbb{E}}$ given by (8.30).

Recall that the $\mathbb{E}$ term (the components are explicitly calculated in Appendix F) arises because pulsar-pulsar correlations are non-Gaussian. If the pulsar-pulsar correlations were Gaussian, then $\mathbb{E}$ would vanish, and [dividing (8.31) by the square of (8.10)] the fractional variation would be

$$\mathbb{E} = 0 \quad \Rightarrow \quad \frac{\sigma_{\chi^2}^2}{\langle \chi^2 \rangle^2} = \frac{2N_{\text{bins}}}{N_{\text{bins}}^2} = \frac{2}{N_{\text{bins}}}. \tag{8.32}$$

This behavior, if $\mathbb{E}$ can be neglected, is that of a standard $\chi^2$-distribution with $N_{\text{bins}}$ degrees of freedom: the fractional deviations of $\chi^2$ away from its mean value are proportional to $N_{\text{bins}}^{-1/2}$. This would also be the case if $\hat{\mathbb{E}}$ could be neglected, with $N_{\text{bins}}$ replaced by $N_{\text{bins}} - 1$.

However, we will see that when there are more than a few pulsars, the $\mathbb{E}$ and $\hat{\mathbb{E}}$ terms are the *dominant* contributors to the variance. Hence, the fractional uncertainties of the $\chi^2$ statistics for the pulsar-pulsar correlations $\rho_{ab}$ have *different* scaling behavior than for the Gaussian case of (8.32). The minimum number of pulsars required for $\mathbb{E}$ or $\hat{\mathbb{E}}$ to dominate depends upon the ratio $\hbar/\mathfrak{h}$. This is apparent from (8.31), since $\mathbb{E}$ and $\hat{\mathbb{E}}$ are proportional to $\mathfrak{h}^8/\hbar^8$ [see (F8) for the factor of $\mathfrak{h}^8$; the $1/\hbar^8$ factor comes from two instances of the inverse covariance matrix needed to form the trace]. For realistic cosmological models, the ratio $\hbar/\mathfrak{h}$ is of order unity [57]. Hence, assuming coefficients of order unity leads to a correct estimate for this minimum number of pulsars. When the number of pulsars exceeds this minimum, then the first term on the rhs of (8.31) may be neglected.

For the remainder of this subsection, we focus on the most stringent statistical tests: each pulsar pair is placed into its own individual angular bin, so that $N_{\text{bins}}$ has its maximum possible value. This corresponds to case (b) discussed in the preamble of this section. The corresponding numbers of bins are

Case (i): $\quad N_{\text{bins}} = N_{\text{pul}}(N_{\text{pul}} + 1)/2 \qquad$ (auto + cross),

Case (ii): $\quad N_{\text{bins}} = N_{\text{pul}} \qquad\qquad\qquad$ (auto only),

Case (iii): $\quad N_{\text{bins}} = N_{\text{pul}}(N_{\text{pul}} - 1)/2 \qquad$ (cross only).

$$(8.33)$$

The unprojected (8.1) and projected $\chi^2$ statistic (8.19) may then be written

$$\chi^2 = \sum_{ab \in \mathcal{S}} \sum_{cd \in \mathcal{S}} \left( \rho_{ab} - h^2 \mu_{ab} \right) \mathbb{C}_{ab,cd}^{-1} \left( \rho_{cd} - h^2 \mu_{cd} \right), \quad (8.34)$$

$$\chi^2(h^2) = \sum_{ab \in \mathcal{S}} \sum_{cd \in \mathcal{S}} \left( \rho_{ab} - \hat{h}^2 \mu_{ab} \right) \mathbb{C}_{ab,cd}^{-1}(h^2) \left( \rho_{cd} - \hat{h}^2 \mu_{cd} \right), \quad (8.35)$$

respectively, where $\mathcal{S}$ denotes the different correlation sets (i.e., auto + cross, auto-only, or cross-only).

The second equation explicitly shows the $h^2$ dependence of the inverse covariance matrix $\mathbb{C}^{-1}$ and $\chi^2$, whereas the first equation does not. This is because the unprojected $\chi^2$ statistic (8.34) assumes that the true value of $h^2$ is known *a priori*, independent of the observed data: $\chi^2$ is only defined for that single true value of $h^2$. In contrast, the projected statistic (8.35) depends upon $h^2$. Here, since the true value of $h^2$ is not known, we explicitly indicate the dependence of $\mathbb{C}^{-1}$ and $\chi^2$ on $h^2$. The optimal estimator $\hat{h}^2$ of the squared GW strain $h^2$ (for the set $\mathcal{S}$) which appears in (8.35) also depends, in general, upon $h^2$ as described in Sec. IX D. Hence, to perform goodness-of-fit tests for the projected $\chi^2$ statistic, we must evaluate $\chi^2(h^2)$ for a range of (assumed) values of $h^2$.

As mentioned earlier, the $\chi^2$ test is meant to supplement "chi by eye" for observational data. This is often presented with a much smaller number of bins than in (8.33). Nevertheless, the equations in this section may still be used to compute the variance $\sigma_{\chi^2}^2$ numerically (for any given set of pulsar sky positions and choice of bins). Thus, a quantitative assessment becomes possible.

We have defined six different $\chi^2$ tests. While their variances differ, we will see that in the large $N_{\text{pul}}$ limit, the situation simplifies. In this limit, the fractional uncertainties $\sigma_{\chi^2}/\langle\chi^2\rangle$ for the auto + cross and cross-correlation-only calculations, hence four of the six variants, approach one another (see, e.g., the right panel of Fig. 16 and the rows labeled by (1) and (3) in Table II). The two auto-correlation-only $\chi^2$ statistics also have similar behavior in the large $N_{\text{pul}}$ limit. So, in this limit, there are really only two types of behavior, not six.

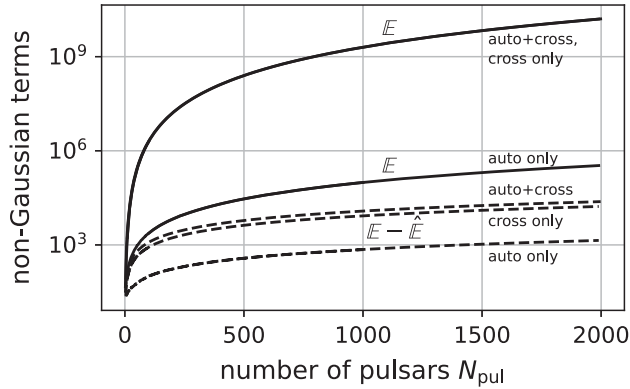
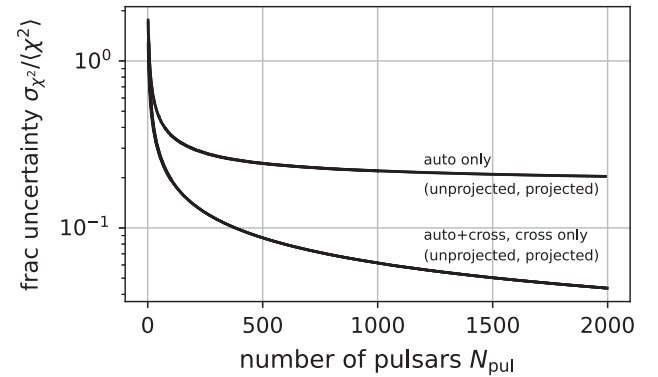

FIG. 16. Left panel: non-Gaussian contributions to the variance of the $\chi^2$ statistics formed using (i) auto + cross correlations, (ii) auto-correlations only, and (iii) cross-correlations only. The solid curves show $\mathbb{E}$ and the dashed curves show $\mathbb{E} - \hat{\mathbb{E}}$, all normalized by $\mathfrak{h}/\hbar = 1$. On this scale, the curves for $\mathbb{E}$ for auto + cross correlations and cross-correlations only are indistinguishable. Right panel: fractional uncertainty $\sigma_{\chi^2}/\langle\chi^2\rangle$ for the six different $\chi^2$ statistics, for a binary-inspiral model [57]. Again, the curves for auto + cross and cross-correlations-only are indistinguishable, as are those for the unprojected and projected $\chi^2$ statistics. These plots demonstrate that there are only two types of limiting behavior for the six statistics.

To determine the variance of the $\chi^2$ statistics, we need to compute $\mathbb{E}$ and $\hat{\mathbb{E}}$, which are traces, (8.15) and (8.30), of the cumulant tensor $\mathbb{E}_{ab,cd,ef,gh}$ or its projected partner. With our covariant/contravariant definitions, as explained before (8.5), the metric used to compute the trace (8.15) is

$$\hat{\mathbb{E}}_{ab,cd,ef,gh} = \sum_{op\in\mathcal{S}}\sum_{qr\in\mathcal{S}}\sum_{st\in\mathcal{S}}\sum_{uv\in\mathcal{S}} P_{ab,op}P_{cd,qr}P_{ef,st}P_{gh,uv}\mathbb{E}_{op,qr,st,uv}. \tag{8.37}$$

The projection operator (8.22) is given by

$$P_{ab,cd} = \delta_{ac}\delta_{bd} - \mu_{\text{bin}}^{-2}\mu_{ab}\sum_{ef\in\mathcal{S}}\mathbb{C}_{cd,ef}^{-1}\mu_{ef}, \tag{8.38}$$

where

$$\mu_{\text{bin}}^2 \equiv \sum_{ab\in\mathcal{S}}\sum_{cd\in\mathcal{S}}\mu_{ab}\mathbb{C}_{ab,cd}^{-1}\mu_{cd}. \tag{8.39}$$

Equation (8.38) for $\mu_{\text{bin}}^2$ is obtained by evaluating (8.18) for the case where each pulsar pair occupies its own bin.

For the calculations that follow, we apply the projection operators in (8.37) to the inverse matrices $\mathbb{C}_{ab,cd}^{-1}$ and $\mathbb{C}_{ef,gh}^{-1}$, rather than to $\mathbb{E}_{op,qr,st,uv}$. This leads to

$$\hat{\mathbb{E}} = \sum_{ab\in\mathcal{S}}\sum_{cd\in\mathcal{S}}\sum_{ef\in\mathcal{S}}\sum_{gh\in\mathcal{S}} \hat{\mathbb{C}}_{ab,cd}^{-1}\hat{\mathbb{C}}_{ef,gh}^{-1}\mathbb{E}_{ab,cd,ef,gh}, \tag{8.40}$$

where

$$\hat{\mathbb{C}}_{ab,cd}^{-1} \equiv \sum_{op\in\mathcal{S}}\sum_{qr\in\mathcal{S}}\mathbb{C}_{op,qr}^{-1}P_{op,ab}P_{qr,cd} \tag{8.41}$$

is the projected version of $\mathbb{C}_{ab,cd}^{-1}$. Thus, we will use (8.40) to evaluate $\hat{\mathbb{E}}$.

Numerical evaluation of the sums in (8.36) and (8.40) is straightforward, and takes a few minutes of CPU time for a few hundred pulsars. Two different sets of indices/subscripts are used. Pulsar indices correspond to the pulsars themselves; pair indices are obtained by constructing an index set corresponding to all pulsar pairs $ab \in \mathcal{S}$. First, the Hellings and Downs correlation matrix $\mu_{ab}$ is evaluated for all members of the pair index set using (2.4). Next $\mathbb{C}_{ab,cd}$ is formed from (2.10) for all pairs in the index set, and inverted. (It can have small eigenvalues, so some care may be needed.) The corresponding four-pulsar-index tensors are then formed, using the pair index set mapping. For $\hat{\mathbb{E}}$, we also need the projection operator $P_{ab,cd}$, which is computed from $\mathbb{C}_{ab,cd}^{-1}$ and $\mu_{ab}$ according to (8.38). This the inverse $\mathbb{C}_{ab,cd}^{-1}$ of the pulsar-pair covariance matrix $\mathbb{C}_{ab,cd}$ from (2.10). Thus,

$$\mathbb{E} \equiv \sum_{ab\in\mathcal{S}}\sum_{cd\in\mathcal{S}}\sum_{ef\in\mathcal{S}}\sum_{gh\in\mathcal{S}} \mathbb{C}_{ab,cd}^{-1}\mathbb{C}_{ef,gh}^{-1}\mathbb{E}_{ab,cd,ef,gh}. \tag{8.36}$$

A similar expression holds for $\hat{\mathbb{E}}$, with $\mathbb{E}_{ab,cd,ef,gh}$ replaced by its projected version is contracted with $\mathbb{C}_{ab,cd}^{-1}$ twice to get $\hat{\mathbb{C}}_{ab,cd}^{-1}$ as defined in (8.41). Next, $\mathbb{E}_{ab,cd,ef,gh}$ is evaluated from (F8) as a product of four $\mu_{ab}$'s. Finally, the four sums over pulsar pairs in (8.36) and (8.40) are carried out using pulsar indices to obtain either $\mathbb{E}$ or $\hat{\mathbb{E}}$.

Plots of $\mathbb{E}$ and $\mathbb{E} - \hat{\mathbb{E}}$ versus $N_{\text{pul}}$ for the three different cases of (i) auto- and cross-correlations, (ii) auto-correlations only, and (iii) cross-correlations only are shown in Fig. 16. For case (i) we used algebraic expressions. For cases (ii) and (iii) we numerically evaluated (8.36) and (8.40) as described above, extrapolating the numerical values to large numbers of pulsars using polynomial fits to the numerical values. The following subsections give details of the calculations for these three different cases.

### 1. Case (i): Using both auto- and cross-correlations

If the $\chi^2$ statistic is formed using both auto- and cross-correlations, then it is easy to obtain simple analytic expressions for $\mathbb{E}$ and $\hat{\mathbb{E}}$ using the techniques of Sec. VII A. Recall that for this case, we replace summations over pairs $ab \in \mathcal{S} = \{ab | a \leq b\}$ with (redundant) independent summations over the individual pulsars

$$\sum_{ab\in\mathcal{S}} \rightarrow \sum_{a=1}^{N_{\text{pul}}}\sum_{b=1}^{N_{\text{pul}}} \equiv \sum_{a,b}. \tag{8.42}$$

In Sec. VII A, we discuss the consequences of replacing the set $\mathcal{S}$ with cardinality $N_{\text{pul}}(N_{\text{pul}}+1)/2$ by a larger set with cardinality $N_{\text{pul}}^2$. Those same considerations apply here.

The advantage of this larger indexing set is that we can explicitly invert the covariance matrix $\mathbb{C}_{ab,cd}$. Since the correlations $\mu_{ab}$ are a symmetric $N_{\text{pul}} \times N_{\text{pul}}$ matrix, it follows from (7.5) that

$$\mathbb{C}_{ab,cd}^{-1} = \frac{1}{4}\hbar^{-4}\left(\mu_{ac}^{-1}\mu_{bd}^{-1} + \mu_{ad}^{-1}\mu_{bc}^{-1}\right), \tag{8.43}$$

noting that this inverse is only defined on the space of symmetric $N_{pul} \times N_{pul}$ matrices. We can use this to compute the quantities needed to characterize the $\chi^2$ statistic.

We first evaluate the projection operator. Substituting (8.43) into (8.39) gives the squared norm

$$\mu_{bin}^2 \equiv \sum_{a,b,c,d} \mu_{ab} \mathbb{C}_{ab,cd}^{-1} \mu_{cd} = \frac{1}{2} \hbar^{-4} N_{pul}, \quad (8.44)$$

where the calculation is identical to (7.11). From (8.38), this implies that the projection operator is

$$P_{ab,cd} = \delta_{ac} \delta_{bd} - \frac{1}{N_{pul}} \mu_{ab} \mu_{cd}^{-1}. \quad (8.45)$$

Note that the second term is symmetric in $ab$ and symmetric in $cd$, but that the first term is not. This is irrelevant for the calculations that follow: we only use $P_{ab,cd}$ as required in (8.41) to project $\mathbb{C}_{ab,cd}^{-1}$, which from (8.43) is explicitly symmetric in both index pairs. However, should one wish to extend the action of $P_{ab,cd}$ to general tensors (which have an antisymmetric part), then one should replace $\delta_{ac} \delta_{bd}$ in (8.45) with the symmetric projector $(\delta_{ac}\delta_{bd} + \delta_{ad}\delta_{bc})/2$.

As a final preparatory step, we use (8.45) to evaluate the projection operation in (8.41). This gives

$$\hat{\mathbb{C}}_{ab,cd}^{-1} = \mathbb{C}_{ab,cd}^{-1} - \frac{1}{2N_{pul}} \hbar^{-4} \mu_{ab}^{-1} \mu_{cd}^{-1} \quad (8.46)$$

for the projected inverse of $\mathbb{C}_{ab,cd}$.

We now use these expressions to compute the non-Gaussian contribution to the variance of $\chi^2$. Substituting (F8) for $\mathbb{E}_{ab,cd,ef,gh}$ into (8.36) and using (8.43), or substituting (F8) into (8.40) and using (8.46) we obtain the following exact expressions:

$$\mathbb{E} = \left(2N_{pul}^3 + 5N_{pul}^2 + 5N_{pul}\right)\left(\frac{\mathfrak{h}}{\hbar}\right)^8, \text{ and}$$

$$\hat{\mathbb{E}} = \left(2N_{pul}^3 + 5N_{pul}^2 - 7N_{pul} - 12 + \frac{12}{N_{pul}}\right)\left(\frac{\mathfrak{h}}{\hbar}\right)^8. \quad (8.47)$$

We have verified that these algebraic expressions agree with numerical results obtained by placing pulsars at random sky points.

These non-Gaussian contributions to the variance of $\chi^2$ are dominant when there are more than a few pulsars. The difference between these dominant contributions to the unprojected and projected statistics is given by (8.47) as

$$\mathbb{E} - \hat{\mathbb{E}} = 12\left(N_{pul} + 1 - \frac{1}{N_{pul}}\right)\left(\frac{\mathfrak{h}}{\hbar}\right)^8. \quad (8.48)$$

When there are many pulsars, so $N_{pul} \gg 1$, these show that $\mathbb{E} \simeq 2N_{pul}^3(\mathfrak{h}/\hbar)^8$ and $\mathbb{E} - \hat{\mathbb{E}} \simeq 12N_{pul}(\mathfrak{h}/\hbar)^8$. Plots of $\mathbb{E}$ and $\mathbb{E} - \hat{\mathbb{E}}$ [in units of $(\mathfrak{h}/\hbar)^8$] as a function of the number of pulsars $N_{pul}$ are shown in Fig. 16.

The numbers of pulsars used in current PTAs is large enough that only the leading term is needed to approximate the variance (also assuming, as we have done in this section, that the number of angular bins is as large as possible). With these assumptions, the fractional fluctuations in the $\chi^2$ statistics are well approximated by keeping only the non-Gaussian $\mathbb{E}$ term on the rhs of (8.31). Using (8.10) and (8.47), we obtain

$$\frac{\sigma_{\chi^2}}{\langle\chi^2\rangle} \simeq \frac{\mathbb{E}^{1/2}}{N_{bins}} \simeq \frac{\hat{\mathbb{E}}^{1/2}}{N_{bins}-1} \simeq \frac{2^{1/2}N_{pul}^{3/2}}{N_{bins}}\left(\frac{\mathfrak{h}}{\hbar}\right)^4 \simeq \frac{2^{5/4}}{N_{bins}^{1/4}}\left(\frac{\mathfrak{h}}{\hbar}\right)^4, \quad (8.49)$$

where the last approximate equality follows from (8.33), which gives $N_{bins}$ in terms of $N_{pul}$. Note that the presence of the $\mathbb{E}$ term leads to a fractional uncertainty that has a different scaling behavior than that of the textbook $\chi^2$ distribution, given in (8.32). See Table II, which compares the fractional uncertainties for the (unprojected) $\chi^2$ statistics for the different correlation sets, for the current sets of PTA collaboration pulsars. Comparing rows (1) and (4) shows the large effect of the non-Gaussian terms.

### 2. Case (ii): Using only auto-correlations

If the $\chi^2$ statistic is constructed from auto-correlations only, it takes a particularly simple form. In this case, $\chi^2$ tests whether the observed auto-correlations $\rho_{aa}$ of pulsars with themselves are all consistent with just one given value. That value (the expected correlation at zero angular separation $\gamma = 0$) is $h^2\mu_{aa} = 2h^2\mu_u(0)$ for the unprojected $\chi^2$ statistic. For the projected $\chi^2$ statistic, the observed auto-correlations are tested for consistency with $\hat{h}^2\mu_{aa} = 2\hat{h}^2\mu_u(0)$. Here, $\hat{h}^2$ is the squared GW strain estimator, formed from auto-correlations only, as described in Sec. VII B.

There are several simplifications that arise when using only auto correlations. First, the covariance matrix simplifies to

$$\mathbb{C}_{aa,bb} = \hbar^4\left(\mu_{ab}\mu_{ab} + \mu_{ab}\mu_{ab}\right) = 2\hbar^4\mu_{ab}^2. \quad (8.50)$$

This is an $N_{pul} \times N_{pul}$ matrix, but note that its elements are proportional to the squares of the elements of the matrix $\mu_{ab}$: it is *not* proportional to the matrix product of $\mu_{ab}$ with itself. Because only some components are needed, the tensor $\mathbb{E}$ given in (F8) reduces to

$$\mathbb{E}_{aa,bb,cc,dd} = 16\mathfrak{h}^8\left(\mu_{ab}\mu_{bc}\mu_{cd}\mu_{da} + \mu_{ac}\mu_{cd}\mu_{db}\mu_{ba} + \mu_{ad}\mu_{db}\mu_{bc}\mu_{ca}\right). \quad (8.51)$$

To further simplify the notation in what follows, we define

$$\mathbb{C}_{ab} \equiv \mathbb{C}_{aa,bb}, \quad P_{ab} \equiv P_{aa,bb}, \quad \mathbb{E}_{abcd} \equiv \mathbb{E}_{aa,bb,cc,dd}, \quad (8.52)$$

where the projection operator $P_{aa,bb}$ is computed from (8.38). Since $\mu_{aa} = 2\mu_u(0)\mathbb{1}_a$, we obtain

$$P_{ab} = \delta_{ab} - \frac{\mathbb{1}_a(\mathbb{C}^{-1}\mathbb{1})_b}{\mathbb{1}^\top \mathbb{C}^{-1}\mathbb{1}}, \quad (8.53)$$

where $\mathbb{C}^{-1} = \mathbb{C}_{ab}^{-1}$ is the matrix inverse of $\mathbb{C} = \mathbb{C}_{ab}$, and $\mathbb{1}$ is a vector with all components equal to one (3.7). Note that the trace of (8.53) correctly gives $N_{\text{pul}} - 1$: the numerator and denominator of the resulting fraction are each the grand sum of $\mathbb{C}^{-1}$, whose ratio is unity.

The non-Gaussian terms given in (8.36) and (8.40) simplify to

$$\mathbb{E} = 16\mathfrak{h}^8 \sum_{a,b,c,d} \mathbb{C}_{ab}^{-1}\mathbb{C}_{cd}^{-1} \left( 2\mu_{ab}\mu_{bc}\mu_{cd}\mu_{da} + \mu_{ad}\mu_{db}\mu_{bc}\mu_{ca} \right), \quad \text{and}$$

$$\hat{\mathbb{E}} = 16\mathfrak{h}^8 \sum_{a,b,c,d} \hat{\mathbb{C}}_{ab}^{-1}\hat{\mathbb{C}}_{cd}^{-1} \left( 2\mu_{ab}\mu_{bc}\mu_{cd}\mu_{da} + \mu_{ad}\mu_{db}\mu_{bc}\mu_{ca} \right), \quad (8.54)$$

where $\hat{\mathbb{C}}_{ab}^{-1}$ is the projected version of $\mathbb{C}_{ab}^{-1}$:

$$\hat{\mathbb{C}}_{ab}^{-1} \equiv \sum_{c,d} \mathbb{C}_{cd}^{-1} P_{ca} P_{db}. \quad (8.55)$$

To obtain the factor of 2 on the rhs's of (8.54), we used the symmetry of $\mathbb{C}_{ab}^{-1}$, $\hat{\mathbb{C}}_{ab}^{-1}$, and $\mathbb{E}_{abcd}$ under interchange of $a$ and $b$.

For this auto-correlation-only case, we do not have a way to simplify the expressions further, so we resort to numerical evaluation, placing simulated pulsars at random points on the sky. In this way, we evaluated $\mathbb{E}$, $\mathbb{E} - \hat{\mathbb{E}}$, and the fractional uncertainty of the $\chi^2$ statistics as a function of the number of pulsars $N_{\text{pul}}$; these are shown in Fig. 17. The black solid and dashed curves show $\mathbb{E}$ and $\mathbb{E} - \hat{\mathbb{E}}$ out to $N_{\text{pul}} = 1000$, normalized so that $\mathfrak{h}/\hbar = 1$. The polynomial fits (indicated by gray dotted curves) are fourth-degree polynomials in $N_{\text{pul}}^{1/2}$ for $\mathbb{E}$ and a first-degree polynomial in $N_{\text{pul}}$ for $\mathbb{E} - \hat{\mathbb{E}}$.

If this polynomial behavior extends to arbitrarily large values of $N_{\text{pul}}$, then the fractional uncertainty in $\chi^2$ will asymptote to a nonzero value. This is because $\mathbb{E} \sim N_{\text{pul}}^2 (\mathfrak{h}/\hbar)^8 \sim N_{\text{bins}}^2 (\mathfrak{h}/\hbar)^8$, which implies $\sigma_{\chi^2}/\langle\chi^2\rangle \approx \mathbb{E}^{1/2}/N_{\text{bins}} \sim \text{const}$ using the first approximate equality from (8.49). Thus, there would be a (nonzero) cosmic variance for $\chi^2$ for auto-correlations only. Spot checks show that the fit has less than 1% deviation out to $N = 2000$, but we have not been able to check larger values.

For $\chi^2$ statistics formed entirely from auto-correlations, we were unable to derive a simple algebraic expression for $\mathbb{E}$, analogous to (8.47). However, if we replace the PTA detector with a detector that is only sensitive to a single set of GW modes, then this is possible in the large-pulsar limit. We describe that here and give details in Appendix G. We hope that this may help others to derive a simple closed form for the PTA detector case.

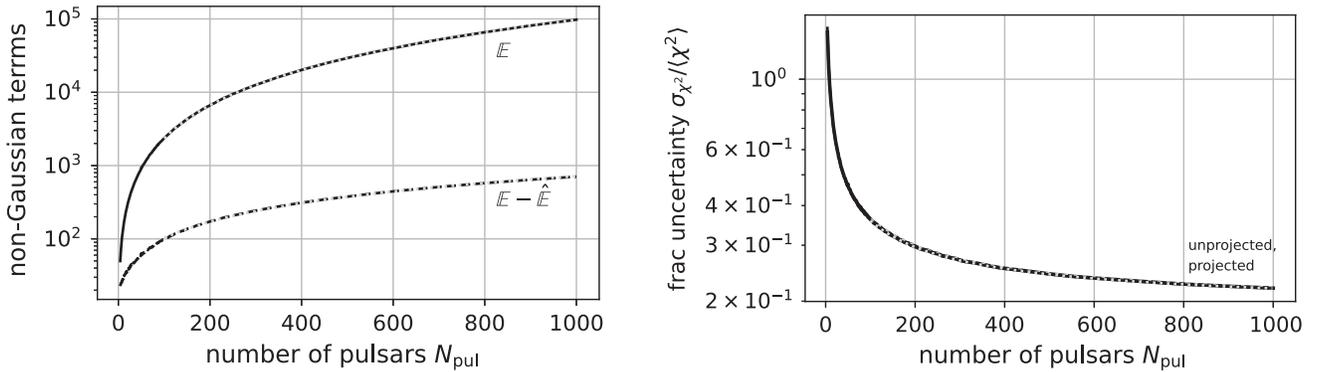

FIG. 17. Auto-correlations only. Left panel: non-Gaussian contributions to the variance of the unprojected and projected $\chi^2$ statistics, showing $\mathbb{E}$ (solid curve) and the difference $\mathbb{E} - \hat{\mathbb{E}}$ (dashed curve). The gray dotted curves show polynomial fits to the numerically determined values out to $N_{\text{pul}} = 1000$. The normalization is $\mathfrak{h}/\hbar = 1$. Right panel: the corresponding fractional uncertainties $\sigma_{\chi^2}/\langle\chi^2\rangle$ for a binary-inspiral model [57]. (On this scale, the unprojected and projected $\chi^2$ statistics are indistinguishable.) If the polynomial fits may be extrapolated to arbitrarily large values of $N_{\text{pul}}$, then the fractional uncertainty in $\chi^2$ asymptotes to a (nonzero) cosmic variance as $N_{\text{pul}} \to \infty$.

Consider (non-PTA) GW detectors whose correlation function (6.10) is defined by $Q_l = \delta_{lL}$ for some non-negative integer $L$. Such a detector is sensitive to exactly $2L + 1$ modes, corresponding to $l = L$ and $m = -L, ..., L$. For this case, $\mathbb{E}$ is small for small values of $N_{\text{pul}}$, and increases as $N_{\text{pul}}$ grows. However, it does not increase without bound. It saturates at a maximum value

$$\mathbb{E} = 2(2L + 1)(4L^2 + 9L + 6)\left(\frac{\mathfrak{h}}{\hbar}\right)^8 \quad \text{when } N_{\text{pul}} \geq (2L + 1)(L + 1). \tag{8.56}$$

Here, $(2L + 1)(L + 1) \equiv N_{\text{DOF}}$ is the number of nonzero eigenvalues of the matrix $\mathbb{C}_{ab}$.

This result can be obtained in two ways. (1) Fitting a polynomial to numerically determined values of $\mathbb{E}$, similar to what is described above for PTA detectors. (2) From an algebraic expression for $\mathbb{E}$ derived in Appendix G in the limit $N_{\text{pul}} \to \infty$, see (G37).

### 3. Case (iii): Using only cross-correlations

We now consider $\chi^2$ statistics formed from cross-correlations only. As in the previous case (auto-correlations only) we have not found a way to calculate the relevant quantities exactly. So, instead, we investigate the behavior of $\mathbb{E}$ and $\hat{\mathbb{E}}$ numerically, placing pulsars at random sky locations, and then computing (8.36) and (8.40).

Plots of $\mathbb{E}$, $\mathbb{E} - \hat{\mathbb{E}}$, and of the fractional uncertainty of the unprojected and projected $\chi^2$ statistics as functions of the number of pulsars $N_{\text{pul}}$ are shown in Fig. 18, together with polynomial fits. The curves are again normalized so that $\mathfrak{h}/\hbar = 1$. The non-Gaussian $\mathbb{E}$ term is well approximated by a third-degree polynomial in $N_{\text{pul}}$ with $\mathbb{E} \simeq 2N_{\text{pul}}^3(\mathfrak{h}/\hbar)^8$ for large $N_{\text{pul}}$. As foreshadowed earlier in Fig. 16, this is the same fit obtained when both auto- and cross-correlations are used, see also (8.47). For large $N_{\text{pul}}$, $\mathbb{E} - \hat{\mathbb{E}}$ is well approximated by a first-degree polynomial in $N_{\text{pul}}$ with leading coefficient $\approx 8.44(\mathfrak{h}/\hbar)^8$. This coefficient is smaller by a factor $\approx 1.42$ times than the corresponding coefficient when both auto- and cross-correlations are used to form $\chi^2$. This is close to $\sqrt{2}$, suggesting that the exact coefficient may be $6\sqrt{2}$.

Since the number of bins is $N_{\text{bins}} = N_{\text{pul}}(N_{\text{pul}} - 1)/2$, it follows that the fractional uncertainty in $\chi^2$ for both the unprojected and projected statistics tends to zero as $N_{\text{bins}}^{-1/4}$ for large $N_{\text{bins}}$ (or $N_{\text{pul}}$). This is shown in the right panel of Fig. 18, and is again similar to the auto + cross correlations case. In fact, as $N_{\text{pul}} \to \infty$, $\mathbb{E}$ and $\sigma_{\chi^2}/\langle\chi^2\rangle$ for the cross-correlations-only $\chi^2$ and the auto+cross-correlations $\chi^2$ are effectively indistinguishable. This is illustrated in both panels of Fig. 16.

### 4. Fractional uncertainties for the $\chi^2$ statistics for the current PTA pulsar locations

The previous sections give numerical results for pulsars located at random sky positions. Here, we consider the specific sets of pulsars currently employed by the different PTAs (see Appendix H and Table IV).

Table II gives the fractional uncertainties for the unprojected $\chi^2$ statistics for the pulsar pairs currently used by these PTAs. The fractional uncertainties for the projected $\chi^2$ statistics would be very close to the unprojected values listed in the table, so are not given. We assume noise-free measurements and put each pulsar pair into its own bin, so that $N_{bins}$ is related to $N_{pul}$ by (8.33). This ensures that the number of bins is as large as possible, giving the sharpest statistical test.

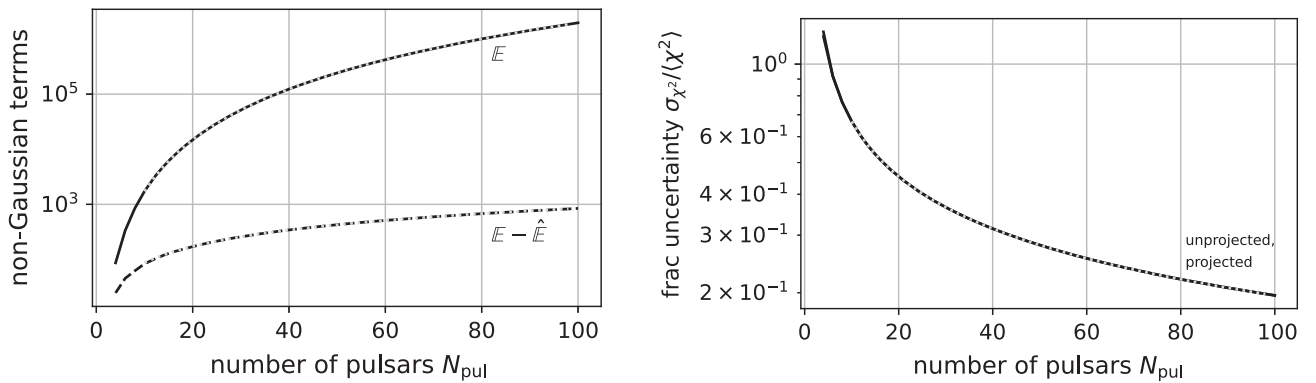

FIG. 18. Similar to Fig. 17 but for cross-correlations only. The polynomial fits to the underlying numerical results are shown by the gray dotted curves. As before, on the scale of the plot in the right panel, there is no discernible difference between the fractional uncertainties for the unprojected and projected $\chi^2$ statistics. The fractional uncertainty for cross-correlations only tends to zero like $N_{\text{pul}}^{-1/2}$, or $N_{\text{bins}}^{-1/4}$, and assumes a binary-inspiral model [57].

TABLE II.   Fractional uncertainties for the different (unprojected) $\chi^2$ statistics discussed in this section, assuming noise-free timing-residual measurements for each PTA's pulsars and the binary-inspiral GW model, see Table III and [57]. The total number of bins for each analysis is shown in parentheses. The values in row (1) are obtained directly from (8.47) for $\mathbb{E}$, whereas those in rows (2) and (3) are obtained by numerically evaluating (8.36) for specific pulsar sky directions) to determine $\mathbb{E}$. The interpretation of row (4), and the way in which it is computed, is discussed in the text.

| PTA: | EPTA | NANOGrav | PPTA | IPTA | Ideal PTA |
|---|---|---|---|---|---|
| Number of pulsars: | 42 | 66 | 26 | 88 | $\infty$ |
| (1) $\sigma_{\chi^2}/\langle\chi^2\rangle$ from auto + cross-correlations: | 0.3060 (903) | 0.2425 (2211) | 0.3930 (351) | 0.2094 (3916) | 0  ($\infty$) |
| (2) $\sigma_{\chi^2}/\langle\chi^2\rangle$ from auto-correlations only: | 0.5168   (42) | 0.4369   (66) | 0.6673   (26) | 0.3943   (88) | > 0? ($\infty$) |
| (3) $\sigma_{\chi^2}/\langle\chi^2\rangle$ from cross-correlations only: | 0.3062 (861) | 0.2426 (2145) | 0.3938 (325) | 0.2095 (3828) | 0  ($\infty$) |
| (4) $\sigma_{\chi^2}/\langle\chi^2\rangle$ (if correlations were Gaussian): | 0.0471 (903) | 0.0301 (2211) | 0.0755 (351) | 0.0226 (3916) | 0  ($\infty$) |

One row and one column of the table are for illustration and pedagogical purposes. Row (4) corresponds to the (incorrect) assumption of Gaussian correlations, and is obtained by taking the square root of (8.32). The contrast with the other rows shows that the non-Gaussian contributions dominate the Gaussian ones. In a similar vein, the final column shows the fractional uncertainty for a fictional PTA containing an infinite number of pulsars, uniformly distributed on the sky. This demonstrates the ultimate potential performance of a PTA, in the distant future.

The table contains a question mark in the final column of row (2). As discussed in Sec. VIII C 2, this is because we have not been able to determine if $\chi^2$ has a (nonzero) "cosmic variance" for auto-correlations only, although our numerical results are consistent with this. In this context, "nonzero cosmic variance" means that the fractional uncertainty in $\chi^2$ approaches a nonzero value in the $N_{pul} \to \infty$ limit.

Appendix G attempts to resolve this question mark, by computing the dominant (non-Gaussian) contribution to $\chi^2$, for auto-correlations only, in the $N_{pul} \to \infty$ limit. We derive an expression that can be evaluated exactly for detectors that respond to a finite number of GW modes and have no pulsar term. However, since a PTA is sensitive to an infinite number of modes and has a pulsar term, it does not answer the question of interest.

The $\chi^2$ statistic can be used to test the "Gaussian ensemble" hypothesis. For a given set of pulsars, numerical simulations of $\chi^2$ using many realizations of the GW background can be employed to estimate the fraction of universes in the Gaussian ensemble that lie within a specified number of standard deviations of the expected mean. (If the distribution of $\chi^2$ were Gaussian—which it is not—then 68% of realizations would lie within $\pm\sigma_{\chi^2}$ of the expected mean, while 99.7% would lie with $\pm 3\sigma_{\chi^2}$ of the expected mean.) If the observed value of $\chi^2$ differs from the expected mean by many standard deviations, then only a tiny fraction of the realizations in the Gaussian ensemble are consistent with the observations. In such cases, it would be reasonable to conclude that the Gaussian ensemble is a poor description of our Universe.

In Fig. 19 we illustrate how this would work in practice, based on numerical simulations containing 40 pulsars (at fixed random sky locations). We constructed hundreds of representative universes drawn from a Gaussian ensemble with squared strain $h^2 = 1$; the figure shows two of these. (The first representative universe is typical, and passes the test. The second representative is an atypical outlier, which fails the test.) For each realization, we construct the projected $\chi^2$ for cross-correlation-only data. While we know the $h^2$ value of the ensemble used in the simulations, this would not be known for real observations. So, following the same philosophy/approach as described at the end of Sec. VII D, we compute $\chi^2(h^2)$, and its expected "window," for a large range of assumed values of $h^2$. The assumption of a Gaussian ensemble is self-consistent if the $\chi^2(h^2)$ values computed from the data passes inside the credible window.

## IX. INCLUDING PULSAR AND MEASUREMENT NOISE

Up to this point, our calculations have assumed noise-free pulsar redshift (or timing-residual) data. Thus, they provide a baseline or "best-case scenario" for analyzing the data, e.g., the best-case precision with which we can recover the expected Hellings and Downs correlation or the smallest fractional uncertainty in estimating the squared GW strain $h^2$.

In this section, we describe how the inclusion of pulsar and measurement noise modifies previous (noise-free) expressions for the optimal Hellings and Downs correlation and its variance (Secs. III, IV, V), for the optimal squared-strain estimator and its variance (Sec. VII), and for the $\chi^2$ statistics and their variances (Sec. VIII). We assume that the noise in each pulsar is *time-stationary* and that this noise is *uncorrelated* between different pulsars. Our extension is not general enough to include noise such as random observatory clock fluctuations, which affect different pulsars in a correlated way.

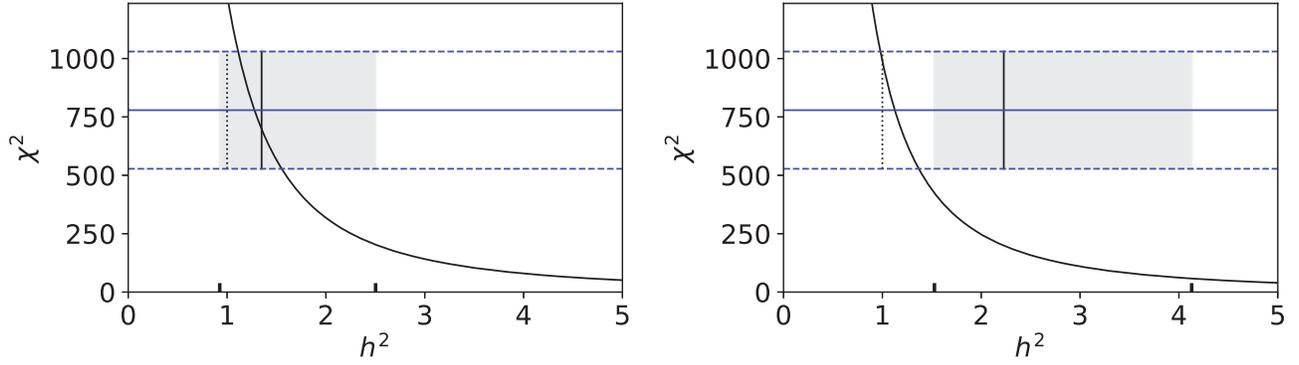

FIG. 19. $\chi^2$ tests for two simulated universes drawn from a noise-free Gaussian ensemble with squared strain $h^2 = 1$ (see text). The horizontal axis shows the (assumed unknown) value of $h^2$; the actual value is indicated by the vertical dotted line. The solid vertical line shows the inferred value $\hat{h}^2$ and the horizontal gray band is the inferred $\pm 1\sigma$ interval, as previously illustrated in Fig. 15. On the vertical axis, the solid curve shows the values of the projected $\chi^2(h^2)$ computed from the simulated observational data. The blue solid and dashed horizontal lines show the $\pm 1\sigma$ window around the expected mean. The left panel shows a typical realization. This "passes" the $\chi^2$ "one-sigma" test, because the black curve enters within the credible window. The right panel shows a rare atypical case. This "fails" the test, because the black curve lies entirely outside the window. The horizontal blue dashed lines representing $\langle \chi^2 \rangle \pm \sigma_{\chi^2}$ are straight because these simulations are noise-free. Including pulsar noise would make $\sigma_{\chi^2}$ dependent on $h^2$, so the upper and lower horizontal boundaries would be curved rather than straight.

Since the noise enters our calculations via the first four moments of the pulsar-pair correlations $\rho_{ab}$, the modifications are straightforward. New noise terms appear in expressions for the expected value $\langle \rho_{ab} \rangle$, for the covariance matrix $\mathbb{C}_{ab,cd} = \langle \rho_{ab}\rho_{cd} \rangle - \langle \rho_{ab} \rangle \langle \rho_{cd} \rangle$, and for the third- and fourth-order cumulants $\mathbb{D}_{ab,cd,ef}$ and $\mathbb{E}_{ab,cd,ef,gh}$. Thus, Secs. IX B, IX C, and IX E focus on these quantities.

If noise is present, and the optimal estimators (for the Helling and Downs correlation or for the squared GW strain) include pulsar auto-correlations, then the noise-free expressions (3.1) and (7.1) are modified. To obtain an unbiased estimator, a nonzero constant term must be added to the existing (proportional to $\rho_{ab}$) terms. This is because $\langle \rho_{ab} \rangle$ acquires an additional (auto-correlated) noise term, as shown in (9.6). The optimal estimators then include constant "bias subtraction" terms, which remove the effects of autocorrelated noise. An example of such a modified squared-strain estimator $\hat{h}^2$ is given in (9.16).

The presence of noise does not modify the definitions of the $\chi^2$ statistics given in (8.1), (8.16), and (8.19), even if those sums include auto-correlations. Note, however, that $\chi^2$ is defined in terms of $\rho_{\text{opt},j}$ and $\hat{h}^2$. Their definitions are modified in the presence of noise, as described above.

Another consequence of including the effects of noise is that the optimal estimator $\hat{h}^2$ and the optimal estimator $\rho_{\text{opt}}$ of the Hellings and Downs correlation now depend on $h^2$, which is the (unknown) quantity that we are trying to estimate. So, the question arises: what value of $h^2$ should we choose to evaluate these estimators? Our solution is to identify the *self-consistent* values of the squared strain. We do this by evaluating the estimator $\hat{h}^2$, for many different (assumed) values of $h^2$, and finding the $h^2$ interval where these agree to within the uncertainty of the estimator. The details of this construction are given in Sec. IX D. We follow the same philosophy/approach that we adopted at the end of Sec. VII D, where (even in the absence of noise) we had to account for the $h^2$ dependence of the variance $\sigma_{\hat{h}^2}^2$.

### A. Pulsar redshifts in the presence of noise

In the presence of noise, the measured redshift $Z_a$ for pulsar $a$ has contributions from the GW signal and from the noise. We express these quantities in the frequency domain as

$$\tilde{Z}_a(f) = \tilde{h}_a(f) + \tilde{n}_a(f), \tag{9.1}$$

where $\tilde{Z}_a(f)$, $\tilde{h}_a(f)$, and $\tilde{n}_a(f)$ denote the Fourier domain representations of the pulsar redshift $Z_a(t)$, the GW contribution $h_a(t)$, and the noise contribution $n_a(t)$. As stated above, we assume that the noise associated with different pulsars $a$ and $b$ is stationary and uncorrelated between pulsars or with the GW background. This implies that the quadratic expectation values of the noise and GW contributions can be written as

$$\langle \tilde{n}_a^*(f)\tilde{n}_b(f') \rangle = 4\pi N_a(f)\delta_{ab}\delta(f - f'),$$
$$\langle \tilde{h}_a^*(f)\tilde{h}_b(f') \rangle = 4\pi H(f)\mu_{ab}\delta(f - f'), \quad \text{and}$$
$$\langle \tilde{n}_a^*(f)\tilde{h}_b(f') \rangle = 0. \tag{9.2}$$

Here $N_a(f)$ is a spectral function for the noise in pulsar $a$, which is analogous to $H(f)$ for the GW background. The angle brackets now denote an average over a Gaussian ensemble of different noise realizations *and* different GW source realizations. [To allow for timing-residual measurements, multiply $H(f)$ and $N_a(f)$ by factors of $(2\pi f)^{-2\alpha}$ as explained in Appendix A. For redshift measurements, set $\alpha = 0$; for timing-residual measurements, set $\alpha = 1$.]

The correlation between redshifts in pulsars $a$ and $b$ (which might denote the same pulsar) is frequency dependent. It is obtained by combining (9.1) and (9.2), giving

$$\langle \tilde{Z}_a^*(f)\tilde{Z}_b(f')\rangle = 4\pi\Gamma_{ab}(f)\delta(f - f'), \qquad (9.3)$$

where

$$\Gamma_{ab}(f) \equiv H(f)\mu_{ab} + N_a(f)\delta_{ab}. \qquad (9.4)$$

As we will see below, to incorporate noise into our previous analyses, $H(f)\mu_{ab}$ is replaced by $\Gamma_{ab}(f)$.

The transition from noise-free expressions to "noise-full" expressions is nontrivial, because $\Gamma_{ab}(f)$ is a function of

$$\rho_{ab} \equiv \overline{Z_aZ_b} \equiv \frac{1}{T}\int_{-T/2}^{T/2}dt\, Z_a(t)Z_b(t) = \int df \int df'\, \mathrm{sinc}\big(\pi(f - f')T\big)\tilde{Z}_a^*(f)\tilde{Z}_b(f'), \qquad (9.5)$$

where $T$ is the total observation time. Note that here, and in the remainder of this section, integrals without limits are taken over $(-\infty, \infty)$. Using (9.3) and (9.4), it immediately follows that

$$\langle\rho_{ab}\rangle = 4\pi\int df\,(2\pi f)^{-2\alpha}\Gamma_{ab}(f) = h^2\mu_{ab} + n_a^2\delta_{ab}, \qquad (9.6)$$

where $h^2$ is given in (2.6) and

$$n_a^2 \equiv 4\pi\int df\,(2\pi f)^{-2\alpha}N_a(f). \qquad (9.7)$$

[We have included a factor of $(2\pi f)^{-2\alpha}$ in some integrals to allow for either redshift or timing-residual measurements as discussed above.] Note that for distinct pulsar pairs $a \neq b$, frequency and not simply numerical factors multiplying $\mu_{ab}$ and $\delta_{ab}$. This complicates matters. We are interested in quantities like $\langle\rho_{ab}\rangle$, $\mathbb{C}_{ab,cd}$, $\mathbb{D}_{ab,cd,ef}$, and $\mathbb{E}_{ab,cd,ef,gh}$. For the noise-free case these contain various factors $h^2$, $\hbar^4$, $\hbar^6$, and $\mathfrak{h}^8$, which are obtained by integrating various products of $H(f)$ with itself over frequency, as shown in (2.6), (2.9), (F9), and (F10).

In the presence of noise, additional terms appear, arising from integrals of various products of $H(f)$, $N_a(f)$, and $N_b(f)$. For example, second-order moments of $\rho_{ab}$ will give rise to three numerical factors involving integrals of products $H(f)H(f')$, $H(f)N_a(f')$, and $N_a(f)N_b(f')$. Similar factors appear in the third- and fourth-order moments of $\rho_{ab}$. In total, 11 new factors arise from the inclusion of noise. These supplement the four factors $h^2$, $\hbar^4$, $\hbar^6$, and $\mathfrak{h}^8$ that are present in the noise-free case.

### B. First moment $\langle\rho_{ab}\rangle$ in the presence of noise

To calculate the first-order moment $\langle\rho_{ab}\rangle$ in the presence of noise, we begin by writing the time-averaged pulsar redshift correlation $\rho_{ab} \equiv \overline{Z_aZ_b}$ in the Fourier domain:

we have $\langle\rho_{ab}\rangle = h^2\mu_u(\gamma_{ab})$, which agrees with the noise-free case. This means that if one restricts attention to a single pulsar pair, then only the auto-correlation measurements $\rho_{aa}$ are affected by the noise. If one looks at correlations between different pulsar pairs, then the noise enters in a more complicated way.

### C. Second moment (covariance) $\mathbb{C}_{ab,cd}$ in the presence of noise

For the covariance matrix $\mathbb{C}_{ab,cd} \equiv \langle\rho_{ab}\rho_{cd}\rangle - \langle\rho_{ab}\rangle\langle\rho_{cd}\rangle$, we need to evaluate the expectation value (Gaussian ensemble average) of the product $\rho_{ab}\rho_{cd}$. This can be written in terms of the four redshift measurements $\tilde{Z}_a, ..., \tilde{Z}_d$ as

$$\rho_{ab}\rho_{cd} = \int df \int df' \int df'' \int df''' \,\mathrm{sinc}\big(\pi(f - f')T\big)\mathrm{sinc}\big(\pi(f'' - f''')T\big)\tilde{Z}_a^*(f)\tilde{Z}_b(f')\tilde{Z}_c^*(f'')\tilde{Z}_d(f'''). \qquad (9.8)$$

We assume that the pulsar noise and the GW background are described by independent central-limit-theorem Gaussian ensembles. This implies that $\tilde{Z}$, which is a sum of these two processes, is also described by a Gaussian ensemble.

Thus, starting from the ensemble average of (9.8), we use Isserlis's theorem [19] to convert the expectation values of products of four redshift measurements in the frequency domain to products of terms containing two redshift measurements. We then use (9.3) to evaluate these expectation values, eliminating two of the four integrals. Finally, we subtract $\langle\rho_{ab}\rangle\langle\rho_{cd}\rangle$, calculated in Sec. IX B, from (9.8). We obtain

$$
\begin{aligned}
\mathbb{C}_{ab,cd} &= (4\pi)^2 \int \mathrm{d}f \int \mathrm{d}f' \operatorname{sinc}^2\big(\pi(f-f')T\big)(4\pi^2 ff')^{-2\alpha}\big(\Gamma_{ac}(f)\Gamma_{bd}(f') + \Gamma_{ad}(f)\Gamma_{bc}(f')\big) \\
&= \hbar^4(\mu_{ac}\mu_{bd} + \mu_{ad}\mu_{bc}) + (\delta_{ac}\delta_{bd} + \delta_{ad}\delta_{bc})N_{ab}^2 + (\delta_{ac}M_a\mu_{bd} + \delta_{bd}M_b\mu_{ac} + \delta_{ad}M_a\mu_{bc} + \delta_{bc}M_b\mu_{ad}),
\end{aligned} \tag{9.9}
$$

where the second line follows from (9.4) by expanding the factors of $\Gamma_{ac}(f)$, $\Gamma_{bd}(f')$, etc., and where

$$
N_{ab}^2 \equiv (4\pi)^2 \int \mathrm{d}f \int \mathrm{d}f' \operatorname{sinc}^2\big(\pi(f-f')T\big)\big(4\pi^2 ff'\big)^{-2\alpha} N_a(f)N_b(f'), \quad \text{and} \tag{9.10}
$$

$$
M_a \equiv (4\pi)^2 \int \mathrm{d}f \int \mathrm{d}f' \operatorname{sinc}^2\big(\pi(f-f')T\big)\big(4\pi^2 ff'\big)^{-2\alpha} N_a(f)H(f'). \tag{9.11}
$$

Under our assumptions, this is the form of the covariance matrix, which includes both the GW and pulsar-noise contributions. Note that (9.9) is valid for all possible choices of the pulsars $a$, $b$, $c$, $d$, including, for example, $a = c$, $b = d$, etc.

Current PTA searches for GW backgrounds assume noise sources similar to those described above. The resulting noise terms appear in the likelihood functions used in Bayesian analyses. They are also used to weight pulsar-pair correlation measurements to form the standard cross-correlation detection statistic [42,50,51].

Pulsar timing data have been noise-dominated (until possibly recently [6–9]). (Indeed, after this paper was completed, some PTAs reported evidence of GW-induced timing correlations consistent with the Hellings and Downs prediction [58–61].) So, current methods to estimate the Hellings and Downs correlation $\rho(\gamma)$ ignore the (GW-induced) off-diagonal entries of the covariance matrix, including only the terms in (9.9) proportional to $N_{ab}^2$. (After this paper was completed, both NANOGrav [58] and EPTA [59] adopted our definition of the Hellings and Downs correlation and the corresponding estimation method. This shifts the means, and increases the variance of the estimates by about a factor of two.) This is a good approximation if the GW signal power is small relative to the noise. But if the GW power is comparable to (or larger than) that of the noise, then it is a mistake to drop the off-diagonal terms. Leaving out these terms, which are proportional to $M_a$ and $\hbar^4$, will lead to incorrect statistical assessments of the data (e.g., confidence intervals that do not have the proper statistical coverage). For this reason, efforts are currently underway [62] to include the full form of the covariance matrix, cf. (9.9), in the optimal pulsar-pair estimators described above.

### D. Estimating the squared GW strain in the presence of noise

In Sec. VII we showed how to estimate the squared GW strain $h^2$ in the absence of noise. Now, we consider how to carry out such estimates if the pulsar noise is not small. For simplicity, we will assume here that: (i) all pulsars have the same noise spectrum, so $N(f) \equiv N_a(f)$, and that this spectrum is completely known, and (ii) the *spectral shape* of the GW spectrum $H(f)$ is known, but *not* its overall normalization (amplitude). This simple model is enough to address the most important questions.

From assumption (i), it follows that the covariance matrix (9.9) simplifies to

$$
\mathbb{C}_{ab,cd} = \hbar^4(\mu_{ac}\mu_{bd} + \mu_{ad}\mu_{bc}) + n^4(\delta_{ac}\delta_{bd} + \delta_{ad}\delta_{bc}) + m^4(\delta_{ac}\mu_{bd} + \delta_{bd}\mu_{ac} + \delta_{ad}\mu_{bc} + \delta_{bc}\mu_{ad}), \tag{9.12}
$$

where

$$
n^2 \equiv 4\pi \int \mathrm{d}f \, N(f), \quad \text{and} \tag{9.13}
$$

$$
n^4 \equiv (4\pi)^2 \int \mathrm{d}f \int \mathrm{d}f' \operatorname{sinc}\big(\pi(f-f')\big) N(f)N(f'), \quad \text{and} \tag{9.14}
$$

$$
m^4 \equiv (4\pi)^2 \int \mathrm{d}f \int \mathrm{d}f' \operatorname{sinc}\big(\pi(f-f')\big) N(f)H(f'). \tag{9.15}
$$

These last three equations are (9.7), (9.10), and (9.11), with $\alpha = 0$ and $n^2$, $n^4$, $m^4$ instead of $n_a^2$, $N_{ab}^2$, $M_a$, since there is no pulsar dependence in these terms. Assumption (ii) fixes the ratio $\hbar^2/h^2$ as indicated in Table III, (i) implies that $n^4$ is proportional to $n^4$, and (i) and (ii) imply that $m^4$ is proportional to $n^2h^2$. Thus, with our assumptions, the covariance matrix $\mathbb{C}_{ab,cd}$ only depends upon a single unknown, which is $h^2$.

The optimal estimator of $h^2$ is given by (7.2) minus a constant term (proportional to $n^2$ below):

$$\hat{h}^2 = \left( \sum_{ab,cd \in \mathcal{S}} \mu_{ab} \mathbb{C}_{ab,cd}^{-1}(h^2) \mu_{cd} \right)^{-1} \sum_{ef,gh \in \mathcal{S}} \mu_{ef} \mathbb{C}_{ef,gh}^{-1}(h^2)(\rho_{gh} - n^2\delta_{gh}). \tag{9.16}$$

The constant term is chosen so that $\hat{h}^2$ is an unbiased estimator of $h^2$, i.e., $\langle \hat{h}^2 \rangle = h^2$, even if auto-correlations are included in the correlation set. The variance of $\hat{h}^2$ has the same form as for the noise-free case, (7.3),

$$\sigma_{\hat{h}^2}^2 = \left( \sum_{ab,cd \in \mathcal{S}} \mu_{ab} \mathbb{C}_{ab,cd}^{-1}(h^2) \mu_{cd} \right)^{-1}, \tag{9.17}$$

but contains the inverse of the "noise-full" covariance matrix (9.12).

These equations lead to an apparently impossible circle of dependence. Both the estimator and its variance are functions of the (unknown) quantity $h^2$, because $\mathbb{C}_{ab,cd}$ (and hence its inverse $\mathbb{C}_{ab,cd}^{-1}$) depend upon $h^2$. This means that to (optimally) estimate $h^2$, we need to know its value.

To explore this complication, we first write (9.16) and (9.17) in more compact matrix/vector notation

$$\hat{h}^2 = \frac{\mu^\top \mathbb{C}^{-1}(h^2)(\rho - n^2\delta)}{\mu^\top \mathbb{C}^{-1}(h^2)\mu} \equiv f(h^2), \tag{9.18}$$

and

$$\sigma_{\hat{h}^2}^2(h^2) = \frac{1}{\mu^\top \mathbb{C}^{-1}(h^2)\mu}, \tag{9.19}$$

where $\delta$ in (9.18) denotes the identity matrix in "pair-indexed vector" form. In the above equations, we explicitly indicate that both the optimal estimator $\hat{h}^2$ and its variance $\sigma_{\hat{h}^2}^2$ depend upon $h^2$. Given observational data, the function $f$ is known; only the correct value of its argument is unknown.

The formulas for $\hat{h}^2$ and its variance are problematic: what value of the unknown quantity $h^2$ should be used on the rhs of (9.18) and (9.19)? The approach we propose is to make the estimate of $\hat{h}^2$ for all possible values of $h^2$, and then to pick the range of estimates which are self-consistent. By this, we mean that the inferred estimate $\hat{h}^2$ agrees with the assumed value $h^2$ within the variance given by (9.19).

The circular aspect is confusing. To obtain an (optimal) estimate $\hat{h}^2$ of the squared GW strain $h^2$, we first need to know $h^2$, which is the quantity that we want to estimate. Fortunately there is a simple graphical solution, starting from our fundamental assumption: that the GW background is described by a Gaussian ensemble. Such an ensemble is completely characterized by its spectrum $H(f)$. We have assumed that the functional form of $H(f)$ is known, apart from an overall scale $h^2$. Thus, our Universe is one realization of the corresponding Gaussian ensemble, *for a particular definite value of* $h^2$. The problem is that we do not know what that value is. So, we embrace this uncertainty, and consider *all possible values of* $h^2$. For each of these possible values we ask: "what value of $h^2$ would the data suggest, and with what uncertainty?" One of those values of $h^2$ is the correct one, but how can we recognize it and estimate the uncertainty?

This approach is illustrated graphically in Fig. 20. To make these plots, we placed 88 pulsars at random sky locations, simulated Gaussian ensembles with three different GW (squared) amplitudes, and then randomly selected a single realization from each ensemble. The horizontal axis shows the possible values of $h^2$; the true value is indicated by the dotted vertical line. The vertical axis shows the optimal estimator $\hat{h}^2$ of $h^2$ (solid black) and its expected one-sigma interval (dashed blue lines) about its expected value $\langle \hat{h}^2 \rangle = h^2$ (solid blue line) based on the observational data $\rho$ using cross-correlations only, assuming that the $h^2$ value on the horizontal axis is the correct value. [The upper end of the one-sigma interval is $h^2 + \sigma_{\hat{h}^2}(h^2)$; the lower end is $h^2 - \sigma_{\hat{h}^2}(h^2)$, if that quantity is positive, else 0.] The range of $h^2$ values that are *self-consistent* lie in the interval where the solid black curve $\hat{h}^2 = f(h^2)$ falls inside the $h^2 \pm \sigma_{\hat{h}^2}(h^2)$ range. This self-consistent interval is denoted in the figure by the gray region, and is also indicated by tick marks on the $h^2$ axis. This is the range of $h^2$ for which the optimally estimated values of $h^2$ are consistent with the assumed value, within one standard deviation. This approach "automatically" identifies the results as an "upper limit" (the interval begins at $h^2 = 0$) or as a "detection" (the interval starts above zero). This is similar in spirit to the well-known method proposed by Feldman and Cousins [63].

Given a set of pulsar sky locations and correlation observations, other approaches to obtain bounds on $h^2$

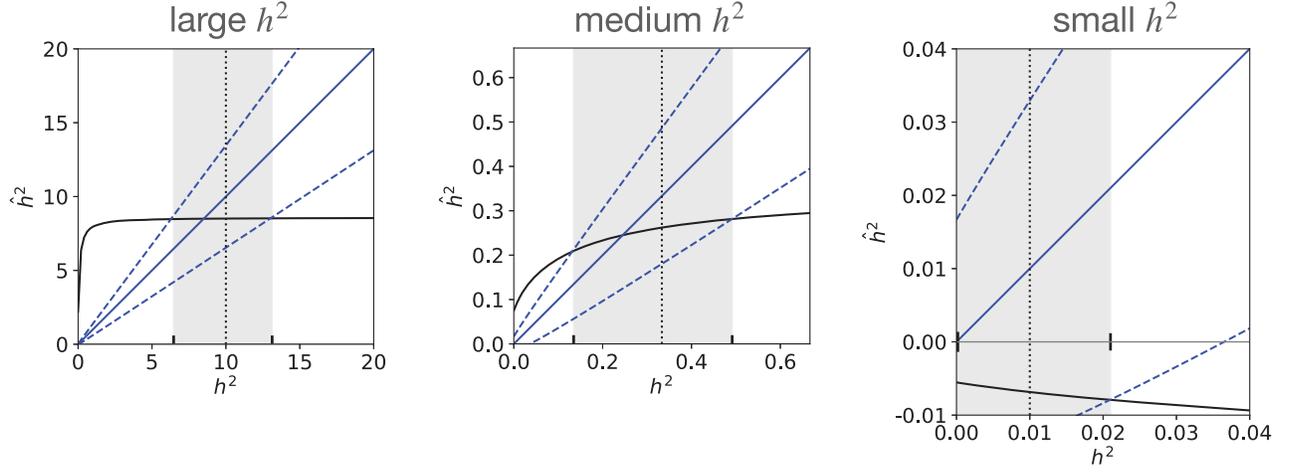

FIG. 20. Self-consistent squared-strain ($h^2$) intervals for simulated GW backgrounds having three different (squared) amplitudes. Each plot shows a single realization of a different Gaussian ensemble, observed with 88 pulsars. The true values of $h^2$ are shown by the vertical dotted lines. From left to right, the simulated GW backgrounds have large, medium, and small amplitudes in comparison to the observational noise, corresponding to $h^2/n^2 = 10, 0.33$, and $0.01$ respectively. The solid black curve shows $\hat{h}^2 = f(h^2)$ for the particular realization of the GW background and noise, constructed using cross-correlations only, while the dashed blue lines bracket the one-sigma intervals $h^2 \pm \sigma_{\hat{h}^2}(h^2)$ about the expected value $\langle \hat{h}^2 \rangle = h^2$ (solid blue diagonal line). The gray regions and corresponding tick marks on the $h^2$ axis are the inferred "one-sigma self-consistent" range of $h^2$. Note that for the weak-signal case, this interval extends down to $h^2 = 0$.

could also be followed. For example, one could simulate many GW backgrounds, with some prior distribution of $h^2$, and obtain a posterior distribution by asking which fraction give rise to correlations consistent with the observed correlations $\rho$. Provided that the priors are not pathological, we expect this would give similar bounds on $h^2$.

### E. Third and fourth moments (cumulants) $\mathbb{D}_{ab,cd,ef}$ and $\mathbb{E}_{ab,cd,ef,gh}$ in the presence of noise

Calculations similar to those of Sec. IX B (first moment) and Sec. IX C (second moment) show how noise modifies $\mathbb{D}_{ab,cd,ef}$ and $\mathbb{E}_{ab,cd,ef,gh}$. These quantities appear in the final expressions for the third- and fourth-order moments of $\rho_{ab}$, as described in Appendix F.

The noise-free expressions are (F7) for $\mathbb{D}_{ab,cd,ef}$ and (F8) for $\mathbb{E}_{ab,cd,ef,gh}$. In the presence of noise, these are replaced by

$$\mathbb{D}_{ab,cd,ef} = 4(4\pi)^3 \int df \int df' \int df'' \, \mathrm{sinc}\big(\pi(f-f')T\big)\mathrm{sinc}\big(\pi(f'-f'')T\big)\mathrm{sinc}\big(\pi(f''-f)T\big)(8\pi^3 ff'f'')^{-2\alpha}$$
$$\times \Big(\Gamma_{b(c}(f)\Gamma_{d)(e}(f')\Gamma_{f)a}(f'') + \Gamma_{a(c}(f)\Gamma_{d)(e}(f')\Gamma_{f)b}(f'')\Big), \tag{9.20}$$

and

$$\mathbb{E}_{ab,cd,ef,gh} = 8(4\pi)^4 \int df \int df' \int df'' \int df''' \, \mathrm{sinc}\big(\pi(f-f')T\big)\mathrm{sinc}\big(\pi(f'-f'')T\big)$$
$$\times \mathrm{sinc}\big(\pi(f''-f''')T\big)\mathrm{sinc}\big(\pi(f'''-f)T\big)(16\pi^4 ff'f''f''')^{-2\alpha}$$
$$\times \Big(\Gamma_{b(c}(f)\Gamma_{d)(e}(f')\Gamma_{f)(g}(f'')\Gamma_{h)a}(f''') + \Gamma_{b(e}(f)\Gamma_{f)(g}(f')\Gamma_{h)(c}(f'')\Gamma_{d)a}(f''')$$
$$+ \Gamma_{b(g}(f)\Gamma_{h)(c}(f')\Gamma_{d)(e}(f'')\Gamma_{f)a}(f''') + a \leftrightarrow b\Big). \tag{9.21}$$

These expressions can be further simplified by using (9.4) to expand the factors of $\Gamma_{bc}(f)$, etc. on the rhs's of (9.20) and (9.21). One then obtains integrals of various products of $H(f)$ and $N_a(f)$, $N_b(f)$, ... with themselves and one another. These recover earlier expressions for $h^2$, $\hbar^4$, $\hbar^6$, $\mathfrak{h}^8$, $n_a^2$, $N_{ab}^2$, $M_a$, and give rise to new expressions, such as

$$Q_a \equiv (4\pi)^3 \int df \int df' \int df'' \operatorname{sinc}(\pi(f-f')T)\operatorname{sinc}(\pi(f'-f'')T)\operatorname{sinc}(\pi(f''-f)T)$$
$$\times (8\pi^3 ff'f'')^{-2\alpha} N_a(f)H(f')H(f'') \tag{9.22}$$

and

$$R_{abc} \equiv (4\pi)^4 \int df \int df' \int df'' \int df''' \operatorname{sinc}(\pi(f-f')T)\operatorname{sinc}(\pi(f'-f'')T)$$
$$\times \operatorname{sinc}(\pi(f''-f''')T)\operatorname{sinc}(\pi(f'''-f)T)(16\pi^4 ff'f''f''')^{-2\alpha} N_a(f)N_b(f')N_c(f'')H(f'''). \tag{9.23}$$

There are a total of eight such new terms.

### F. Properties of the noise contribution to the covariance matrix

Does the inclusion of pulsar noise in the covariance matrix affect the analyses described in the previous sections? For a finite number of pulsars and sufficiently loud noise, the answer is yes. But in the limit of a sufficiently large number of pulsar pairs that faithfully cover the sky, the noise contributions (for independent pulsar noise) average to zero, leaving only the GW contribution.

To see this, we can repeat the calculation of Sec. IV including the additional noise terms. As before, the relevant entries of the covariance matrix are those where the rows and columns correspond to pulsar pairs $ab$ and $cd$ all separated by (approximately) the same angle $\gamma$. We showed in Secs. IV A and IV D that in the limit of a large number of such pulsar pairs distributed uniformly on the sky, $\mathbb{1}$ is an eigenvector of the matrix block $C_{jj}$ (with $\gamma_j \equiv \gamma$) for the noise-free contributions to $\mathbb{C}$. Then, according to (3.24), the variance of the optimal estimator in this narrow bin is the average value of the entries of this block.

So, is this also the case when we include the noise terms in (9.9)? Examining those terms by first symmetrizing over the indices $c$ and $d$, and then expanding $\mu_{ac}$, etc. in terms of the corresponding Hellings and Downs functions $\mu_u(\gamma_{ac})$ and Kronecker deltas $\delta_{ac}$, etc. using (2.4), we obtain

$$\mathbb{N}_{ab,cd} \equiv (\delta_{ac}\delta_{bd} + \delta_{ad}\delta_{bc})N_{ab}^2 + (\delta_{ac}M_a\mu_{bd} + \delta_{bd}M_b\mu_{ac} + \delta_{ad}M_a\mu_{bc} + \delta_{bc}M_b\mu_{ad})$$
$$= 2[N_{ab}^2 + (M_a + M_b)\mu_u(0)]\delta_{a(c}\delta_{d)b} + 2M_a\delta_{a(c}\mu_u(\gamma_{d)b}) + 2M_b\delta_{b(c}\mu_u(\gamma_{d)a}). \tag{9.24}$$

But since all of the above terms involve either a single Kronecker delta or a product of two Kronecker deltas, they give zero in the limit of an infinite number of pulsar pairs distributed uniformly on the sky, as shown explicitly in Appendix C.

Thus, the noise terms do not contribute to the limiting behavior of the full covariance matrix, and

$$\sigma_{\mathrm{opt}}^2(\gamma) = \sigma_{\mathrm{opt,signal}}^2(\gamma) + \sigma_{\mathrm{opt,noise}}^2(\gamma) = \sigma_{\cos}^2(\gamma), \tag{9.25}$$

just as we found for the noise-free case.

## X. HOW TO APPLY THE OPTIMAL ESTIMATOR

It is helpful to think about how the methods described in this paper would be used in practice. Imagine that we are handed a set of pulsar-pair correlation measurements $\rho_{ab}$ associated with an angular bin at angle $\gamma$. From these measured correlations, we want to construct an optimal estimator of the correlation at angle $\gamma$. We proceed as described in this paper. First, we compute the weights $w_{ab}$ from (3.9) or (3.18). For this, we need the (inverse of the) matrix block $C_{jj}$ of $\mathbb{C}$ appropriate for that angular-separation bin $\gamma_j \approx \gamma$. *But this covariance matrix cannot be obtained or estimated from the data that we have—a theoretical model is needed.*

For the Gaussian ensemble, in the limit of noise-free observations, the theoretical model of the covariance matrix is given by (2.10). The values of $\mu_{ac}$, $\mu_{bd}$, etc. on the rhs of (2.10) are fully determined by the sky locations of the pulsars via Eqs. (2.4) and (1.1). The value of $\hbar^4$, which also appears on the rhs of (2.10), is not needed: it cancels out of the numerator and denominator of (3.9) when evaluating the weights, thus allowing us to determine $\rho_{\mathrm{opt},j}$ from the measured correlations $\rho_{ab}$ [64].

The important point to note is that in following this procedure *we have assumed that the Gaussian ensemble model is correct.* Logically, there are only two possibilities: (a) the data/Universe is well described by the Gaussian ensemble, or (b) the data/Universe is *not* well described by the Gaussian ensemble. In case (a), the optimal estimator that we form, and the variance that we attribute to it, will both be meaningful and consistent. However, in case (b), neither the weights that we have derived nor the variance of the optimal estimator that we compute, have any justification.

Consider a specific example: a (non-Gaussian!) ensemble of universes containing GW sources which radiate at distinct nonoverlapping frequencies. Such an ensemble is constructed in Sec. III B of [11], where it is demonstrated that the cosmic variance is zero, so the pulsar-averaged correlations always follow the $\mu_u(\gamma)$ shape exactly. But if we followed the procedure described above, we would obtain the same estimated variance for the optimal estimator as in the Gaussian confusion-noise case, where the pulsar-averaged correlation curves do not follow $\mu_u(\gamma)$ precisely. For this example, the pulsar-averaged correlations computed from the data would follow the expected mean much more closely than expected for the Gaussian ensemble.

The conclusion is that the construction provided here is mainly useful as a consistency test, given that we have a finite set of observations that are not uniform on the sky. It can be used to determine the number of pulsar pairs to place in each angular bin, and also used to weight the correlations of those pairs. Finally, it can be used to estimate (for the Gaussian ensemble) the variance that would be expected for that set of pulsar pairs. If the correlation curve that is obtained has fluctuations of about $\pm 1\sigma_{opt}$ around the $h^2\mu_u(\gamma)$ curve, then we conclude that the Universe is consistent with the Gaussian ensemble predictions. If the fluctuations are significantly smaller or significantly larger, then the opposite conclusion is reached. These possibilities can be tested with the $\chi^2$ goodness-of-fit statistics that we define and discuss in Sec. VIII A.

In this sense, the results of this paper are useful because they predict the total Gaussian ensemble variance (pulsar plus cosmic) for a finite number of pulsars at specific sky locations, whereas the variance $\sigma_{cos}^2$ computed in Ref. [11] is for the limit of an infinite number of uniformly distributed pulsars.

## XI. CONCLUSION

The standard definition of the Hellings and Downs correlation is for a single pair of pulsars. We have extended and generalized that, defining the Hellings and Downs correlation for a set of pulsar pairs whose angular separations $\gamma$ lie in some (perhaps narrow) range.

Assuming that the GW background is described by a Gaussian ensemble, we have found the optimal way to weight the correlations of the individual pulsar pairs, so that the Hellings and Downs correlation has the smallest possible variations from the mean. These weights are normalized to produce an average that agrees with the single-pair correlation, enabling direct comparison. Moreover, in the absence of pulsar and instrumental noise, it is straightforward to compute the expected variance of these weighted sums.

As pulsar pairs are added at some angle, the variance of the correlation decreases. This variance starts at the total variance $\sigma_{tot}^2$ of [11] for a single pulsar pair. It decreases smoothly as pulsar pairs are added, eventually approaching the cosmic variance $\sigma_{cos}^2$ of that same reference. The plots in Fig. 6 illustrate this transition. Physically, there is a simple way to understand this. Once there are enough pulsar pairs at a given angular separation, adding additional pulsar pairs does not provide any additional information about the GW-induced correlations. This is because those pairs lie close on the sky to other pairs: their correlations provide redundant information.

We have also calculated the covariance of the optimal estimators for different angular separation bins. In the limit of an infinite number of pulsar pairs lying in narrow bins $j$ and $k$, we obtain an analytic expression (4.11) for the cosmic covariance matrix. This shows that the likely fluctuations away from the Hellings and Downs curve are strongly correlated (or anticorrelated) across three broad angular separation regions. We also computed the cosmic covariance matrix in harmonic space (4.15), where it takes diagonal form: a product of Legendre polynomials evaluated at $\cos\gamma_j$ and $\cos\gamma_k$, multiplied by the square of the harmonic space coefficients for the Hellings and Downs correlation.

To test if the observed deviations away from the expected Hellings and Downs curve are consistent with expectations, we proposed two sets of $\chi^2$ goodness-of-fit statistics. The first set assumes knowledge of the squared GW strain $h^2$ independent of the observational data, while the second set estimates $h^2$ from the pulsar-pair correlation data themselves. Somewhat surprisingly, we found that if the correlation set used to construct these statistics includes cross-correlations, then the fractional variance in the corresponding estimator of $h^2$ tends to zero in the limit of an infinite number of pulsars distributed uniformly on the sky. This means there is no cosmic variance for these $h^2$ estimators, despite the nonzero cosmic variance in the recovery of the expected Hellings and Downs correlation. We traced this behavior to the particular form of the pulsar response function, which is sensitive to an infinite number of modes.

We believe that the PTA GW sources in our Universe are well described by a Gaussian ensemble, and we have a good idea of their spectrum. This means that, for a specific set of pulsars at specific sky locations, we can predict the best-case variance in the Hellings and Downs correlation. This best case assumes that the intrinsic pulsar and experimental noise are small, and so provides a lower bound on the uncertainty of observational results. A simple recipe for making this prediction, given the pulsar sky locations, is given at the end of Sec. V B. Modifications needed to extend our formalism to include pulsar and measurement noise were described in Sec. IX.

In future PTA searches, using improved telescopes and larger pulsar populations, *we expect that the variance of the Hellings and Downs correlation will reveal exactly the cosmic variance*. In practice, this means that the optimal estimator of the Hellings and Downs correlation will not agree exactly in shape with the classic Hellings and Downs curve $\mu_u(\gamma)$. The cosmic variance is an estimator of this difference, and observational data can be tested for consistency with it, for example, by using $\chi^2$ statistics like those we have proposed.

## ACKNOWLEDGMENTS

The authors thank Curt Cutler for helpful discussions about the variance of the squared strain, and EPTA, NANOGrav, and PPTA for providing lists of their pulsar names and sky positions. B. A. is grateful to the members of the IPTA Detection Committee for their help in understanding many aspects of PTA data analysis, and J. D. R. acknowledges support from NSF Physics Frontiers Center Award PFC-2020265 and start-up funds from Texas Tech University.

## OUTLINE OF APPENDICES

The appendices contain calculations and arguments which are useful for the main body of the paper. Here, we outline their contents.

Appendix A shows how our results for the Hellings and Downs correlation of pulsar redshifts can be easily modified to treat pulsar timing residuals.

The relationship between the mean and the variance for the Gaussian ensemble depends upon certain integrals of the power spectrum. We discuss this dependence in Appendix B and calculate the relationship for a simple binary-inspiral model of the GW background.

In Appendix C, we examine the Kronecker delta terms which appear in the covariance matrix. We show that if there are many pulsar pairs distributed uniformly on the sky, then these terms constitute a set of measure zero and do not affect the variance.

In Appendix D, we consider the case where there are large numbers of pulsar pairs at a small (integer) number $\kappa$ of distinct separation angles. We show that the quantity relevant for the variance of the Hellings and Downs correlation estimator may be calculated from $\kappa(\kappa-1)/2$ quantities, obtained from the two-point function given in Appendix G in [11] via (4.8).

We exploit this result in Appendix E to show how (with a binned analysis employing a large number of pulsar-pair correlations at two different angles) it is possible to "beat" the cosmic variance. This does not come for free: angular resolution is lost.

In Appendix F, we calculate the third- and fourth-order moments of the pulsar-pair correlations $\rho_{ab}$. These are needed to evaluate the variance of the $\chi^2$ statistics defined in Secs. VIII A. Since $\rho_{ab}$ is quadratic in the Gaussian-distributed redshift (or timing-residual) measurements, the third- and fourth-order moments of $\rho_{ab}$ turn into sixth- and eighth-order moments of a Gaussian distribution. We evaluate these using Isserlis's theorem [19], obtaining triple and quadruple integrals of the GW power spectrum. We compute these integrals for both redshift and timing-residual measurements, assuming the same simple binary-inspiral background used in Appendix B. Results are given in Fig. 23 and Table III.

In Appendix G, we calculate the non-Gaussian contributions to the variance of the $\chi^2$ statistics in the many pulsar limit, for the special case where only auto-correlations are included. These analytic expressions can be evaluated exactly for GW detectors that (a) respond to only a finite number of harmonic modes and (b) have no pulsar terms. In this case, the expression becomes a finite sum. However, PTA detectors do not satisfy either condition, and the sums contain an infinite number of terms. For this reason, we have not been able to determine if the fractional variance of the auto-correlation-only $\chi^2$ statistics is finite or tends to zero as $N_{pul} \to \infty$ for PTA detectors. Our partial results may help others to resolve this question.

Finally, in Appendix H, we list the sky positions of the 88 pulsars currently monitored by three active PTA collaborations [37]. These are used for the example plots of Sec. V B, shown in Figs. 3 and 9, and for constructing Tables I and II.

## APPENDIX A: MEAN AND VARIANCE OF TIMING-RESIDUAL CORRELATIONS (VERSUS REDSHIFT CORRELATIONS)

Much of the literature and experimental/observational work for PTAs is based on pulsar timing residuals rather than redshifts. Pulsar timing residuals may be obtained by integrating the pulse redshifts with respect to time, which inserts a factor of $1/2\pi i f$ into frequency domain formulas. This makes it straightforward to modify formulas for the first moment, second moment, etc., and the variance and covariance of pulsar timing redshifts so that they can also be applied to pulsar timing residuals.

Following this procedure, formulas identical to (2.3) and (2.10) are obtained for the first moment and covariance matrix of timing-residual correlations, by replacing $H(f)$ with $H(f)/4\pi^2 f^2$ in the definitions of $h^2$ and $\hbar^4$ given in (2.6) and (2.9). To ensure that our formulas can cover both possibilities, we introduce a constant $\alpha$ into them.

Set $\alpha = 0$ to describe pulsar redshift correlations, or set $\alpha = 1$ to describe pulsar timing-residual correlations.

## APPENDIX B: RELATIVE SCALING BETWEEN THE CORRELATION MEAN AND VARIANCE FOR NOISE-FREE DATA

The GW Gaussian ensemble is completely specified by the real spectral function $H(f) \geq 0$ defined by (2.6) and (2.7). For example, $H(f)$ sets the overall scale $h^2$ of the expected Hellings and Downs correlation via (2.3). The spectral function $H(f)$ also determines the scale $\hbar^4$ of the covariance matrix $\mathbb{C}_{ab,cd}$ via (2.9) and (2.10).

This implies that $\hbar^4$ sets the scale of the total variance and the cosmic variance, since those are determined entirely by $\mathbb{C}_{ab,cd}$. It follows immediately that the ratio of the variance of the optimal estimator (for a given number of pulsar pairs and binning procedure) to (say) the cosmic variance is identical for *any* Gaussian ensemble: the quantities being compared only depend upon $\hbar^4$. Thus, plots in this paper that only contain variances are *universal*: they apply to any Gaussian ensemble.

In contrast, the ratio of the variance to the (squared) mean is model dependent: it depends upon the ratio $\hbar^4/h^4 \leq 1$ and hence upon $H(f)$. To compare the (square of the) mean and the variance, we must specify the spectrum $H(f)$, which defines the Gaussian ensemble. Hence, plots in this paper that contain variances and means (for example, Fig. 9) are *model dependent*. To make them, we must assume some form of the spectrum $H(f)$.

One simple source model is constructed in Sec. III A in [11]. This "confusion-noise model" consists of a large set of GW sources, uniformly distributed in space, *all radiating GWs at exactly the same frequency, but with different random phases*. The corresponding spectral function $H(f)$ is the sum of a delta function centered at that frequency, plus another equal one at the corresponding negative frequency. In Appendix C 3 in [11] it is shown that for this model $\hbar^4/h^4 \in [1/2, 1)$; if the frequency is an integer multiple of the inverse observation time, then $\hbar^4/h^4 = 1/2$. Note that the same values for this ratio are obtained for pulsar redshift correlations ($\alpha = 0$) and for pulsar timing-residual correlations ($\alpha = 1$). This source model is easy to construct and understand, so we use this value $\hbar^4/h^4 = 1/2$ in Fig. 1. However, the model is unrealistic because in our Universe the sources are distributed over frequency.

Here, we construct a more realistic spectral model, which we then use for Fig. 9 and elsewhere. This more realistic model assumes that the GW background is produced by large numbers of compact binary systems, whose energy loss is dominated by GW emission in the Newtonian limit (orbital velocities $\ll$ speed of light). In this case, $H(f) \propto |f|^{-7/3}$. Physically, this spectrum extends to very low frequencies (corresponding to periods longer than thousands of years).

Although it continues to much lower frequencies, for the purpose of computing PTA correlations, the spectrum is effectively "cut off" below a characteristic frequency $f_0 > 0$. Hence, we take

$$H(f) = \begin{cases} q|f|^{-7/3} & \text{for } |f| > f_0, \text{ and} \\ 0 & \text{for } |f| \leq f_0, \end{cases} \quad \text{(B1)}$$

where $q$ is a constant. The cutoff frequency $f_0$ is of order $1/T$, where $T$ is the total observation time. It arises because PTA data analysis in effect "removes" low-frequency components during the fitting process that subtracts (quadratic and higher-order terms in) the pulsar's intrinsic spindown to obtain timing residuals [65–67]. For further details, see the "transmission functions" illustrated in Fig. 2 of Ref. [67].

The ratio of the variance to the squared mean is proportional to $\hbar^4/h^4$. For our simple spectral model, this ratio depends only [68] upon the low-frequency cutoff $f_0$. This can be seen by substituting (B1) into (2.6) and (2.9) to obtain

$$\frac{\hbar^4}{h^4} = \frac{\int_{\pi f_0 T}^{\infty} \mathrm{d}x \int_{\pi f_0 T}^{\infty} \mathrm{d}y \left[\frac{1}{2}\operatorname{sinc}^2(x-y) + \frac{1}{2}\operatorname{sinc}^2(x+y)\right](xy)^{-7/3-2\alpha}}{\left(\int_{\pi f_0 T}^{\infty} \mathrm{d}x \, x^{-7/3-2\alpha}\right)^2}. \quad \text{(B2)}$$

Here, we have changed to dimensionless variables $x = \pi T f$ and $y = \pi T f'$, and used $H(f) = H(-f)$ to write the integrals in one-sided form. The constant $\alpha$ that appears in (B2) is explained in Appendix A. Set $\alpha = 0$ to describe the ratios of pulsar redshift correlations, or set $\alpha = 1$ to describe the ratios of pulsar timing-residual correlations.

The behavior of $\hbar^4/h^4 < 1$ as given in (B2) has interesting limits. As $f_0 T \to 0$, the integrands and integrals are dominated by values of $x$ and $y$ which are small enough that both sinc functions approach unity. The numerator of (B2) then approaches the denominator, and $\hbar^4/h^4 \to 1$. For values of $f_0 T$ which are somewhat larger, the first sinc function approaches unity, whereas the second from (4.3) or (9.24), do not contribute to the average value of the covariance matrix components over a row, $\langle \mathbb{C}_{ab,cd} \rangle_{cd \in \gamma}$ defined by (4.4), where $\gamma$ denotes the angular separation of the $ab$ and $cd$ pairs. In the limit where the number of pulsar pairs goes to infinity, these terms constitute a set of measure zero, and make negligible contributions to the average. [If each pulsar pair is placed in its own bin, then the Kronecker delta terms are not a set of measure zero and must be included; see for example (G8).]

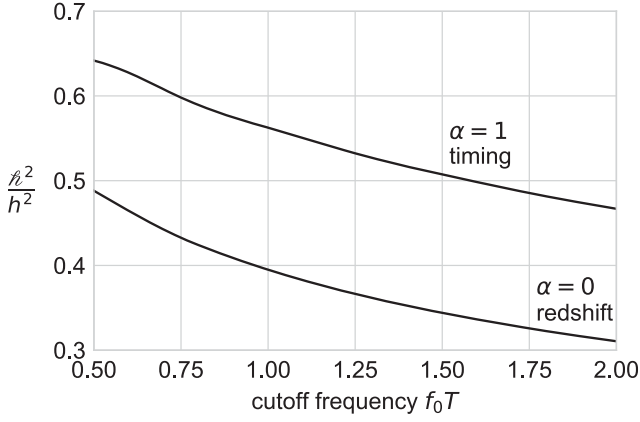

FIG. 21. The quantity $\hbar^2/h^2$ determines the ratio of the standard deviation $\sigma$ in the Hellings and Downs correlation to its mean $\langle \rho \rangle$. The ratio depends upon the spectrum $H(f)$ of the Gaussian ensemble. Here, this corresponds to a set of binary-inspiral sources (see text). The different values of the cutoff frequency $f_0$ are expressed in units of the inverse observation time $1/T$. The $\alpha = 0$ curve is for redshift correlations, and the $\alpha = 1$ curve is for timing-residual correlations (see Appendix A).

We begin by considering the terms proportional to $\delta_{a(c}\delta_{d)b}$. The factors $\mu_u^2(0)$, $N_{ab}^2$, or $(M_a + M_b)$ multiplying these terms are all finite, and hence will not affect the following analysis. Such terms are easy to evaluate, since there is only one value of $c$ and $d$ that will match the fixed values of $a$ and $b$:

$$2\langle \delta_{a(c}\delta_{d)b} \rangle_{cd \in \gamma} = 1/n_{\text{pairs}}. \qquad (C2)$$

This is manifestly independent of the fixed pulsar pair $ab$, and vanishes in the limit of an infinite number of pulsar pairs distributed uniformly on the sky: it is a set of measure zero.

The terms proportional to a single Kronecker delta are also sets of measure zero. This may be understood by looking at just one of four subterms

$$s_{ab} \equiv \langle \delta_{ac}\mu_u(\gamma_{db}) \rangle_{cd \in \gamma}. \qquad (C3)$$

sinc function falls off quickly, and $\hbar^4/h^4 \to 1/2$. For the cases of interest, where $f_0 T$ is of order unity, the first sinc function is also less than one (although it contributes more than the second sinc function).

The ratio $\hbar^2/h^2$ [the square root of (B2)] is evaluated numerically and shown in Fig. 21 as a function of the effective lower cutoff frequency $f_0 T$. A reasonable choice is $f_0 = 1/T$, which is justified by Fig. 2 of Ref. [67]. Taking $f_0 T = 1$ in Fig. 21 implies $\hbar^2/h^2 \approx 0.5622$ for timing residuals and $\hbar^2/h^2 \approx 0.3905$ for redshifts (see Table III). We use the first of these values for the plots in Fig. 9. For the other plot in the paper (Fig. 1) that shows a mean value, we use $\hbar^4/h^4 = 1/2$, corresponding to the single-frequency confusion-noise model described above and in [11].

## APPENDIX C: PROOF THAT TERMS INVOLVING KRONECKER DELTAS DO NOT CONTRIBUTE TO THE COSMIC VARIANCE

Here, we show that terms in the covariance matrix $\mathbb{C}_{ab,cd}$ involving one or two Kronecker delta terms, such as

$$2\hbar^4 \left[ \mu_u^2(0)\delta_{a(c}\delta_{d)b} + \mu_u(0)\left( \delta_{a(c}\mu_u(\gamma_{d)b}) + \delta_{b(c}\mu_u(\gamma_{d)a}) \right) \right] \quad \text{or}$$
$$2\left[ N_{ab}^2 + (M_a + M_b)\mu_u(0) \right]\delta_{a(c}\delta_{d)b} + 2M_a\delta_{a(c}\mu_u(\gamma_{d)b}) + 2M_b\delta_{b(c}\mu_u(\gamma_{d)a}) \qquad (C1)$$

We can again ignore the factors $\mu_u(0)$, $M_a$, and $M_b$ since they are finite multiplicative constants. This average also vanishes in the limit of an infinite number of pulsar pairs distributed uniformly on the sky. To see this, note that (C3) can be bounded above by replacing $\mu_u(\gamma_{db})$ with its maximum value $\mu_u(0)$ and bounded below by replacing $\mu_u(\gamma_{db})$ with its minimum value. Looking first at the upper bound, one must have

$$s_{ab} \leq \mu_u(0)\langle \delta_{ac} \rangle_{cd \in \gamma}. \qquad (C4)$$

Since $\delta_{ac}$ is independent of $d$, the averaging over $d$ (within a cone of angle $\gamma$ about $c$) has no effect; that is, the $cd$ averaging reduces to an average just over $c$. Now, imagine explicitly carrying out this average as a sum of $N$ terms divided by $N$. Cover the surface of the sphere with $N$ equal-area uniformly distributed patches, which are the discrete values of $c$. In the sum over $c$, the Kronecker delta in (C4) gets a contribution from only a single term (the term for which the $c$ patch contains $a$). Hence, we obtain

$$s_{ab} \leq \mu_u(0)/N, \qquad (C5)$$

so $s_{ab}$ is bounded above by $\mu_u(0)/N$. A similar argument shows that $s_{ab}$ is bounded below by a constant/$N$. Since both bounds converge to zero for large $N$, we conclude that $s_{ab}$ vanishes as $N \to \infty$.

## APPENDIX D: COMPUTATIONS WITH MATRICES OF A CERTAIN FORM

We consider real, symmetric, positive-definite matrices built up from smaller matrices, of the form

$$
\mathbb{C} = \begin{pmatrix}
C_{11} & C_{12} & \cdots & C_{1\kappa} \\
\hline
C_{21} & C_{22} & \cdots & C_{2\kappa} \\
\hline
\vdots & \vdots & \ddots & \vdots \\
\hline
C_{\kappa 1} & C_{\kappa 2} & \cdots & C_{\kappa\kappa}
\end{pmatrix}. \quad \text{(D1)}
$$

Here, each of the $C_{jk}$ is a (possibly nonsquare) matrix, and $\kappa$ is a positive integer. We assume that each of these $\kappa^2$ matrices has the property that *the sum of any row of $C_{jk}$ has the same value as the sum of any other row of $C_{jk}$.* (Since $\mathbb{C}$ is symmetric, this implies that the sum of any column of $C_{jk}$ equals the sum of any other column of $C_{jk}$.)

Note that the symbol $\mathbb{C}$ is the same as we gave to our definition of the covariance (2.2). However, in many cases of interest, (D1) is a submatrix of the full covariance matrix so that $\kappa$ does not range over the full angular range from $0°$ to $180°$. Thus, in this appendix (Appendix D), $\mathbb{C}$ may refer either to the full covariance matrix (2.2) or to some submatrix of it along the diagonal. This also holds for Appendix E.

We introduce the symbol $d_k$ (for $k = 1, ..., \kappa$) to denote the number of columns in the matrix $C_{jk}$. Note that this is independent of $j$. Note also that since $\mathbb{C}$ is symmetric, the number of rows of the matrix $C_{jk}$ is given by $d_j$. Hence, the matrix $C_{jk}$ has dimension ($d_j$ rows) × ($d_k$ columns). The sum

$$
N = \dim(\mathbb{C}) = \sum_{j=1}^{\kappa} d_j \quad \text{(D2)}
$$

is the number of rows or columns of the matrix $\mathbb{C}$.

Finally, we define the quantity $s_{jk}$ to be the *average value of the entries of the block $C_{jk}$.* Note that since the sum of any row of $C_{jk}$ gives the same number, independent of the row, that sum must be given by $d_k s_{jk}$.

Equation (D1) is the form taken by the covariance matrix $\mathbb{C}$ of the Hellings and Downs correlation, when there are $\kappa$ distinct values of the pulsar-pair separation angle $\gamma_j$, for $j = 1, ..., \kappa$, and there are $d_j$ distinct pulsar pairs at each angular separation. We have shown in Sec. IV that when there are large numbers of pulsar pairs at each angle, uniformly distributed on the sky, then the average value of any row of the $C_{jk}$ matrix is row independent.

In this appendix, we obtain a simple formula for the "sum of interest," which is the quantity $\mu^\top \mathbb{C}^{-1} \mu$. Here, $\mu$ is an $N$-dimensional column vector containing values of the Hellings and Downs curve corresponding to the angular separation of the pairs: for each $j = 1, \cdots, \kappa$, $\mu$ has $d_j$ identical entries $\mu_u(\gamma_j)$.

We begin by constructing a set of $\kappa$ column vectors $e_1, e_2, ..., e_\kappa$ of dimension $N$. Note that these do not form a basis for $\mathbb{C}$, since that would require $N$ vectors, and here we have far fewer. However, we will see that the $e_j$ form a basis for the subspace of interest, needed to evaluate the sum of interest.

The first of these vectors has $d_1$ equal nonzero elements followed by $N - d_1$ zeros. The second of these vectors has $d_1$ zeros followed by $d_2$ equal nonzero elements, followed by $N - d_1 - d_2$ zeros, and so on. Thus, we have

$$
\begin{aligned}
e_1 &= d_1^{-1/2}(1, ..., 1, 0, ..., 0, ......, 0, ..., 0)^\top, \\
e_2 &= d_2^{-1/2}(0, ..., 0, 1, ..., 1, ......, 0, ..., 0)^\top, \\
&\cdots \\
e_\kappa &= d_\kappa^{-1/2}(0, ..., 0, 0, ..., 0, ......, 1, ..., 1)^\top. \quad \text{(D3)}
\end{aligned}
$$

The ($N$-dimensional) dot product of any two of these vectors follows immediately from (D3), and is

$$
e_j^\top e_k = \delta_{jk}, \quad \text{(D4)}
$$

where $\delta_{jk}$ is the Kronecker delta. Hence, the dot product of distinct vectors vanishes.

Because the sum of any row of $C_{jk}$ yields the same number, the action of $\mathbb{C}$ on any one of the vectors $e_k$ produces a linear sum of the full set: the vectors $e_1, e_2, ..., e_\kappa$ form an invariant subspace under the action of $\mathbb{C}$. Multiplying $e_k$ given in (D3) on the left by $\mathbb{C}$ given by (D1) yields

$$
\mathbb{C} e_k = \sum_{j=1}^{\kappa} S_{jk} e_j, \quad \text{(D5)}
$$

where

$$
S_{jk} \equiv (d_j d_k)^{1/2} s_{jk} \quad \text{for } j, k = 1, ..., \kappa. \quad \text{(D6)}
$$

The square root factors in (D6) arise from counting the contributions arising from each entry of the column vector $\mathbb{C} e_k$.

To evaluate the sum of interest, we exploit the fact that $s_{jk}$ and $S_{jk}$ are symmetric $\kappa \times \kappa$ matrices. This allows us to transform the $N \times N$ dimensional problem to a smaller $\kappa \times \kappa$ dimensional problem. Let $S^{-1}$ denote the matrix inverse of $S$, which satisfies (and is also defined by)

$$
\sum_{\ell=1}^{\kappa} S_{j\ell} S_{\ell k}^{-1} = \delta_{jk}. \quad \text{(D7)}
$$

From (D6) and (D7), it immediately follows that

$$
S_{jk}^{-1} = (d_j d_k)^{-1/2} s_{jk}^{-1} \quad \text{for } j, k = 1, ..., \kappa, \quad \text{(D8)}
$$

where $s^{-1}$ denotes the matrix inverse of $s$.

With the matrix inverse $S^{-1}$ we define a set of $N$-dimensional column vectors

$$v_j = \sum_{k=1}^{\kappa} S_{kj}^{-1} e_k \quad \text{for } j = 1, \ldots, \kappa. \tag{D9}$$

We now show that $\mathbb{C}^{-1} e_j = v_j$. Consider the action of $\mathbb{C}$ on $v_j$. Multiplying (D9) on the left by $\mathbb{C}$ and using (D5) we obtain

$$\begin{aligned}
\mathbb{C} v_j &= \sum_{k=1}^{\kappa} S_{kj}^{-1} \sum_{\ell=1}^{\kappa} S_{\ell k} e_\ell \\
&= \sum_{\ell=1}^{\kappa} \left( \sum_{k=1}^{\kappa} S_{\ell k} S_{kj}^{-1} \right) e_\ell \\
&= e_j, \tag{D10}
\end{aligned}$$

where on the second line we use the definition of the matrix inverse (D7) to replace the quantity in curved brackets with the Kronecker delta. Multiplying (D10) on the left by $\mathbb{C}^{-1}$, it follows immediately that

$$\mathbb{C}^{-1} e_j = v_j. \tag{D11}$$

Taking the inner product of (D11) from the left with $e_k^\top$ and using (D9) and (D4) gives us the inner product

$$e_k^\top \mathbb{C}^{-1} e_j = S_{kj}^{-1}, \tag{D12}$$

from which we can now easily evaluate the quantity of interest.

We write the $N$-dimensional column vector $\mu$ as a linear combination of the $e_j$:

$$\mu = \sum_{j=1}^{\kappa} \mu_j \sqrt{d_j} e_j, \tag{D13}$$

where $\mu_j \equiv \mu_{\mathrm{u}}(\gamma_j)$ is the value of the Hellings and Downs curve at angle $\gamma_j$. The quantity of interest is then evaluated as

$$\begin{aligned}
\mu^\top \mathbb{C}^{-1} \mu &= \sum_{j=1}^{\kappa} \sum_{k=1}^{\kappa} \mu_j \mu_k \sqrt{d_j} \sqrt{d_k} e_j^\top \mathbb{C}^{-1} e_k \\
&= \sum_{j=1}^{\kappa} \sum_{k=1}^{\kappa} \mu_j \mu_k (d_j d_k)^{1/2} S_{jk}^{-1} \\
&= \sum_{j=1}^{\kappa} \sum_{k=1}^{\kappa} \mu_j \mu_k s_{jk}^{-1} \\
&= \hat{\mu}^\top s^{-1} \hat{\mu}, \tag{D14}
\end{aligned}$$

where on the second line we use (D12) to evaluate the matrix elements of $\mathbb{C}^{-1}$ and on the third line we use (D8) to replace the matrix elements of $S^{-1}$ by those of $s^{-1}$.

The $\kappa$-dimensional column vector $\hat{\mu}$ which appears in the final line has components

$$\hat{\mu} = (\mu_1, \mu_2, \ldots, \mu_\kappa)^\top. \tag{D15}$$

Thus, the $N \times N$ dimensional computation of $\mu^\top \mathbb{C}^{-1} \mu$ has been reduced to a $\kappa \times \kappa$ dimensional problem, where $\kappa$ is the number of distinct angular bins.

An important feature of this result is that when there are enough pulsar pairs in a particular angular separation bin, then the variance is not significantly reduced by adding more pulsar pairs to that angular separation bin. Here, "enough pairs" means that the average value of the covariance matrix rows at that angle are approximately constant. This is because (D14) depends only upon the matrix $s_{jk}$. That matrix is determined by the average value of each block of the covariance matrix. Once there are enough pulsar pairs in a given angular separation bin, adding more pulsar pairs to that bin does not significantly change the average value of the block.

## APPENDIX E: "BEATING" THE COSMIC VARIANCE (A SIMPLE EXAMPLE)

Here, with a simple example [69], we show that the variance of the binned optimal estimator with two distinct angles can be *smaller* than the cosmic variance for a point estimator at a single angle.

We take an angular-separation bin containing a large number of pulsar pairs with separation angles $\gamma_1$ and $\gamma_2$, respectively, uniformly distributed on the sky. Let $\gamma_1 \approx 54°$ lie at the leftmost minimum of the cosmic variance $\sigma_{\cos}^2$, as shown in the left panel of Fig. 6, and $\gamma_2 \approx 49°$ lie at a zero of the expected Hellings and Downs correlation, so $\mu_2 \equiv \mu_{\mathrm{u}}(\gamma_2) = 0$. We will demonstrate that the joint optimal estimator has a *smaller* variance than the minimum value of the cosmic variance $\sigma_{\cos}^2(\gamma_1)$. This is possible because the binned estimator uses information from correlation measurements made at *both* angular separations $\gamma_1$ and $\gamma_2$. By sacrificing some angular resolution, the variance of the optimal estimator drops below the minimum cosmic variance. Such behavior is visible in Fig. 7.

We assume that there are $n_{\mathrm{pairs}}$ pulsar pairs at each of the two angles, so that the covariance matrix $\mathbb{C}$ for the two separation angles is $2n_{\mathrm{pairs}} \times 2n_{\mathrm{pairs}}$ in size. As described in Appendix D, and shown in (D14), in the limit as $n_{\mathrm{pairs}} \to \infty$, the variance of the optimal estimator can be calculated from the $2 \times 2$ matrix

$$s = s_{jk} = \begin{pmatrix} s_{11} & s_{12} \\ s_{12} & s_{22} \end{pmatrix}. \tag{E1}$$

The four values that appear in $s$ are the average values of the entries in each of the four corresponding $n_{\mathrm{pairs}} \times n_{\mathrm{pairs}}$ blocks $C_{jk}$, where $j, k = 1, 2$.

An explicit expression for $s_{jk}$ was derived in (4.8)

$$s_{jk} = \hbar^4 \int_0^\pi d\beta \sin\beta \Big( \mu_{++}(\gamma_j, \beta) \mu_{++}(\gamma_k, \beta) + \mu_{\times\times}(\gamma_j, \beta) \mu_{\times\times}(\gamma_k, \beta) \Big). \tag{E2}$$

As shown in (4.32), for large $n_{\text{pairs}}$, the variances of the individual (single-angle) optimal estimators approach the cosmic variances, so $s_{11} = \sigma_1^2 = \sigma_{\cos}^2(\gamma_1)$ and $s_{22} = \sigma_2^2 = \sigma_{\cos}^2(\gamma_2)$.

Since (E2) has the form of a positive-definite inner product, the Schwarz inequality ensures that $s_{12}^2 \leq s_{11} s_{22}$. Thus, we can parametrize the off-diagonal elements of $s$ as

$$s_{12} = r \sigma_1 \sigma_2, \tag{E3}$$

where the Schwarz inequality ensures that the real number $r \in [-1, 1]$.

The variance $\sigma_{\text{opt}}^2$ of the optimal binned estimator is expressed in terms of $\mu^T \mathbb{C}^{-1} \mu$ by (3.11). This is evaluated in (D14), in terms of the matrix $s$ of (E1). In that way, we obtain

$$\begin{aligned}
\frac{\mu_{\text{bin}}^2}{\sigma_{\text{opt}}^2} = \mu^\top \mathbb{C}^{-1} \mu &= \hat{\mu}^\top s^{-1} \hat{\mu} \\
&= \frac{1}{s_{11} s_{22} - s_{12}^2} \left( \mu_1^2 s_{22} + \mu_2^2 s_{11} - 2\mu_1 \mu_2 s_{12} \right) \\
&= \frac{1}{1 - r^2} \left( \frac{\mu_1^2}{\sigma_1^2} + \frac{\mu_2^2}{\sigma_2^2} - 2r \frac{\mu_1 \mu_2}{\sigma_1 \sigma_2} \right). 
\end{aligned} \tag{E4}$$

To obtain the second line, we explicitly inverted the $2 \times 2$ matrix in (E1), and set $\hat{\mu} = (\mu_1, \mu_2)^\top$ where $\mu_1 \equiv \mu_{\text{u}}(\gamma_1)$ and $\mu_2 \equiv \mu_{\text{u}}(\gamma_2)$. For the final line, we used the definitions of $\sigma_1^2$, $\sigma_2^2$, and $r$ in terms of $s_{11}$, $s_{22}$, and $s_{12}$.

In signal processing and data analysis, the quantities

$$S_{\text{opt}} \equiv h^2 \mu_{\text{bin}}/\sigma_{\text{opt}}, \quad S_1 \equiv h^2 \mu_1/\sigma_1, \text{ and } S_2 \equiv h^2 \mu_2/\sigma_2, \tag{E5}$$

which appear in (E4) are called expected SNRs. We include the factors of $h^2$ on the rhs of (E5) so that the numerators have the interpretation of correlation $\rho$. Larger (absolute) SNR values indicate that the mean can be determined with higher fractional precision. Because the mean of the Hellings and Downs correlation varies with angle $\gamma$, it is these SNRs (rather than the variance) that reflect the additional precision obtained by combining data from multiple angles. We will see that a loss in precision is not possible (the converse Santa Claus Principle in action [70]).

A simple geometric argument shows that combining data from the two angular-separation bins always increases the SNR. First, multiply (E4) by $h^4$ and use (E5) to obtain

$$S_{\text{opt}}^2 = \frac{1}{1 - r^2} \left( S_1^2 + S_2^2 - 2r S_1 S_2 \right). \tag{E6}$$

Then introduce new variables $x$ and $y$ so that

$$\frac{S_1}{S_{\text{opt}}} = \frac{x - y}{\sqrt{2}}, \qquad \frac{S_2}{S_{\text{opt}}} = \frac{x + y}{\sqrt{2}}. \tag{E7}$$

Equation (E6) satisfied by the three SNRs may now be written

$$\frac{x^2}{1 + r} + \frac{y^2}{1 - r} = 1, \tag{E8}$$

which is the equation of an ellipse with major radius $\sqrt{1 + |r|}$ and minor radius $\sqrt{1 - |r|}$, rotated by $45°$ with respect to axes defined by $S_1$ and $S_2$. This is shown in Fig. 22. Note that $0 \leq$ minor radius $\leq 1 \leq$ major radius $\leq \sqrt{2}$. This means that the ellipse always lies within the square, and hence $S_{\text{opt}}^2 \geq S_1^2$ and $S_{\text{opt}}^2 \geq S_2^2$. Thus, the optimal (binned) estimator always has an SNR that is greater than or equal to the SNRs for the individual (narrow-bin) estimators corresponding to the angular separations $\gamma_1$ and $\gamma_2$.

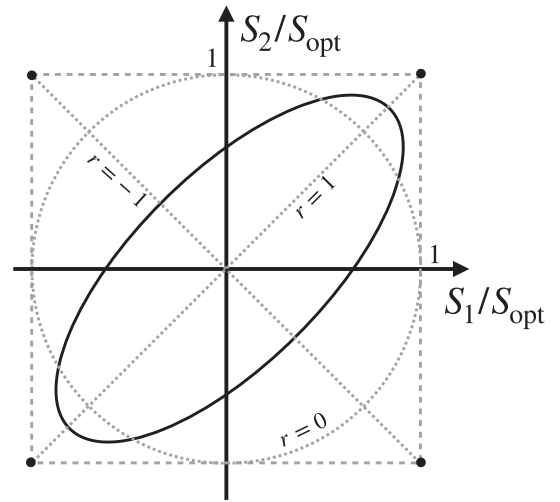

FIG. 22. Equation (E6) constrains the ratios of SNRs, $S_1/S_{\text{opt}}$ and $S_2/S_{\text{opt}}$, to lie on an ellipse. The shape of the ellipse is determined by $r \in [-1, 1]$, and is also shown for $r = 0$ (dotted unit circle) and for $r = \pm 1$ (degenerate dotted line segments at $\pm 45°$). The ellipse *always* lies within the square, proving geometrically that $S_{\text{opt}}^2 \geq S_1^2$ and $S_{\text{opt}}^2 \geq S_2^2$.

Some insight can be gained from two special cases: (i) If the correlation measurements at the two discrete angles are uncorrelated with one another (i.e., $r = 0$), then the ellipse is simply the unit circle, and the SNRs add in quadrature:

$$S_{\text{opt}}^2 = S_1^2 + S_2^2. \tag{E9}$$

(ii) For perfectly correlated or anticorrelated correlation measurements $r = \pm 1$, and $S_2 = \pm S_1$. The ellipses degenerate into lines as shown in Fig. 22, and one has

$$S_{\text{opt}}^2 = S_1^2 = S_2^2. \tag{E10}$$

In this second case, the SNR of the optimal (binned) estimator has the same value as that for the (narrow-bin) estimator for either angle separately, since no new information is obtained by considering the second angle.

Finally, to show that $\sigma_{\text{opt}}^2$ can be less than the global minimum of the cosmic variance, we take $\gamma_1 \approx 54°$ at a minima of the cosmic variance curve, and $\gamma_2 \approx 49°$ at a zero of the expected Hellings and Downs correlation, so $\mu_2 = 0$. Pick $\mu_{\text{bin}}$ to lie anywhere between $\mu_2 = 0$ and $\mu_1$. We have proved that $S_{\text{opt}}^2 \geq S_1^2$, which implies from (E5) that

$$\sigma_{\text{opt}}^2 \leq \frac{\mu_{\text{bin}}^2}{\mu_1^2} \sigma_1^2 < \sigma_1^2. \tag{E11}$$

Since $\sigma_1^2$ is at the minimum of the cosmic variance, $\sigma_{\text{opt}}^2$ lies below that. As mentioned at the start of this appendix, this behavior is apparent in Fig. 7. [For producing the plot in Fig. 7, we choose the normalization condition on the weights to be $\mu_{\text{bin}} \equiv \mu^\top \mathbb{1}/n_{\text{pairs}} = (\mu_1 + \mu_2)/2 = \mu_1/2$, leading to $\sigma_{\text{opt}}^2 \leq \sigma_1^2/4$ or $\sigma_{\text{opt}} \leq \sigma_{\cos}/2$.]

## APPENDIX F: THIRD AND FOURTH MOMENTS OF THE CORRELATION

To calculate the variance of the $\chi^2$ test statistics defined in Sec. VIII A, we need the third and fourth moments of the correlations $\rho_{ab}$ between pairs of pulsars. These may be computed using the methods of Appendix C in [11]. We used similar methods to compute the first and second moments in Secs. II A and II B. For convenience, we repeat those expressions here:

$$\langle \rho_{ab} \rangle = h^2 \mu_{ab}, \quad \text{and}$$
$$\langle \rho_{ab}\rho_{cd} \rangle = \langle \rho_{ab} \rangle \langle \rho_{cd} \rangle + \mathbb{C}_{ab,cd}, \tag{F1}$$

where

$$\mathbb{C}_{ab,cd} = \hbar^4 \big( \mu_{ac}\mu_{bd} + \mu_{ad}\mu_{bc} \big), \tag{F2}$$

and

$$h^2 \equiv 4\pi \int \mathrm{d}f \, (2\pi f)^{-2\alpha} H(f), \quad \text{and} \tag{F3}$$

$$\hbar^4 \equiv (4\pi)^2 \int \mathrm{d}f \int \mathrm{d}f' \, \mathrm{sinc}^2\big(\pi(f-f')T\big) \big(4\pi^2 ff'\big)^{-2\alpha} H(f)H(f'). \tag{F4}$$

The constant $\alpha$ is explained in Appendix A. For redshift correlations, $\alpha = 0$ and $\hbar$ is a dimensionless strain, whereas for timing-residual correlations, $\alpha = 1$ and $\hbar$ has units of time × (dimensionless) strain.

Since $\rho_{ab}$ is quadratic in the pulsar redshift measurements $Z_a$, calculating third- and fourth-order moments of $\rho_{ab}$ means evaluating expectation values of terms containing six or eight factors of $Z$. Since $Z$ is a Gaussian random variable, applying Isserlis's theorem [19] to evaluate these expectation values, we obtain

$$\langle \rho_{ab}\rho_{cd}\rho_{ef} \rangle = \langle \rho_{ab} \rangle \langle \rho_{cd} \rangle \langle \rho_{ef} \rangle + \langle \rho_{ab} \rangle \mathbb{C}_{cd,ef} + \langle \rho_{cd} \rangle \mathbb{C}_{ef,ab} + \langle \rho_{ef} \rangle \mathbb{C}_{ab,cd} + \mathbb{D}_{ab,cd,ef}, \quad \text{and} \tag{F5}$$

$$\begin{aligned}
\langle \rho_{ab}\rho_{cd}\rho_{ef}\rho_{gh} \rangle = {} & \langle \rho_{ab} \rangle \langle \rho_{cd} \rangle \langle \rho_{ef} \rangle \langle \rho_{gh} \rangle \\
& + \langle \rho_{ab} \rangle \langle \rho_{cd} \rangle \mathbb{C}_{ef,gh} + \langle \rho_{ab} \rangle \langle \rho_{ef} \rangle \mathbb{C}_{gh,cd} + \langle \rho_{ab} \rangle \langle \rho_{gh} \rangle \mathbb{C}_{cd,ef} \\
& + \langle \rho_{cd} \rangle \langle \rho_{ef} \rangle \mathbb{C}_{gh,ab} + \langle \rho_{cd} \rangle \langle \rho_{gh} \rangle \mathbb{C}_{ab,ef} + \langle \rho_{ef} \rangle \langle \rho_{gh} \rangle \mathbb{C}_{ab,cd} \\
& + \langle \rho_{ab} \rangle \mathbb{D}_{cd,ef,gh} + \langle \rho_{cd} \rangle \mathbb{D}_{ef,gh,ab} + \langle \rho_{ef} \rangle \mathbb{D}_{gh,ab,cd} + \langle \rho_{gh} \rangle \mathbb{D}_{ab,cd,ef} \\
& + \mathbb{C}_{ab,cd}\mathbb{C}_{ef,gh} + \mathbb{C}_{ab,ef}\mathbb{C}_{cd,gh} + \mathbb{C}_{ab,gh}\mathbb{C}_{cd,ef} + \mathbb{E}_{ab,cd,ef,gh},
\end{aligned} \tag{F6}$$

where

$$\mathbb{D}_{ab,cd,ef} = 4\hbar^6 \big( \mu_{b(c}\mu_{d)(e}\mu_{f)a} + \mu_{a(c}\mu_{d)(e}\mu_{f)b} \big), \quad \text{and} \tag{F7}$$

$$\mathbb{E}_{ab,cd,ef,gh} = 8\mathfrak{h}^8 \big( \mu_{b(c}\mu_{d)(e}\mu_{f)(g}\mu_{h)a} + \mu_{b(e}\mu_{f)(g}\mu_{h)(c}\mu_{d)a} + \mu_{b(g}\mu_{h)(c}\mu_{d)(e}\mu_{f)a} + a \leftrightarrow b \big), \tag{F8}$$

and

$$\hbar^6 \equiv (4\pi)^3 \int df \int df' \int df'' \, \mathrm{sinc}\big(\pi(f - f')T\big)\mathrm{sinc}\big(\pi(f' - f'')T\big)\mathrm{sinc}\big(\pi(f'' - f)T\big)$$
$$\times \big(8\pi^3 ff'f''\big)^{-2\alpha} H(f)H(f')H(f''), \quad \text{and} \tag{F9}$$

$$\mathfrak{h}^8 \equiv (4\pi)^4 \int df \int df' \int df'' \int df''' \, \mathrm{sinc}\big(\pi(f - f')T\big)\mathrm{sinc}\big(\pi(f' - f'')T\big)$$
$$\times \mathrm{sinc}\big(\pi(f'' - f''')T\big)\mathrm{sinc}\big(\pi(f''' - f)T\big)\big(16\pi^4 ff'f''f'''\big)^{-2\alpha} H(f)H(f')H(f'')H(f'''). \tag{F10}$$

The quantities $\mathbb{C}_{ab,cd}$, $\mathbb{D}_{ab,cd,ef}$, and $\mathbb{E}_{ab,cd,ef,gh}$ are the unique (up to scale) quadratic, cubic, and quartic combinations of the $\mu$'s that are symmetric with respect to each pair of indices $ab$, $cd$, ... separately, as well as under interchange of the pairs $ab \leftrightarrow cd$, etc. The scaling factors $h^2$, $\hbar^2$, $\hbar^2$, and $\mathfrak{h}^2$, which are determined by the GW spectrum, are bounded by $\mathfrak{h}^2 \le \hbar^2 \le \hbar^2 \le h^2$. The quantities $\mathbb{D}_{ab,cd,ef}$ and $\mathbb{E}_{ab,cd,ef,gh}$ are the third- and fourth-order *cumulants* [71] of $\rho_{ab}$. These cumulants vanish for a Gaussian distribution, showing that the correlations $\rho_{ab}$ are not Gaussian. This is no surprise as the $\rho_{ab}$ are formed by summing products of Gaussian-distributed redshift (or timing-residual) measurements $Z_a$ and $Z_b$.

Using the property $H(f) = H(-f)$, the integrals for $\hbar^6$ and $\mathfrak{h}^8$ can be put into single-sided form. It is helpful to first introduce a pair of functions of two variables $f_m$ and $f_n$:

$$P_{mn} \equiv \mathrm{sinc}\big(\pi(f_m + f_n)T\big) \quad \text{and} \quad M_{mn} \equiv \mathrm{sinc}\big(\pi(f_m - f_n)T\big), \tag{F11}$$

where $P$ and $M$ respectively denote "plus" and "minus." In terms of these functions, we can write the integrals in (2.6), (2.9), (F9), and (F10) as

$$h^2 = 8\pi \int_0^\infty df_1 \, (2\pi f_1)^{-2\alpha} H(f), \tag{F12}$$

$$\hbar^4 = \frac{1}{2}(8\pi)^2 \int_0^\infty df_1 \int_0^\infty df_2 \, (4\pi^2 f_1 f_2)^{-2\alpha} H(f_1)H(f_2)\big[M_{12}M_{21} + P_{12}P_{21}\big], \tag{F13}$$

$$\hbar^6 = \frac{1}{2^2}(8\pi)^3 \int_0^\infty df_1 \int_0^\infty df_2 \int_0^\infty df_3 \, (8\pi^3 f_1 f_2 f_3)^{-2\alpha} H(f_1)H(f_2)H(f_3)\big[M_{12}M_{23}M_{31} + 3M_{12}P_{23}P_{31}\big], \quad \text{and} \tag{F14}$$

$$\mathfrak{h}^8 = \frac{1}{2^3}(8\pi)^4 \int_0^\infty df_1 \int_0^\infty df_2 \int_0^\infty df_3 \int_0^\infty df_4 \, (16\pi^4 f_1 f_2 f_3 f_4)^{-2\alpha} H(f_1)H(f_2)H(f_3)H(f_4)$$
$$\times \big[ M_{12}M_{23}M_{34}M_{41} + P_{12}P_{23}P_{34}P_{41} + 4M_{12}M_{23}P_{34}P_{41} + 2M_{12}P_{23}M_{34}P_{41} \big]. \tag{F15}$$

We now evaluate these integrals for two cases of interest.

The first case of interest is the stochastic GW background model of Appendix B, for which $H(f)$ is given by (B1). This vanishes below frequency $f_0$, and behaves as a power-law with slope $-7/3$ above that frequency. As explained there, the effective low-frequency cutoff $f_0$ is of order $1/T$, where $T$ is the total observation time. To evaluate (F12), (F13), (F14), and (F15) for this spectrum, we set up a hypercubic grid in each dimension, which is uniformly spaced in the variable $u$ defined by $\mathrm{d}u = f^{-7/3-2\alpha}\mathrm{d}f$. The integral is then approximated by a sum over this grid. Representing the functions $M$ and $P$ by square symmetric matrices (whose rows and columns label the grid points), the sum may be written as the trace of the corresponding matrix products, which can be quickly and efficiently evaluated [72]. The results are shown in Fig. 23, and Table III gives the values for the nominal case $f_0 T = 1$.

TABLE III. The coefficients $\hbar^2$, $\hbar^2$, and $\mathfrak{h}^2$ for the binary-inspiral GW background model of (B1) for a (nominal) low-frequency cutoff $f_0 = 1/T$. These are accurate to four decimal places.

| | $h^2/h^2$ | $\hbar^2/h^2$ | $\hbar^2/h^2$ | $\mathfrak{h}^2/h^2$ |
|---|---|---|---|---|
| redshift ($\alpha = 0$) | 1 | 0.3905 | 0.3229 | 0.3002 |
| timing ($\alpha = 1$) | 1 | 0.5622 | 0.4933 | 0.4665 |

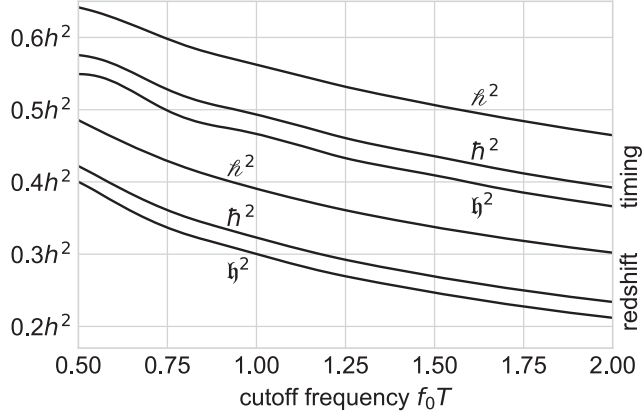

FIG. 23. The coefficients $\hbar^2$, $\hbar^2$, and $\mathfrak{h}^2$ for the binary-inspiral GW background model of (B1), expressed as a fraction of the mean-squared GW amplitude $h^2$. The upper three curves show the timing-residual case $\alpha = 1$, and the lower three curves show the redshift case $\alpha = 0$.

The second case of interest assumes that the spectrum $H(f)$ is a Dirac delta function with support at a frequency which is commensurate with the observation time. In this case, the functions defined in (F11) become $P_{mn} = 0$ and $M_{mn} = 1$, so we obtain $\hbar^4/h^4 = 1/2$, $\hbar^6/h^6 = 1/4$, and $\mathfrak{h}^8/h^8 = 1/8$, or, equivalently, $\hbar^2/h^2 = 2^{-1/2} \approx 0.7071$, $\hbar^2/h^2 = 2^{-2/3} \approx 0.6300$, and $\mathfrak{h}^2/h^2 = 2^{-3/4} \approx 0.5946$, for either timing ($\alpha = 1$) or redshift ($\alpha = 0$).

## APPENDIX G: EXPRESSIONS FOR $\mathbb{E}$ AND $\hat{\mathbb{E}}$ IN THE LARGE $N_{\mathrm{pul}}$ LIMIT FOR AUTO-CORRELATIONS ONLY

In this appendix, we derive analytic expressions for $\mathbb{E}$ and $\hat{\mathbb{E}}$ in the large $N_{\mathrm{pul}}$ limit. We assume that $\chi^2$ is formed from auto-correlations only, as described in Sec. VIII C 2.

Our final expressions may be easily evaluated for GW detectors that respond to only a finite number of harmonic modes ($l \leq L_{\max}$) and have no pulsar terms. These non-PTA detectors are described in Sec. VI.

We do not know how to evaluate these expressions for detectors that are sensitive to all modes, or if pulsar terms are present, both of which are true for PTAs. In these cases, the sums over modes $l$ are unbounded: we have not found a way to evaluate or characterize them. Thus, we regard these results as partial results, and present them here hoping that others will be able to make further progress.

We begin by finding a large-$N_{\mathrm{pul}}$ expression for the correlation function of a general GW detector, as defined by Eqs. (6.9) and (6.10):

$$\mu_{ab} = (1 + \eta^2 \delta_{ab})U_{ab}, \quad U_{ab} = \sum_l Q_l P_l(\cos\gamma_{ab}). \quad \text{(G1)}$$

A PTA detector corresponds to $Q_l = (2l + 1)C_l$, with $\eta = 1$ to include pulsar terms, and with the $C_l$ given by (4.14).

To include the pulsar terms in the large-$N_{\mathrm{pul}}$ limit, observe that

$$\sum_a \simeq \frac{N_{\mathrm{pul}}}{4\pi}\int d\hat{p}_a. \quad \text{(G2)}$$

Here, and throughout this appendix, we use $\simeq$ to denote an approximate equality that becomes exact in the limit of many pulsars ($N_{\mathrm{pul}} \to \infty$) uniformly distributed on the sky. The normalization of (G2) may be checked by applying this relation to 1: one obtains $N_{\mathrm{pul}}$ on both sides. It follows immediately that the Kronecker delta $\delta_{ab}$ has the continuum form

$$\delta_{ab} \simeq \frac{4\pi}{N_{\mathrm{pul}}}\delta^2(\hat{p}_a, \hat{p}_b), \quad \text{(G3)}$$

where $\delta^2(\hat{p}_a, \hat{p}_b)$ is the two-dimensional Dirac delta function on the 2-sphere $S^2$. The normalization of (G3) can be verified by summing both sides over all pulsars using (G2), giving unity on both sides.

Lastly, we expand the two-dimensional Dirac delta function as a sum of Legendre polynomials. Because the spherical harmonics are a complete orthonormal set on $S^2$, the Dirac delta function is

$$\delta^2(\hat{p}_a, \hat{p}_b) = \sum_{l=0}^{\infty}\sum_{m=-l}^{l} Y_{lm}(\hat{p}_a)Y_{lm}^*(\hat{p}_b). \quad \text{(G4)}$$

(Note that in the remainder of this section, double sums of this form are indicated by $\sum_{lm}$.) Using the addition theorem (6.6) to carry out the sum over $m$ in (G4) and using (G3) gives

$$\delta_{ab} \simeq \frac{4\pi}{N_{\mathrm{pul}}}\sum_l \frac{2l + 1}{4\pi}P_l(\cos\gamma_{ab}). \quad \text{(G5)}$$

Thus, defining $U(0) \equiv U_{aa} = \sum_l Q_l$, and using (G1) and (G5) we can write

$$\mu_{ab} \simeq \sum_l\left(Q_l + \eta^2 U(0)\frac{2l + 1}{N_{\mathrm{pul}}}\right)P_l(\cos\gamma_{ab})$$
$$= \sum_{lm} 4\pi\left(\frac{Q_l}{2l + 1} + \frac{\eta^2 U(0)}{N_{\mathrm{pul}}}\right)Y_{lm}(\hat{p}_a)Y_{lm}^*(\hat{p}_b)$$
$$= \sum_{lm}\Lambda_l Y_{lm}(\hat{p}_a)Y_{lm}^*(\hat{p}_b), \quad \text{(G6)}$$

where the second line follows from the addition theorem for spherical harmonics, (6.6), and the third line defines the expansion coefficients

$$\Lambda_l \equiv 4\pi \left( \frac{Q_l}{2l+1} + \frac{\eta^2 U(0)}{N_{\text{pul}}} \right). \tag{G7}$$

Although we are examining the case $N_{\text{pul}} \to \infty$, we cannot drop the term proportional to $1/N_{\text{pul}}$, because the $Q_l/(2l+1)$ term may equal zero after some finite value of $l$, or may decrease rapidly for increasing $l$. So, for any finite number of pulsars, there is a critical value of $l$ at which the term proportional to $1/N_{\text{pul}}$ dominates $\Lambda_l$.

Equations (G6) and (G7) are an eigenvector/eigenvalue decomposition of the $N_{\text{pul}} \times N_{\text{pul}}$ correlation matrix $\mu_{ab}$ in the large $N_{\text{pul}}$ limit. The normalized eigenvectors are spherical harmonics $Y_{lm}(\hat{\Omega})$ evaluated at the pulsar directions $\hat{p}_a$ and $\hat{p}_b$, with eigenvalues $\Lambda_l$, which are independent of $m$.

We now do an analogous harmonic expansion of $\mathbb{C}_{ab}$. For a general GW detector, (8.50) shows that its elements are proportional to the squares of the elements of (G1), which is written down explicitly in (7.17):

$$\mathbb{C}_{ab} = 2\hbar^4 \left( U_{ab}^2 + (2\eta^2 + \eta^4) U^2(0) \delta_{ab} \right). \tag{G8}$$

Each of the terms on the rhs is a function only of the angular separation $\gamma_{ab}$ between the two pulsars and not their individual directions $\hat{p}_a$ and $\hat{p}_b$. Hence, each term in $\mathbb{C}_{ab}$ may be written as an expansion in terms of Legendre polynomials of $\cos \gamma_{ab}$. We use (G5) for the term proportional to $\delta_{ab}$. The $U_{ab}^2$ term may be expressed as

$$U_{ab}^2 = \sum_l D_l P_l(\cos \gamma_{ab}), \tag{G9}$$

where the $D_l$ may be found using the orthogonality of the Legendre polynomials (4.19). This gives

$$\begin{aligned}
D_l &= \frac{2l+1}{2} \int_{-1}^{1} dx \, U_{ab}^2(x) P_l(x) \\
&= \frac{2l+1}{2} \sum_{l'} \sum_{l''} Q_{l'} Q_{l''} \int_{-1}^{1} dx \, P_{l'}(x) P_{l''}(x) P_l(x) \\
&= (2l+1) \sum_{l'} \sum_{l''} Q_{l'} Q_{l''} \begin{pmatrix} l & l' & l'' \\ 0 & 0 & 0 \end{pmatrix}^2,
\end{aligned} \tag{G10}$$

where the Wigner 3j symbols [53,54] are defined by an integral over the two-sphere $S^2$, given in the second equality of (G16). Thus, substituting (G5) and (G9) into (G8), and using the addition theorem (6.6) to write the Legendre polynomials in terms of spherical harmonics yields

$$\mathbb{C}_{ab} \simeq \sum_{lm} \lambda_l Y_{lm}(\hat{p}_a) Y_{lm}^*(\hat{p}_b), \tag{G11}$$

where the expansion coefficients are

$$\lambda_l \equiv 8\pi \hbar^4 \left( \frac{D_l}{2l+1} + \frac{(2\eta^2 + \eta^4) U^2(0)}{N_{\text{pul}}} \right). \tag{G12}$$

When there are many pulsars, (G11) is an eigenvector/eigenvalue decomposition of $\mathbb{C}_{ab}$, completely analogous to (G6) for $\mu_{ab}$.

In the many-pulsar limit, we can also expand $\mathbb{C}_{ab}^{-1}$ in a spherical harmonic basis, obtaining

$$\mathbb{C}_{ab}^{-1} \simeq \left( \frac{4\pi}{N_{\text{pul}}} \right)^2 \sum_{\{l|\lambda_l \neq 0\}} \sum_{m=-l}^{l} \lambda_l^{-1} Y_{lm}(\hat{p}_a) Y_{lm}^*(\hat{p}_b). \tag{G13}$$

The sum is restricted to those values of $l$ for which $\lambda_l$ is nonvanishing. (For PTAs, $\lambda_l$ is always nonzero, but for non-PTA detectors without a pulsar term, some $\lambda_l$ may vanish. In such cases, $\mathbb{C}_{ab}^{-1}$ is actually a pseudoinverse, as discussed in Sec. VIC.)

It is straightforward to verify that (G13) is the inverse of $\mathbb{C}_{ab}$ in the many-pulsar limit. Begin with $\sum_b \mathbb{C}_{ab}^{-1} \mathbb{C}_{bc}$. Use (G2) to replace the sum over $b$ with an integral. Evaluate the integral, and use orthonormality of the spherical harmonics to eliminate two of the sums. Use (G4) to express the remaining sum of spherical harmonics as $\delta^2(\hat{p}_a, \hat{p}_c)$, and lastly, use (G3) to transform that into $\delta_{ac}$.

We now use these expressions to evaluate $\mathbb{E}$ as given by (8.54). The two $\mathbb{C}^{-1}$ matrices are replaced by (G13), the four $\mu_{ab}$ are replaced by (G6), and the four sums over $a$, $b$, $c$, $d$ are replaced by (G2). This gives

$$\mathbb{E} \simeq 16\hbar^8 \sum_{\{l_1|\lambda_{l_1} \neq 0\}} \sum_{m_1} \sum_{\{l_2|\lambda_{l_2} \neq 0\}} \sum_{m_2} \sum_{l_3 m_3} \sum_{l_4 m_4} \sum_{l_5 m_5} \sum_{l_6 m_6} \lambda_{l_1}^{-1} \lambda_{l_2}^{-1} \Lambda_{l_3} \Lambda_{l_4} \Lambda_{l_5} \Lambda_{l_6} (2X + Y), \tag{G14}$$

where

$$\begin{aligned}
X &\equiv (-1)^{m_1 + m_2 + m_3 + m_4 + m_5 + m_6} I_{l_1 m_1; l_3 m_3; l_6, -m_6} I_{l_1, -m_1; l_3, -m_3; l_4 m_4} I_{l_2 m_2; l_4, -m_4; l_5 m_5} I_{l_2, -m_2; l_5, -m_5; l_6 m_6}, \\
Y &\equiv (-1)^{m_1 + m_2 + m_3 + m_4 + m_5 + m_6} I_{l_1 m_1; l_3 m_3; l_6, -m_6} I_{l_1, -m_1; l_4, -m_4; l_5 m_5} I_{l_2 m_2; l_5, -m_5; l_6, m_6} I_{l_2, -m_2; l_3, -m_3; l_4 m_4}
\end{aligned} \tag{G15}$$

are products of integrals of spherical harmonics, obtained from the integrals that replaced the sums over $a$, $b$, $c$, $d$.

The $I$'s are defined as integrals of the product of three spherical harmonics, which may be written in terms of the Wigner 3j symbols [53,54] as

$$
\begin{aligned}
I_{lm;l'm';l''m''} &\equiv \int d\hat{\Omega}\, Y_{lm}(\hat{\Omega}) Y_{l'm'}(\hat{\Omega}) Y_{l''m''}(\hat{\Omega}) \\
&= \sqrt{\frac{(2l+1)(2l'+1)(2l''+1)}{4\pi}} \begin{pmatrix} l & l' & l'' \\ 0 & 0 & 0 \end{pmatrix} \begin{pmatrix} l & l' & l'' \\ m & m' & m'' \end{pmatrix}.
\end{aligned}
\tag{G16}
$$

Where they are needed to avoid ambiguity, we add commas within the subscripts of the $I$'s, for example, in (G15).

The integrals $I_{lm;l'm';l''m''}$ have several important properties. The first is obvious from the definition (G16): they are symmetric under any permutation of the three pairs $lm$, $l'm'$ and $l''m''$. A second obvious property is that they vanish unless $m + m' + m'' = 0$; this is a consequence of the azimuthal dependence of $Y_{lm}(\theta,\phi) \propto e^{im\phi}$. A third relation is obtained by considering the inversion transformation $\hat{\Omega} \to -\hat{\Omega}$, which maps a point $\hat{\Omega}$ on the sphere to its antipodal point [73]. Under this transformation, a point with spherical coordinates $(\theta,\phi)$ is mapped to a point with coordinates $(\pi - \theta, \phi + \pi)$. Since the volume element $d\hat{\Omega}$ is invariant under this transformation, the integral in (G16) must be unchanged if we replace $d\hat{\Omega}$ by $d(-\hat{\Omega})$ or, equivalently, if we leave the volume element unchanged but replace the arguments of the three spherical harmonic functions by $-\hat{\Omega}$. Since $Y_{lm} \propto e^{im\phi} P_l^m(\cos\theta)$, under the transformation $Y_{lm}(-\hat{\Omega}) = Y_{lm}(\pi - \theta, \pi + \phi) = e^{im\pi}(-1)^{m+l} Y_{lm}(\theta,\phi) = (-1)^l Y_{lm}(\theta,\phi) = (-1)^l Y_{lm}(\hat{\Omega})$. We thus obtain

$$
I_{lm;l'm';l''m''} = (-1)^{l+l'+l''} I_{lm;l'm';l''m''} \Rightarrow I_{lm;l'm';l''m''} = 0 \quad \text{for } l + l' + l'' \text{ odd.}
\tag{G17}
$$

Thus follows the third property: if $l + l' + l''$ is odd, then $I_{lm;l'm';l''m''}$ vanishes.

We note that the second equality in (G16) does not suffice to define the Wigner 3j symbols for two reasons. First, if $l + l' + l''$ is odd, the left-hand side of the equality and the leftmost 3j symbol vanish, but the right 3j symbol may be nonzero. Furthermore, if $m = m' = m'' = 0$, then the overall sign is not determined. In such cases, the symbol is non-negative if $l + l' + l''$ is divisible by four, otherwise $l + l' + l''$ is nonpositive.

The powers of $(-1)$ in (G15) arise because some of the spherical harmonics are complex-conjugated: we have used $Y_{lm}^*(\hat{\Omega}) = (-1)^m Y_{l,-m}(\hat{\Omega})$ to obtain

$$
\int d\hat{\Omega}\, Y_{lm}^*(\hat{\Omega}) Y_{l'm'}(\hat{\Omega}) Y_{l''m''}(\hat{\Omega}) = (-1)^m I_{l,-m;l'm';l''m''}.
\tag{G18}
$$

Similar relations exist for integrals containing any product of three complex-conjugated or unconjugated $Y_{lm}$.

If all three spherical harmonics are complex-conjugated, the integral is invariant:

$$
I_{lm;l'm';l''m''} = I_{l,-m;l',-m';l'',-m''}.
\tag{G19}
$$

To see this, consider the integral

$$
\int d\hat{\Omega} Y_{lm}^*(\hat{\Omega}) Y_{l'm'}^*(\hat{\Omega}) Y_{l''m''}^*(\hat{\Omega}) = (-1)^{m+m'+m''} I_{l,-m;l',-m';l'',-m''} = I_{l,-m;l',-m';l'',-m''},
\tag{G20}
$$

where the first equality follows from $Y_{lm}^*(\hat{\Omega}) = (-1)^m Y_{l,-m}(\hat{\Omega})$ and the second equality follows because $I$ vanishes unless the sum of the three $m$'s is zero. But we can easily transform the lhs of (G20) to the same quantity without the complex conjugates. The integral over the azimuthal angle that appears on the lhs of (G20) is

$$
\int_0^{2\pi} d\phi\, e^{-i(m+m'+m'')\phi} = \int_0^{-2\pi} -du\, e^{i(m+m'+m'')u} = \int_{2\pi}^0 -du\, e^{i(m+m'+m'')u} = \int_0^{2\pi} du\, e^{i(m+m'+m'')u}.
\tag{G21}
$$

The first equality in (G21) follows from a change of variables $\phi = -u$, the second equality is because the argument is periodic with period $2\pi$, and the final sign change is from flipping the direction of integration. This is the same integral that would appear on the lhs of (G20) if the complex conjugates were removed, thus proving (G19).

The expression (G14) for $\mathbb{E}$ may be computed for any detector response that has only a finite number of nonzero

$\Lambda_l$ or $\lambda_l$, but naive counting indicates that the number of terms in the sums is of order $L_{\max}^{12}$. Computationally, this is very expensive, but the cost can be greatly reduced by combining and eliminating terms.

As a first step, we can simplify the expression (G14) for $\mathbb{E}$ by noting that the Wigner 3j symbols $\begin{pmatrix} l & l' & l'' \\ m & m' & m'' \end{pmatrix}$ vanish unless the following three conditions ("selection rules" [53,54]) are satisfied:

(i) $\quad m \in \{-l, -l+1, ..., l\}, \qquad m' \in \{-l', -l'+1, ..., l'\}, \qquad m'' \in \{-l'', -l''+1, ..., l''\},$

(ii) $\quad m + m' + m'' = 0,$

(iii) $\quad |l - l'| \le l'' \le l + l'.$

$$\tag{G22}$$

Because (G14) arises from a sum over spherical harmonics, condition (i) is always satisfied. Note that the Wigner 3j symbol which appears in (G16) is invariant under any even permutation of the three "columns." Under an odd permutation, it is invariant if the sum of the top row is even; otherwise it changes sign.

Using the symmetries and selection rules, we can reduce the number of terms in the sum to order $L_{\max}^9$. Consider the index structure of the four $I$'s that appear in (G15) for $X$. Selection rule (ii) implies:

$$
\begin{aligned}
m_1 + m_3 - m_6 &= 0, \\
-m_1 - m_3 + m_4 &= 0, \\
m_2 - m_4 + m_5 &= 0, \\
-m_2 - m_5 + m_6 &= 0.
\end{aligned}
\tag{G23}
$$

Since the final equation is implied by the previous ones, only three of these equations are independent. For $X$ these imply

$$m_4 = m_6 = m_1 + m_3, \qquad m_5 = m_1 - m_2 + m_3. \tag{G24}$$

In addition, selection rule (iii) applied allows us to restrict the summations over $l_4$, $l_5$, and $l_6$ to a smaller range of values. For $X$ this implies

$$\sum_{l_4} = \sum_{l_4 = |l_1 - l_3|}^{l_1 + l_3}, \qquad \sum_{l_5} = \sum_{l_5 = |l_2 - l_4|}^{l_2 + l_4}, \qquad \sum_{l_6} = \sum_{l_6 = |l_1 - l_3|}^{l_1 + l_3}. \tag{G25}$$

A similar set of conditions and restrictions hold for $Y$. These imply that

$$m_4 = m_2 + m_3, \quad m_5 = m_1 + m_2 + m_3, \quad m_6 = m_1 + m_3, \tag{G26}$$

and

$$\sum_{l_4} = \sum_{l_4 = |l_2 - l_3|}^{l_2 + l_3}, \qquad \sum_{l_5} = \sum_{l_5 = |l_1 - l_4|}^{l_1 + l_4}, \qquad \sum_{l_6} = \sum_{l_6 = |l_1 - l_3|}^{l_1 + l_3}. \tag{G27}$$

Thus, when evaluating $\mathbb{E}$, the summations over $m_4$, $m_5$, and $m_6$ can be eliminated, and the range of the summations over $l_4$, $l_5$, and $l_6$ can be restricted. These significantly reduce the computational cost of evaluating $\mathbb{E}$.

An explicit expression for $\mathbb{E}$ with the reduced number of summations is

$$\mathbb{E} \simeq 2A + B, \tag{G28}$$

where

$$A \equiv 16\hbar^8 \sum_{\{l_1|\lambda_{l_1}\neq 0\}} \sum_{\{l_2|\lambda_{l_2}\neq 0\}} \sum_{l_3} \sum_{l_4=|l_1-l_3|}^{l_1+l_3} \sum_{l_5=|l_2-l_4|}^{l_2+l_4} \sum_{l_6=|l_1-l_3|}^{l_1+l_3} \lambda_{l_1}^{-1} \lambda_{l_2}^{-1} \Lambda_{l_3} \Lambda_{l_4} \Lambda_{l_5} \Lambda_{l_6}$$

$$\times \frac{(2l_1+1)(2l_2+1)(2l_3+1)(2l_4+1)(2l_5+1)(2l_6+1)}{(4\pi)^2} \begin{pmatrix} l_1 & l_3 & l_6 \\ 0 & 0 & 0 \end{pmatrix} \begin{pmatrix} l_1 & l_3 & l_4 \\ 0 & 0 & 0 \end{pmatrix} \begin{pmatrix} l_2 & l_4 & l_5 \\ 0 & 0 & 0 \end{pmatrix} \begin{pmatrix} l_2 & l_5 & l_6 \\ 0 & 0 & 0 \end{pmatrix}$$

$$\times \sum_{m_1=-l_1}^{l_1} \sum_{m_2=-l_2}^{l_2} \sum_{m_3=-l_3}^{l_3} \begin{pmatrix} l_1 & | & l_3 & | & l_6 \\ m_1 & | & m_3 & | & -m_1-m_3 \end{pmatrix} \begin{pmatrix} l_1 & | & l_3 & | & l_4 \\ -m_1 & | & -m_3 & | & m_1+m_3 \end{pmatrix}$$

$$\times \begin{pmatrix} l_2 & | & l_4 & | & l_5 \\ m_2 & | & -m_1-m_3 & | & m_1-m_2+m_3 \end{pmatrix} \begin{pmatrix} l_2 & | & l_5 & | & l_6 \\ -m_2 & | & -m_1+m_2-m_3 & | & m_1+m_3 \end{pmatrix} \tag{G29}$$

and

$$B \equiv 16\hbar^8 \sum_{\{l_1|\lambda_{l_1}\neq 0\}} \sum_{\{l_2|\lambda_{l_2}\neq 0\}} \sum_{l_3} \sum_{l_4=|l_2-l_3|}^{l_2+l_3} \sum_{l_5=|l_1-l_4|}^{l_1+l_4} \sum_{l_6=|l_1-l_1|}^{l_1+l_1} \lambda_{l_1}^{-1} \lambda_{l_2}^{-1} \Lambda_{l_3} \Lambda_{l_4} \Lambda_{l_5} \Lambda_{l_6}$$

$$\times \frac{(2l_1+1)(2l_2+1)(2l_3+1)(2l_4+1)(2l_5+1)(2l_6+1)}{(4\pi)^2} \begin{pmatrix} l_1 & l_3 & l_6 \\ 0 & 0 & 0 \end{pmatrix} \begin{pmatrix} l_1 & l_4 & l_5 \\ 0 & 0 & 0 \end{pmatrix} \begin{pmatrix} l_2 & l_5 & l_6 \\ 0 & 0 & 0 \end{pmatrix} \begin{pmatrix} l_2 & l_3 & l_4 \\ 0 & 0 & 0 \end{pmatrix}$$

$$\times \sum_{m_1=-l_1}^{l_1} \sum_{m_2=-l_2}^{l_2} \sum_{m_3=-l_3}^{l_3} (-1)^{m_1+m_2} \begin{pmatrix} l_1 & | & l_3 & | & l_6 \\ m_1 & | & m_3 & | & -m_1-m_3 \end{pmatrix} \begin{pmatrix} l_1 & | & l_4 & | & l_5 \\ -m_1 & | & -m_2-m_3 & | & m_1+m_2+m_3 \end{pmatrix}$$

$$\times \begin{pmatrix} l_2 & | & l_5 & | & l_6 \\ m_2 & | & -m_1-m_2-m_3 & | & m_1+m_3 \end{pmatrix} \begin{pmatrix} l_2 & | & l_3 & | & l_4 \\ -m_2 & | & -m_3 & | & m_2+m_3 \end{pmatrix}. \tag{G30}$$

To avoid any ambiguity in their second rows, we have added vertical separators to the Wigner 3j symbols. Note that for detectors that respond to an infinite number of $l$-modes, or for detectors that include a pulsar term, the summations over $l$ extend to infinity.

We now prepare to calculate the non-Gaussian contribution $\hat{\mathbb{E}}$ to the $\chi^2$ statistic as given in (8.54). This requires the projection operator $P_{ab}$ and the projected inverse metric $\hat{\mathbb{C}}_{ab}^{-1}$; see (8.53) and (8.55). Using (G13) for $\mathbb{C}_{ab}^{-1}$, it is easy to see that in the large pulsar limit, $\mathbb{1}$ is an eigenvector of $\mathbb{C}^{-1}$:

$$\sum_b \mathbb{C}_{ab}^{-1} \mathbb{1}_b \simeq \frac{4\pi}{N_{\text{pul}}} \frac{1}{\lambda_0} \mathbb{1}_a \Rightarrow \mathbb{1}^\top \mathbb{C}^{-1} \mathbb{1} \simeq \frac{4\pi}{\lambda_0}, \tag{G31}$$

where $\mathbb{1}$ is a vector containing all ones (3.7). The first relation may be verified by using (G13) to replace $\mathbb{C}_{ab}^{-1}$ and (G2) to replace the summation over $b$ with an integral. Carrying out the integral, and using the orthonormality of the spherical harmonics, all terms vanish apart from the one with $l=m=0$. This term gives the first relation, which in

turn implies the second relation. Now, substituting (G31) into (8.53) leads to

$$P_{ab} \simeq \delta_{ab} - \frac{1}{N_{\text{pul}}} \mathbb{1}_a \mathbb{1}_b. \tag{G32}$$

(Note that this has the correct trace $N_{\text{pul}} - 1$.) Substituting this projection operator into (8.55) and using (G31) again, yields

$$\hat{\mathbb{C}}_{ab}^{-1} \simeq \mathbb{C}_{ab}^{-1} - \frac{4\pi}{\lambda_0} \frac{1}{N_{\text{pul}}^2} \mathbb{1}_a \mathbb{1}_b. \tag{G33}$$

This completes our preparation for the calculation that follows.

To compute the non-Gaussian contribution $\hat{\mathbb{E}}$ in the many-pulsar limit, we determine the difference between the two expressions given in (8.54). Since $\mathbb{C}^{-1}$ and $\hat{\mathbb{C}}^{-1}$ are related by (G33) (twice), the difference consists of two terms. The first term (1) contains $\mathbb{1}_a \mathbb{1}_b \mathbb{1}_c \mathbb{1}_d$, and the second term (2) contains $\mathbb{1}_a \mathbb{1}_b \mathbb{C}_{cd}^{-1} + \mathbb{1}_c \mathbb{1}_d \mathbb{C}_{ab}^{-1}$. To evaluate these, we replace the summations with integrals using (G2) (four times), replace $\mu_{ab}$ with a sum of harmonics using (G6) (four times), and in term (2) we replace $\mathbb{C}^{-1}$ with the harmonic expansion (G13). The four integrals may now be carried out. In term (1), each gives Kronecker deltas of the form $\delta_{l_1 l_2} \delta_{m_1 m_2}$, which allow all but the final sum to be evaluated. In term (2), two of the integrals evaluate to give Kronecker deltas. The remaining two integrals over three spherical harmonics each give rise to terms defined by (G16). In simplifying the product of those two terms, it is helpful to keep in mind that one may freely change the signs of the $m$'s, because they are summed over both signs. Making use of (G17), (G19), and (G20), one obtains

$$\mathbb{E} - \hat{\mathbb{E}} \simeq -\frac{48\hbar^8}{(4\pi\lambda_0)^2} \sum_l (2l+1)\Lambda_l^4 + \frac{32\hbar^8}{4\pi\lambda_0} \sum_{\{l_1|\lambda_{l_1}\neq 0\}} \sum_{m_1} \sum_{l_2 m_2} \sum_{l_3 m_3} \lambda_{l_1}^{-1} \Lambda_{l_2} \Lambda_{l_3}^2 (2\Lambda_{l_3} + \Lambda_{l_2}) I_{l_1 m_1; l_2 m_2; l_3 m_3}^2, \qquad (G34)$$

where the first term on the rhs arises from (1) and the second from (2). As earlier, we can use the Wigner 3j selection rules to eliminate the sum over $m_3$ in the last term (since $m_3 = -m_1 - m_2$), and also restrict the range of the sum over $l_3$ to run from $|l_1 - l_2|$ to $l_1 + l_2$. Making use of (G16) to evaluate the $I^2$ factor, we obtain

$$\mathbb{E} - \hat{\mathbb{E}} \simeq -\frac{48\hbar^8}{(4\pi\lambda_0)^2} \sum_l (2l+1)\Lambda_l^4 + \frac{32\hbar^8}{4\pi\lambda_0} \sum_{\{l_1|\lambda_{l_1}\neq 0\}} \sum_{l_2} \sum_{l_3=|l_1-l_2|}^{l_1+l_2} \lambda_{l_1}^{-1} \Lambda_{l_2} \Lambda_{l_3}^2 (2\Lambda_{l_3} + \Lambda_{l_2})$$

$$\times \frac{(2l_1+1)(2l_2+1)(2l_3+1)}{4\pi} \begin{pmatrix} l_1 & l_2 & l_3 \\ 0 & 0 & 0 \end{pmatrix}^2 \sum_{m_1=-l_1}^{l_1} \sum_{m_2=-l_2}^{l_2} \begin{pmatrix} l_1 & \Big| & l_2 & \Big| & l_3 \\ m_1 & \Big| & m_2 & \Big| & -m_1-m_2 \end{pmatrix}^2. \qquad (G35)$$

While we have not indicated it, the sum over $l_3$ may be further restricted to those values for which $l_1 + l_2 + l_3$ is even. This follows from (G17).

To end this section, we consider a simple example: a (non-PTA) detector which is sensitive to only a single harmonic mode ($Q_l = \delta_{lL}$), in the absence of pulsar terms ($\eta = 0$). For this case, the above expressions for $\mathbb{E}$ and $\hat{\mathbb{E}}$ simplify considerably. Using (G7), (G10), and (G12), we obtain

$$\Lambda_l = \frac{4\pi}{2L+1}\delta_{lL} \quad \text{and} \quad \lambda_l = 8\pi\hbar^4 \frac{D_l}{2l+1} = 8\pi\hbar^4 \begin{pmatrix} l & L & L \\ 0 & 0 & 0 \end{pmatrix}^2 \Rightarrow \lambda_0 = \frac{8\pi\hbar^4}{2L+1}. \qquad (G36)$$

Symmetry properties of the above Wigner 3j symbol then imply that $\lambda_l \neq 0$ only if $l = 0, 2, \ldots, 2L$. Thus, the total number of nonzero eigenvalues of the covariance matrix $\mathbb{C}_{ab}$ is given by $N_{\text{DOF}} = (2L+1)(L+1)$.

Substituting the expressions in (G36) for $\Lambda_l$ and $\lambda_l$ into (G28), (G29), (G30), and (G35) leads to the following simpler expressions:

$$\mathbb{E} \simeq 4\left(\frac{\hbar}{\hbar}\right)^8 \sum_{l_1=0}^{L} \sum_{l_2=0}^{L} (4l_1+1)(4l_2+1)$$

$$\times \sum_{m_1=-2l_1}^{2l_1} \sum_{m_2=-2l_2}^{2l_2} \sum_{m_3=-L}^{L} \left[ 2 \begin{pmatrix} 2l_1 & \Big| & L & \Big| & L \\ m_1 & \Big| & m_3 & \Big| & -m_1-m_3 \end{pmatrix}^2 \begin{pmatrix} 2l_2 & \Big| & L & \Big| & L \\ m_2 & \Big| & -m_1-m_3 & \Big| & m_1-m_2+m_3 \end{pmatrix}^2 \right.$$

$$+ (-1)^{m_1+m_2} \begin{pmatrix} 2l_1 & \Big| & L & \Big| & L \\ m_1 & \Big| & m_3 & \Big| & -m_1-m_3 \end{pmatrix} \begin{pmatrix} 2l_1 & \Big| & L & \Big| & L \\ m_1 & \Big| & m_2+m_3 & \Big| & -m_1-m_2-m_3 \end{pmatrix}$$

$$\left. \times \begin{pmatrix} 2l_2 & \Big| & L & \Big| & L \\ m_2 & \Big| & m_3 & \Big| & -m_2-m_3 \end{pmatrix} \begin{pmatrix} 2l_2 & \Big| & L & \Big| & L \\ m_2 & \Big| & m_1+m_3 & \Big| & -m_1-m_2-m_3 \end{pmatrix} \right], \qquad (G37)$$

and

$$\mathbb{E} - \hat{\mathbb{E}} \simeq \frac{12}{2L+1} \left(\frac{\mathfrak{h}}{\hbar}\right)^8 \left[ -1 + 2\sum_{l_1=0}^{L}(4l_1+1) \sum_{m_1=-2l_1}^{2l_1} \sum_{m_2=-L}^{L} \begin{pmatrix} 2l_1 & L & L \\ m_1 & m_2 & -m_1-m_2 \end{pmatrix}^2 \right], \qquad (G38)$$

where we "doubled" the indices $l_1$ and $l_2$ since the summand vanishes if either of these is odd.

By evaluating the rhs of (G37) and (G38), we find

$$\mathbb{E} \simeq 2(2L+1)(4L^2+9L+6)\left(\frac{\mathfrak{h}}{\hbar}\right)^8 \quad \text{and} \quad \mathbb{E} - \hat{\mathbb{E}} \simeq 12\left(\frac{4L^2+6L+1}{2L+1}\right)\left(\frac{\mathfrak{h}}{\hbar}\right)^8. \qquad (G39)$$

These are amazingly simple formulas, considering the summations of products of the various Wigner 3j symbols that enter (G37) and (G38). We have not found an analytical approach to evaluate the sums leading to (G39).

## APPENDIX H: SKY LOCATIONS OF PULSARS CURRENTLY MONITORED BY PTA COLLABORATIONS

Table IV gives the sky locations of the 88 pulsars currently monitored by the European Pulsar Timing Array (EPTA) collaboration, the North American Nanohertz Observatory for Gravitational Waves (NANOGrav) collaboration, and the Parkes Pulsar Timing Array (PPTA) collaboration [37]. We used these to create Figs. 3, 8, and 9, and for constructing Tables I and II: see Secs. V B, VII C, and VIII C for details. The sky locations are in equatorial coordinates, with right ascension (ra) and declination (dec) given in degrees. They are related to spherical polar coordinates (in radians) via $\theta = \pi/2 - (\pi/180°)\text{dec}$ and $\phi = (\pi/180°)\text{ra}$.

TABLE IV. The 88 pulsars currently employed by the European Pulsar Timing Array, the North American Nanohertz Observatory for Gravitational Waves, and the Parkes Pulsar Timing Array collaborations. The final column identifies which PTAs employ this pulsar. The sky location is given by right ascension and declination in degrees. The International Pulsar Timing Array (IPTA) pulsars are the full list.

| Pulsar name | ra (deg) | dec (deg) | PTA |
|---|---|---|---|
| J0023 + 0923 | 5.8 | 9.4 | N |
| J0030 + 0451 | 7.6 | 4.9 | EN |
| J0034 − 0534 | 8.6 | −5.6 | E |
| J0218 + 4232 | 34.5 | 42.5 | E |
| J0340 + 4130 | 55.1 | 41.5 | N |
| J0406 + 3039 | 61.6 | 30.7 | N |
| J0437 − 4715 | 69.3 | −47.3 | NP |
| J0509 + 0856 | 77.3 | 8.9 | N |
| J0557 + 1551 | 89.4 | 15.8 | N |
| J0605 + 3757 | 91.3 | 38.0 | N |
| J0610 − 2100 | 92.6 | −21.0 | EN |

*(Table continued)*

TABLE IV. *(Continued)*

| Pulsar name | ra (deg) | dec (deg) | PTA |
|---|---|---|---|
| J0613 − 0200 | 93.4 | −2.0 | ENP |
| J0614 − 3329 | 93.5 | −33.5 | N |
| J0621 + 1002 | 95.3 | 10.0 | E |
| J0636 + 5128 | 99.0 | 51.5 | N |
| J0645 + 5158 | 101.5 | 52.0 | N |
| J0709 + 0458 | 107.3 | 5.0 | N |
| J0711 − 6830 | 108.0 | −68.5 | P |
| J0740 + 6620 | 115.2 | 66.3 | N |
| J0751 + 1807 | 117.8 | 18.1 | E |
| J0900 − 3144 | 135.2 | −31.7 | E |
| J0931 − 1902 | 142.8 | −19.0 | N |
| J1012 + 5307 | 153.1 | 53.1 | E |
| J1012 − 4235 | 153.1 | −42.6 | N |
| J1017 − 7156 | 154.5 | −71.9 | P |
| J1022 + 1001 | 155.7 | 10.0 | ENP |
| J1024 − 0719 | 156.2 | −7.3 | ENP |
| J1045 − 4509 | 161.5 | −45.2 | P |
| J1125 + 7819 | 171.5 | 78.3 | N |
| J1125 − 6014 | 171.5 | −60.2 | P |
| J1312 + 0051 | 198.2 | 0.9 | N |
| J1446 − 4701 | 221.6 | −47.0 | P |
| J1453 + 1902 | 223.4 | 19.0 | N |
| J1455 − 3330 | 223.9 | −33.5 | EN |
| J1545 − 4550 | 236.5 | −45.8 | P |
| J1600 − 3053 | 240.2 | −30.9 | ENP |
| J1603 − 7202 | 240.9 | −72.0 | P |
| J1614 − 2230 | 243.7 | −22.5 | N |
| J1630 + 3734 | 247.7 | 37.6 | N |
| J1640 + 2224 | 250.1 | 22.4 | EN |
| J1643 − 1224 | 250.9 | −12.4 | ENP |
| J1705 − 1903 | 256.4 | −19.1 | N |
| J1713 + 0747 | 258.5 | 7.8 | ENP |
| J1719 − 1438 | 259.8 | −14.6 | N |
| J1721 − 2457 | 260.3 | −25.0 | E |
| J1730 − 2304 | 262.6 | −23.1 | ENP |
| J1732 − 5049 | 263.2 | −50.8 | P |
| J1738 + 0333 | 264.7 | 3.6 | EN |
| J1741 + 1351 | 265.4 | 13.9 | N |
| J1744 − 1134 | 266.1 | −11.6 | ENP |
| J1745 + 1017 | 266.4 | 10.3 | N |

*(Table continued)*

TABLE IV. (Continued)

| Pulsar name | ra (deg) | dec (deg) | PTA |
| --- | --- | --- | --- |
| J1747 − 4036 | 267.0 | −40.6 | N |
| J1751 − 2857 | 267.9 | −29.0 | EN |
| J1801 − 1417 | 270.5 | −14.3 | E |
| J1802 − 2124 | 270.5 | −21.4 | EN |
| J1804 − 2717 | 271.1 | −27.3 | E |
| J1811 − 2405 | 272.8 | −24.1 | N |
| J1824 − 2452A | 276.1 | −24.9 | P |
| J1832 − 0836 | 278.1 | −8.6 | NP |
| J1843 − 1113 | 280.9 | −11.2 | EN |
| J1853 + 1303 | 283.5 | 13.1 | EN |
| J1857 + 0943 | 284.4 | 9.7 | ENP |
| J1903 + 0327 | 285.8 | 3.5 | N |
| J1909 − 3744 | 287.4 | −37.7 | ENP |
| J1910 + 1256 | 287.5 | 12.9 | EN |
| J1911 + 1347 | 288.0 | 13.8 | EN |
| J1911 − 1114 | 288.0 | −11.2 | E |
| J1918 − 0642 | 289.7 | −6.7 | EN |
| J1923 + 2515 | 290.8 | 25.3 | N |

TABLE IV. (Continued)

| Pulsar name | ra (deg) | dec (deg) | PTA |
| --- | --- | --- | --- |
| J1939 + 2134 | 294.9 | 21.6 | ENP |
| J1944 + 0907 | 296.0 | 9.1 | N |
| J1946 + 3417 | 296.6 | 34.3 | N |
| J1955 + 2908 | 298.9 | 29.1 | EN |
| J2010 − 1323 | 302.7 | −13.4 | EN |
| J2017 + 0603 | 304.3 | 6.1 | N |
| J2019 + 2425 | 304.9 | 24.4 | E |
| J2033 + 1734 | 308.4 | 17.6 | EN |
| J2043 + 1711 | 310.8 | 17.2 | N |
| J2124 − 3358 | 321.2 | −34.0 | ENP |
| J2129 − 5721 | 322.3 | −57.4 | P |
| J2145 − 0750 | 326.5 | −7.8 | ENP |
| J2214 + 3000 | 333.7 | 30.0 | N |
| J2229 + 2643 | 337.5 | 26.7 | EN |
| J2234 + 0611 | 338.6 | 6.2 | N |
| J2241 − 5236 | 340.4 | −52.6 | P |
| J2302 + 4442 | 345.7 | 44.7 | N |
| J2317 + 1439 | 349.3 | 14.7 | EN |
| J2322 + 2057 | 350.6 | 21.0 | EN |

(Table continued)